\documentclass[]{aa}
\usepackage[T1]{fontenc}
\usepackage[utf8]{inputenc}

\usepackage{graphicx}
\usepackage{natbib}
\usepackage{ulem}
\usepackage{textcomp}
\usepackage{gensymb}
\usepackage{longtable}
\usepackage{threeparttable}
\usepackage{multicol}
\usepackage{multirow}
\usepackage{float}
\usepackage{todonotes}
\usepackage[Symbol]{upgreek}
\usepackage{amsmath}
\usepackage{etoolbox}
\usepackage{txfonts}
\usepackage{url}
\usepackage{xcolor}

\usepackage[most]{tcolorbox}
\usepackage{hyperref}
\usepackage{xcolor}
\newcommand{\enzo}{{\it {\small ENZO}}}
\newcommand{\CRaTer}{{\it {\small CRaTer}}}
\newcommand{\MUSIC}{{\it {\small MUSIC}}}

\begin{document}
 
\title{Simulating the transport of relativistic electrons and magnetic fields injected by radio galaxies in the intracluster medium}

\author{F. Vazza\inst{1,2,3}, D. Wittor\inst{2,1}, G. Brunetti\inst{3},  M. Br\"{u}ggen\inst{2}}

\offprints{%
 E-mail: franco.vazza2@unibo.it}
\institute{Dipartimento di Fisica e Astronomia, Universit\'{a} di Bologna, Via Gobetti 93/2, 40122, Bologna, Italy
\and  Hamburger Sternwarte, University of Hamburg, Gojenbergsweg 112, 21029 Hamburg, Germany
\and Istituto di Radioastronomia, INAF, Via Gobetti 101, 40122, Bologna, Italy}

\authorrunning{F. Vazza, D. Wittor, G. Brunetti \& M. Br\"{u}ggen}
\titlerunning{Simulated Radio Galaxies in ENZO}

\date{Accepted ???. Received ???; in original form ???}

\abstract{Radio galaxies play an important role in the seeding of cosmic rays and magnetic fields in galaxy clusters. 
Here, we simulate the evolution of relativistic electrons injected into the intracluster medium by radio galaxies. Using passive tracer particles added to magnetohydrodynamical adaptive-mesh simulations, we calculate the evolution of the spectrum of relativistic electrons taking into account energy losses and re-acceleration mechanisms associated with the dynamics of the intracluster medium. 
Re-acceleration can occur at shocks via diffusive shock acceleration, and in turbulent flows via second-order Fermi re-acceleration.
This study confirms that relativistic electrons from radio galaxies can efficiently fill the intracluster medium over scales of several $100 \rm ~Myr$, and that they create a stable reservoir of fossil electrons that remains available for further re-acceleration by shock waves and turbulent gas motions. Our results also show that late evolution of radio lobes and remnant radio galaxies is significantly affected by the dynamics of the surrounding  
intracluster medium. Here the diffusive re-acceleration couples the evolution of relativistic particles to the gas perturbations. In the near future, deep radio observations, especially at low frequencies, can probe such mechanisms in galaxy clusters.}
\maketitle

\label{firstpage}

 \section{Introduction}\label{sec::intro}
Galaxy clusters are the largest reservoir of relativistic plasma in the Universe, in which relativistic particles\footnote{In this paper we will use the terms relativistic particles and cosmic-ray particles interchangeably.} are subject to acceleration and energy losses and are confined by turbulent magnetic fields \citep[e.g.][]{bbp97,sa99,bj14,Bykov19}.
Gamma-ray observations \citep[e.g.][]{fermi14} and radio observations \citep[e.g.][]{2019SSRv..215...16V} have provided important constraints on the complex life cycle of relativistic plasma in the intracluster medium (ICM).

Radio-bright, bipolar outflows from active galactic nuclei (AGN) are commonly found in clusters of galaxies.
Radio galaxies play an important role as they have the potential to release large quantities
of relativistic plasma and magnetic fields into the ICM 
 \citep[e.g.][]{volk99}.
Many recent numerical simulations have investigated the dynamics of relativistic jets as they expand into the ambient medium, and the role of magneto-hydrodynamical (MHD) instabilities in determining the jet morphology and stability \citep[e.g.][]{1982A&A...113..285N,1998A&A...333.1117B,2007MNRAS.382..526P,2005SSRv..121...21M,2010MNRAS.402....7M,2014MNRAS.443.1482H,2016A&A...596A..12M,2020arXiv200913540B}.

Cosmological simulations predict that radio-mode feedback is crucial for shaping the thermodynamic properties of the gas in galaxy groups and clusters at low redshifts \citep[e.g.][]{2008ApJ...687L..53P,2009MNRAS.398...53B,mcc10,2010MNRAS.401.1670F,2017MNRAS.470.1121T}. At higher redshifts ($z \geq 2-3$) most of the mass growth of supermassive black holes (SMBH) occurs in the radiatively efficient quasar mode, with an accretion rate larger than a few percent of the Eddington rate \citep[e.g.][]{2007MNRAS.380..877S}.

Observationally, the duty cycle of radio galaxies in the ICM is still poorly constrained. Observations of nearby massive galaxy clusters by \citet{2020MNRAS.496.2613B} concluded that only about half of such systems show clear evidence for radio-mode AGN feedback. Several deep LOFAR observations have discovered  steep-spectrum, filamentary and distorted radio structures, often connecting old tails of radio galaxies with diffuse radio emission across a wide range of spatial scales
\citep[]{2018MNRAS.473.3536W,2020ApJ...897...93B,2020A&A...634A...4M}. Very recent deep JVLA observations \citet{2021arXiv210105305G} have also unveiled complex substructures in the wake of radio galaxy NGC 1272 in the Perseus clusters. Its morphology and spectrum support the possibility that fossil plasma ejected by AGN jets is being re-accelerated by shear and compressive motions in the ICM and that the fossil plasma fuels the the mini-halo in the Perseus cluster. 

The emission from tailed radio galaxies is sometimes linked to diffuse radio emission from the ICM, both in the form of radio halos \citep[][]{2018MNRAS.473.3536W} or radio relics \citep[e.g.][]{2005ApJ...627..733M,2014ApJ...794...24O,2014ApJ...785....1B,2017NatAs...1E...5V,2019MNRAS.489.3905S}.  In most cases, a volume-filling distribution of fossil relativistic electrons is required to explain the observed radio power \citep[e.g.][]{gb01,2005ApJ...627..733M,ka12, 2013MNRAS.435.1061P,va14relics,va15relics, bj14,2018JKAS...51..185K,2020A&A...634A..64B}. The recent ASKAP/EMU observation of the merging A3391-3395 pairs of galaxy clusters has also detected a three times higher sky density of giant radio galaxies \citep[][]{mb21} than previously assumed. This suggests that radio jets can have a larger role in seeding cosmic rays far from their source.

Radio galaxies can also play a role in the genesis of magnetic fields in galaxy clusters \citep[][]{Kronberg..1999ApJ,Furlanetto&Loeb..ApJ2001,xu09}, as they can seed primordial magnetic fields \citep[e.g.][for a recent review]{2020arXiv201010525V}. Outside of the virial regions of clusters, the impact of galaxies on extragalactic magnetic fields is expected to be detectable with the next generation of deep radio surveys, both in total intensity and polarisation  \citep[e.g.][]{va17cqg,2018Galax...6..128L}. 
The dynamical interaction between radio jets and the ICM occurs on a wide range of spatial
scales ($1$~kpc $\leq L \leq 100$~kpc). 
A number of simulations have given valuable insights into the interaction between AGN and the surrounding ICM, mostly based on hydrodynamics and starting from the point where the relativistic plasma has reached approximate pressure
equilibrium with the surrounding ICM and starts to be dominated by buoyancy \citep[e.g.][]{2001ApJ...554..261C,2002Natur.418..301B,mb07,2006MNRAS.373L..65H,mcc2010, gaspari11b,2012MNRAS.427.1614Y,2019ApJ...871....6Y}. 

On the other hand, a few papers have focused on the direct injection of magnetic fields by AGN, and on their impact on the evolution of the ICM. 
In particular,
\citet[][]{xu09} and \cite{2011ApJ...739...77X} included magnetised outflows from radio galaxies in cosmological {\enzo} simulations, and studied the build-up of cluster magnetic fields from the injection of individual AGN jets. 
\citet{2012ApJ...750..166M}, simulated the effect of "cluster weather" on radio lobes by injecting magnetised jets into a galaxy cluster extracted from a (Smoothed Particle Hydrodynamics) cosmological simulation, and resimulating a cluster cutout region with a grid-based MHD method, in order to evolve radio jets and their magnetic properties at high resolution. 
Using the AREPO code, \citet{2020arXiv200812784B} performed high-resolution simulations of jets from AGN, and produced realistic X-ray and radio properties (albeit assuming a distribution of magnetic fields and relativistic electrons in post-processing).

In this new work we devote more computational resources to the simulation of the properties of non-thermal components injected by jets, while neglecting the effect of gas cooling. Our main focus is on the evolution of radio jets, as well as their magnetic fields and relativistic electrons. In particular, we model the evolution of relativistic electrons subject to ageing and acceleration processes. We rely on passive tracers, which in turn, do not allow us to study the interplay of a relativistic plasma with the ICM, as is done in other work \citep[][]{oj10,2012ApJ...750..166M,nolting19a,nolting19b}. 
However, our simulations offer an unprecedented view of the large-scale circulation and advection of fossil electrons on Megaparsec scales, in a regime scarcely affected by radiative cooling and other processes neglected in this work. They also allow us to test several re-acceleration processes together with a realistic model of the evolving ICM.

Our paper is structured as follows:
in Sec.~2 we present the cosmological simulations and numerical methods employed in this paper, while in Sec.~3 we give our results on the evolution of radio sources (3.1), on the impact of radio galaxies on the ICM (3.2), and on the the energy evolution relativistic electrons (3.3). Limitations of our approach are discussed in Sec.~4 and our conclusions are given in Sec.~5. 

 \section{Simulations}\label{sec::simu}

 \subsection{\enzo-simulations}

We used the cosmological \enzo-MHD  \citep{enzo14} code to produce realistic simulations of the formation of a galaxy group and of the thermal and non-thermal feedback from radio galaxies within it. 

We use nested grids to provide a uniform resolution at the highest refinement level. Each simulation covers a root-grid of (50 Mpc$/h$)$^3$ and is sampled with $128^3$ cells. Using \MUSIC \ \citep{music}, four additional nested regions with increasing spatial resolution were nested, until the innermost (4 Mpc/$h$)$^3$
region where the cluster  forms is uniformly covered at a $\Delta x=24.4$~kpc$/h$ uniform resolution. During the simulation, two additional level of mesh refinement were added using a local gas/DM overdensity criterion ($\Delta \rho/\rho \geq 3$), allowing the simulation to reach a maximum resolution of $\approx 8.86$~kpc, typically in the cluster core region.
As a result of our nested grid approach, the mass resolution for dark matter in our cluster formation region is of  $m_{\rm DM}=2.82 \cdot 10^{6} ~M_{\odot}$ per dark matter particle, for the highest resolution particles that are used to fill the innermost AMR level since the start of the simulation. 

The MHD solver employs the Local Lax-Friedrichs (LLF) Riemann solver to compute the fluxes in the Piece-wise Linear Method (PLM).  We initialised a simple uniform magnetic field at $z=50$ with a value of $B_0=0.1  \ \mathrm{nG}$ in each direction. We point to recent work \citep[e.g. Sec 2.1 in][]{wittor20} for further details on this method. \\
All simulations in this work are non-radiative and only differ in the number and timing of episodes of feedback from SMBH particles. 

The jets and relativistic electrons are injected into a galaxy group with a final  virial mass of  $M_{100} \approx 1.5 \cdot 10^{14} M_\odot$ at $z=0.0$, $M_{100} \approx 1.1 \cdot 10^{14} M_\odot$ at $z=0.5$ and $M_{100} \approx 5 \cdot 10^{13} M_\odot$ at $z=1$. The green colors in Fig.~1 help in visualising its main evolutionary events. 
At $z=1$ the group still has to assemble about $2/3$ of its mass, which is later accreted mostly from a group companion merging at $z \sim 0.8$ (from the upper right corner in the panels). Subsequent minor mergers occur across the entire lifetime of the group, while a more prominent second major merger occurs between $z=0.3$ and $z=0.1$, following the accretion of a second massive companion (entering from the right in Fig. 1 at a redshift of $z=0.5$). 

Throughout this paper, we used the following cosmological parameters: $h = 0.678$, $\Omega_{\Lambda} = 0.692$, $\Omega_{\mathrm{M}} = 0.308$ and $\Omega_{\mathrm{b}} = 0.0478$, based on the results from the \citet{2016A&A...594A..13P}.

\begin{table}
\begin{center}
\caption{Input parameters for the SMBH models in our simulations, and reference parameters for the AGN jets, measured as an average at the end of each jet's life ($t_{\rm jet}$).}
\footnotesize
\centering \tabcolsep 2pt
\begin{tabular}{c|c|c}
  Parameter  & Run2 & Run1 \\\hline
 
     $z_{\rm jet}$ & 0.5 & 1.0 \\
      $M_{\rm BH}$  [$\rm M_{\odot}$]& $10^{9} $ & $10^{7}$\\
      $\alpha_{\rm Bondi}$  & $10$ & $10^{4} $\\
      $\dot{M_{\rm BH}}$ [$\rm M_{\odot}/yr$] & $10^{-7}$ & $ 10^{-9}$\\
      $t_{\rm jet}$ [Myr] & $10$ & $200$\\
      $L_{\rm BH}$  [erg/s] &$1.7\cdot 10^{45}$ & $1.7 \cdot 10^{44}$  \\
      $E_{\rm kin,jet}$  [erg]& $2.6 \cdot 10^{56}$ & $5.6 \cdot 10^{57}$  \\
      $E_{\rm th,jet}$  [erg]& $1.2 \cdot 10^{57}$ & $5.1 \cdot 10^{57}$  \\
      $E_{\rm mag,jet}$  [erg]& $1.6 \cdot 10^{56}$ & $3.7 \cdot 10^{57}$ \\
      $B_{\rm av,jet}$  [$\rm \mu G$] & $1.4$ & $3.6$ \\
      $B_{\rm max,jet}$  [$\rm \mu G$] & $3.9$ & $43.2$ \\
      $v_{\rm av,jet}$ [$\rm km/s$] & $630$ & $1267$   \\
  \end{tabular}
  \end{center}
\label{table:tab1}
\end{table}

\begin{figure}
    \centering
    \includegraphics[width=0.495\textwidth]{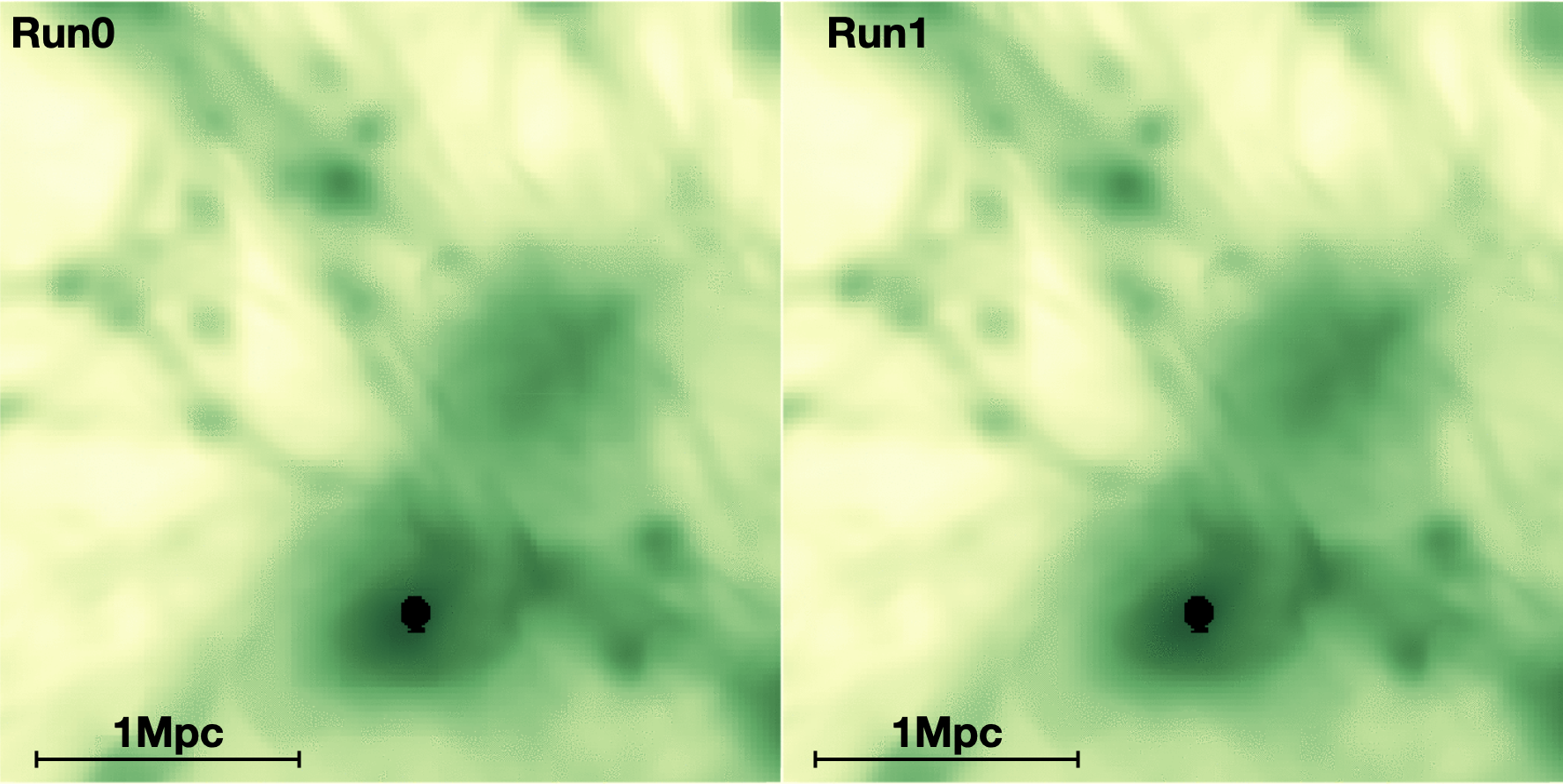}
   \includegraphics[width=0.495\textwidth]{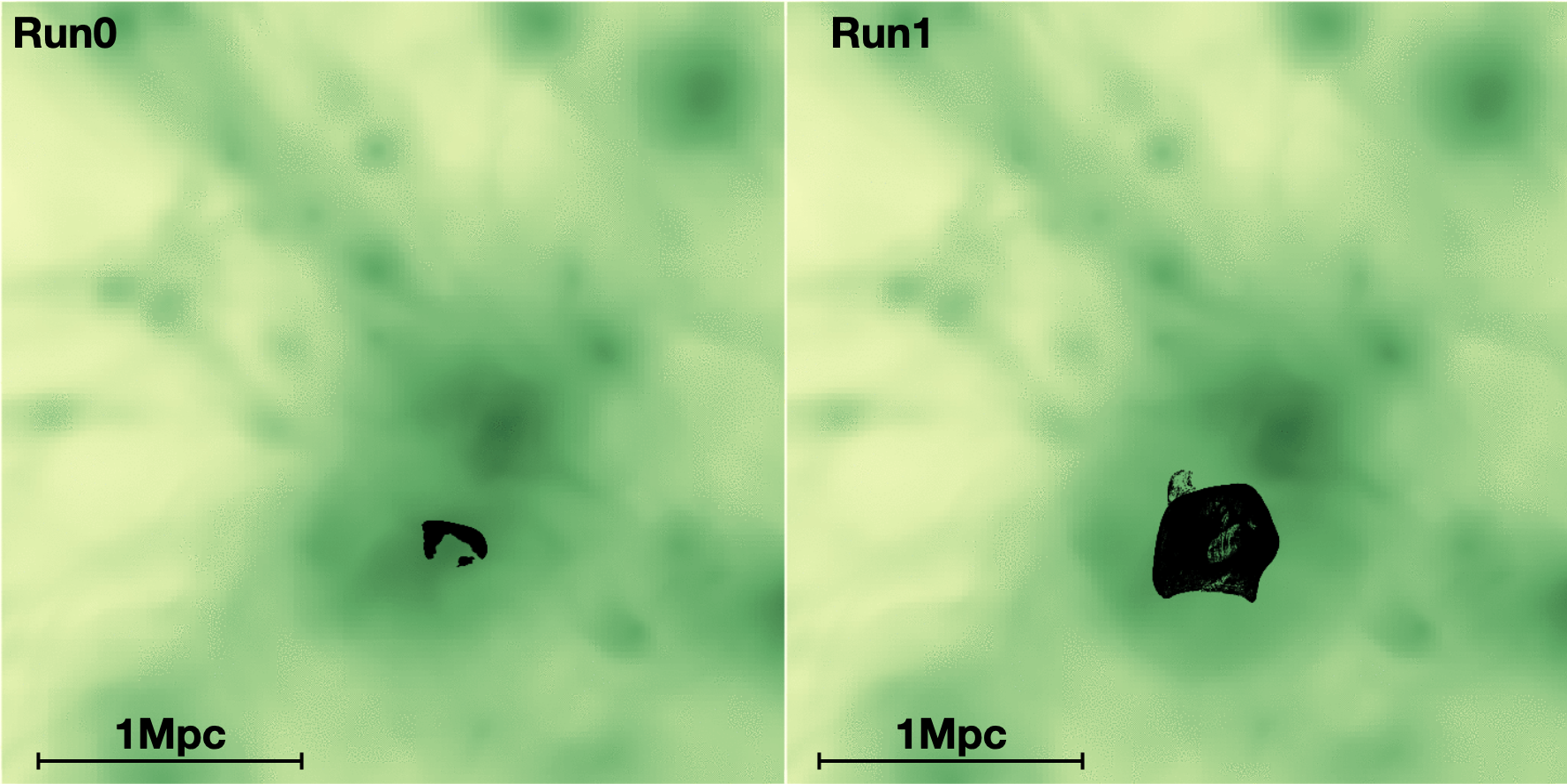}
   \includegraphics[width=0.495\textwidth]{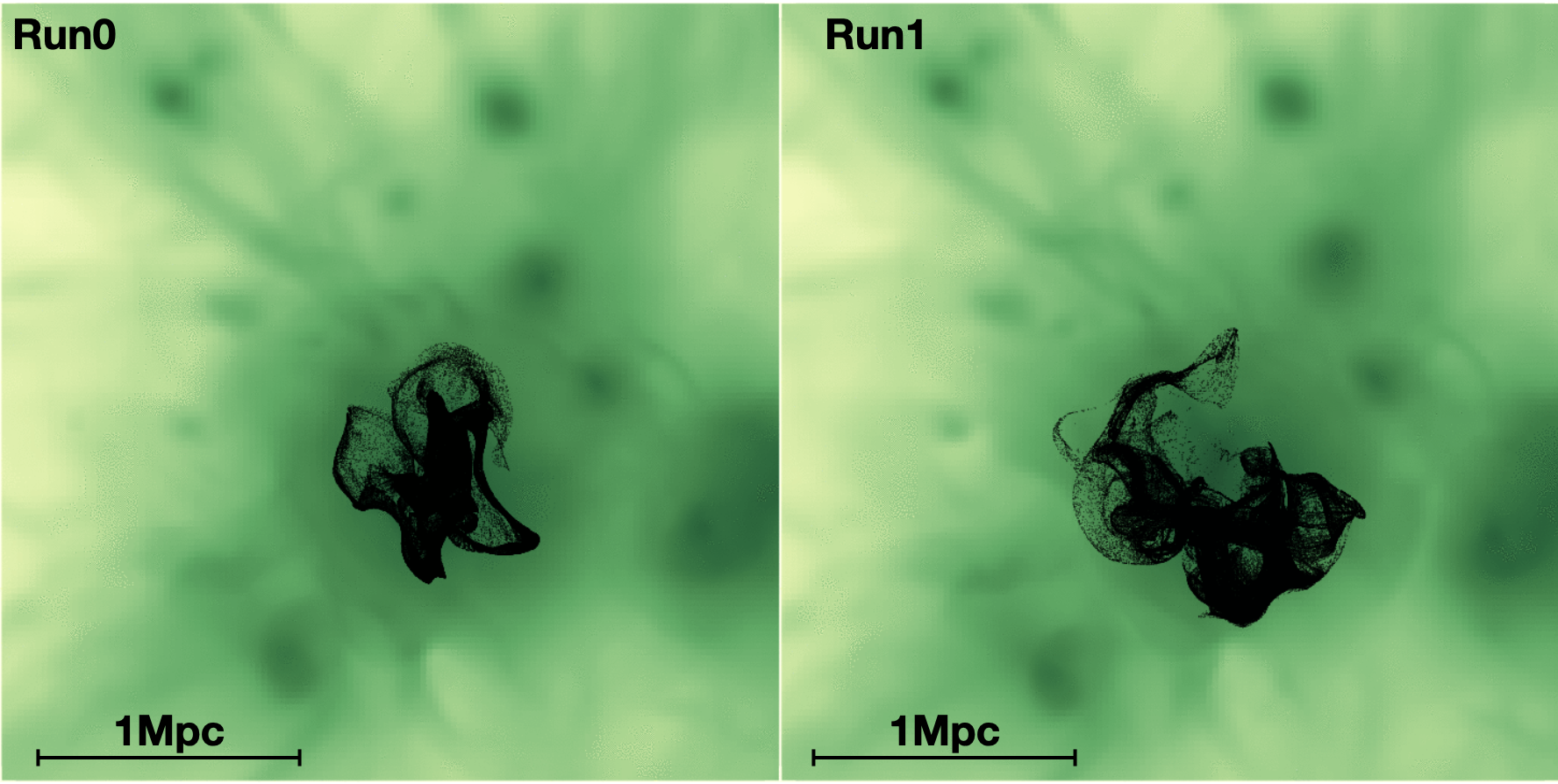}
   \includegraphics[width=0.495\textwidth]{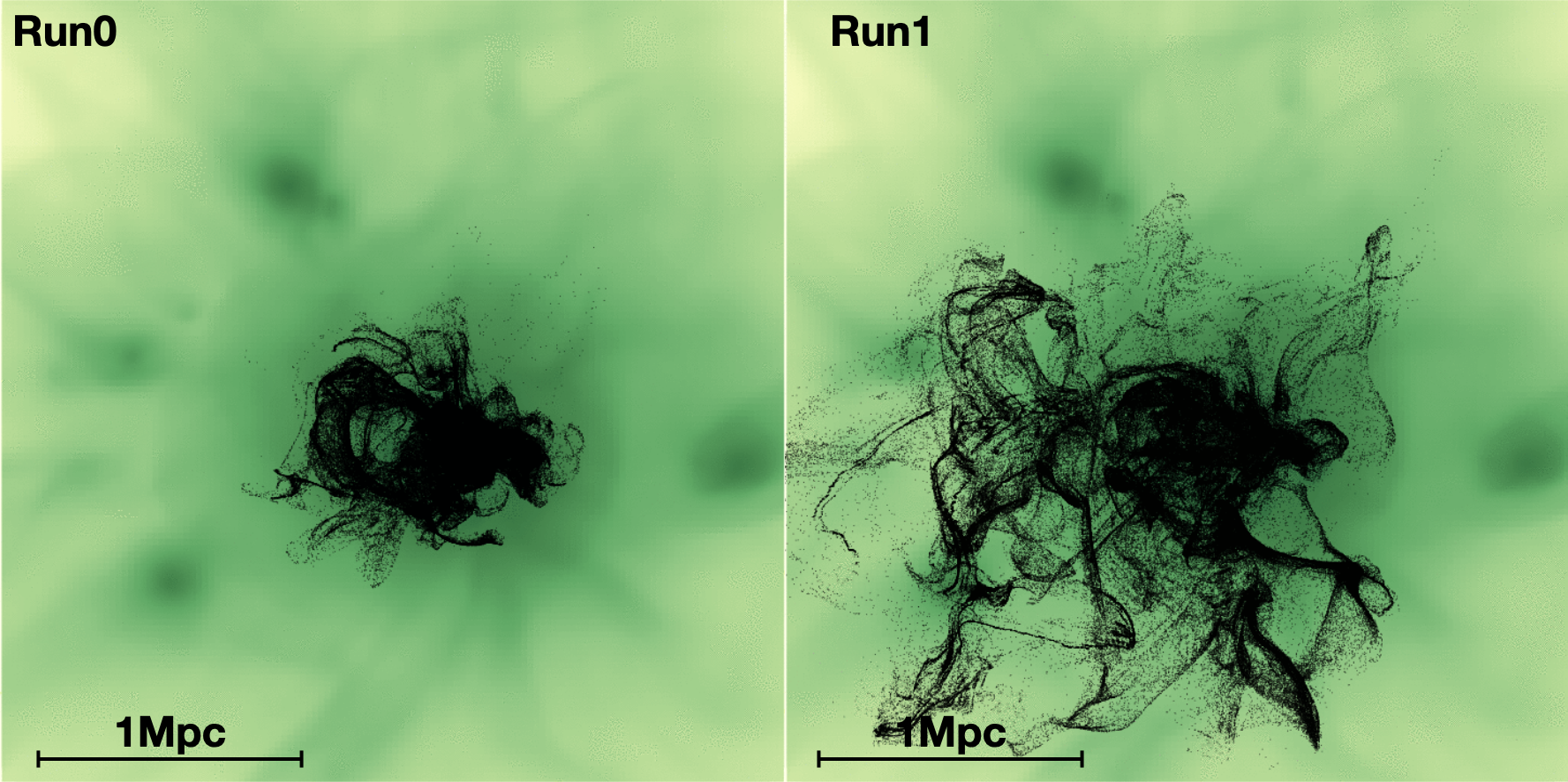}
    
    \caption{Maps of all tracers evolved in our run Run0 (left) and Run1 (right) at $z=1$, $z=0.8$, $z=0.5$ and $z=0.1$. The colours represent the projected gas density (with a $log_{10}$ stretching). The line of sight is parallel to the jet axis, to give a sense of the lateral expansion of tracers relative to the jet (which points towards the observer in this case).}
    \label{fig:tracer_denis}
\end{figure}

\subsubsection{Modelling the growth of supermassive black holes and radio jets in {\enzo}}

The simulation of SMBHs follows the formulation by \citet[][]{2011ApJ...738...54K}, released in the public version of {\enzo} and supplemented with a few ad-hoc prescriptions to include the magnetic feedback from radio jets. Given the absence of radiative cooling, our runs lack a self-regulating mechanism to switch the feedback cycle on and off \citep[e.g.][for a few examples]{gaspari11b,2015ApJ...813L..17R,2019MNRAS.489..802R}. The thermal structure of the ICM around the SMBH is thus not entirely realistic, which also implies that typical parameters for the launching of jets from AGN must be adjusted to reproduce realistic radio galaxies.

After extensive testing of the possibility of generating {\it kinetically dominated} jets as implemented in other simulations \citep[][]{dubois10,gaspari11b}, we resorted to purely {\it thermal} energy feedback around the SMBH. This is motivated by the fact that even the interaction of a fast jet with a $T_{\rm ICM}\geq 10^7 ~\rm K$ cluster core will not produce realistic shock heating and dynamics. This is in conflict with usual models for AGN jets in cool-core atmospheres (where $T \leq 10^6 ~\rm K$ during the jet launching stage e.g. \citealt{2000ApJ...535..650B,2002Natur.418..301B}).
While a pure thermal feedback would  also be in conflict with recent Sunyaev-Zeldovich measurements of AGN-blown cavities that show a substantial amount of non-thermal support inside these cavities \citep[][]{2019ApJ...871..195A}, a large fraction of the feedback energy in our model is also provided by the magnetic and kinetic energy produced by our bipolar magnetic feedback model (see below).
This approach allowed us to relax the otherwise tight constraint on the time-step in our simulation, set by the typically large jet velocities ($v_j \sim 10^4 \rm km/s$) in the vicinity of the SMBH.

In summary, our approach is suitable to investigate the impact of radio galaxies on the gas in large-scale structures, provided that we focus on $\gg 10$~kpc scales and that dynamical features on smaller scales can be neglected.
  

Following the implementation by \citet{2011ApJ...738...54K}, each SMBH particle is allowed to accrete gas, based on the spherical Bondi-Hoyle model.  Due to the lack of spatial resolution, it is necessary to simulate the mass growth of SMBH with ad-hoc parametric formulations. For example, the commonly used Eddington-limited Bondi–Hoyle accretion  generally has to be boosted by an efficiency parameter $\alpha \geq 1$, so that

\begin{equation}
    \dot{M}_{\rm BH}=\rm min\left(\frac{4 \pi \alpha G^2 M_{\rm BH}^2\rho_B}{c_s^3} ; \frac{4 \pi G M_{\rm BH} m_p}{\epsilon_r \sigma_T c}\right),
\end{equation}
where $M_{\rm BH}$ is the SMBH's mass, $c_s$ is the sound speed of the gas 
at the SMBH's location (which we assume to be relative to a fixed $10^6 ~\rm K$ temperature of the accretion disc), $\rho_B$ is the local gas density,  $m_p$ is the proton mass, $\sigma_{\rm T}$ is the Thomson scattering cross-section, and $\epsilon_r$ is the radiative efficiency of the SMBH, so that  $L_{\rm BH} = \epsilon_r \dot{M}_{\rm BH} c^2$ is the bolometric luminosity of the SMBH. 

 Given our resolution, we cannot resolve the Bondi radius and simulate how the SMBH gravitationally influence their surroundings. Neither can we resolve the multi-phase interstellar medium around the host galaxy, given that we do not include cooling and star formation. Hence we adopted a factor, $\alpha_{\rm Bondi}>1$, \citep[e.g.][]{2009MNRAS.398...53B,gaspari12,2016Natur.534..218T} that parametrises the true mass accretion rate onto the SMBH. 
After testing we used two different values of $\alpha_{\rm Bondi}$ depending on the epoch of jet injection, which  also depends on the spatial resolution at the location of SMBH seed particles.  For models in which radio jets start at $z=1$ (Run1), our computational grid is not fully refined at the location of the SMBHs, which combined with the lack of radiative cooling requires a high value for $\alpha$, i.e. $\alpha_{\rm Bondi} \sim 10^3-10^4$, in order to produce jets leading to a realistic radio morphology. 
For the simulation in which our radio jets are launched at $z=0.5$ (Run2) the cluster has largely assembled and the computational grid is refined down to the highest AMR level, at least around the SMBH region. In this case, a modest
boost factor ($\alpha_{\rm Bondi} \sim 5-10$) is found to produce a jet morphology comparable to real radio galaxies in a dense atmosphere, as in, e.g. \citet{2009MNRAS.398...53B}.


  The {\enzo} model for SMBH feedback by \citet{2011ApJ...738...54K} deposits energy at the maximum numerical resolution in the form of {\it thermal} energy. 
 A SMBH particle releases thermal feedback on the surrounding gas, as an extra thermal energy output from each black hole particle, assuming an efficiency $\epsilon_r \epsilon_{\rm BH} = E_{\rm jet}/(\Delta M \Delta t c^2)$  between the accreted mass, $\Delta M$, during the timestep ($\Delta t$)  and the feedback energy $E_{\rm jet}$. The thermal feedback energy is distributed to the neighbouring 27 gas cells around the SMBH. 
We assumed {\enzo}'s default efficiency parameters,  $\epsilon_r=0.1$ and $\epsilon_{\mathrm{BH}}=0.05$ \citep[e.g.][]{2011ApJ...738...54K,enzo14}, which also yielded a good match with observed galaxy clusters scaling relations as found in previous work \citep[][]{va13feedback,va17cqg}.

The SMBHs also inject magnetic fields in the form of magnetic dipoles ($2 \times 2$ cells at the highest resolution level) located at $\pm 1$ cell along the $z$-direction from the SMBH).
This very simple topology is made necessary by the limitation of spatial resolution.  More sophisticated choices, e.g., considering helical magnetic fields (\citealt{2007MNRAS.378..662R}) require higher spatial resolution, i.e. $\leq 0.1 \rm ~kpc$ \citep[e.g.][]{2020ApJ...896...86C}. 
The imposition of a fixed jet alignment along the $z$ axis is not very physical. Yet, the jets are launched only once and the simulation lacks of physical and spatial detail in the gas accretion onto the SMBH. Hence, any launching direction seems equally likely, and therefore we pre-defined a fixed launching axis in order to make the post-processing analysis easier.

The injected magnetic energy is normalised to a fixed fraction of the total feedback energy $E_{\rm B,jet} = \epsilon_{\rm B,jet} E_{\rm jet}$, with $\epsilon_{\rm B,jet}=0.1$.
However, this does not directly correspond to the magnetisation of jets since the thermal energy is released isotropically around the SMBH, while the magnetic energy is directed{\footnote{Note that we oriented the poles always along the same coordinate axis.}}. For this reason, the typical magnetic energy of our jets (see e.g. Tab. 1) is a factor $\sim 10$ larger than the kinetic and thermal energy assigned to cells at the poles of SMBHs, where jets are launched, making them initially {\it magnetically dominated} \citep[e.g.][]{xu09,2019A&A...621A.132M}. 


In this work, we restrict to the results of three runs, and in particular of two different combinations of feedback parameters (beside the reference simulation without jets), which produce plausible radio jet morphologies at $z \sim 1$ (Run 1) and at $z \sim 0.5$ (Run2). The kinetic jet power of these two runs is across the $\sim 10^{45} \rm ~erg/s$ border, which based on correlations derived by radio surveys \citep[e.g.][]{2015ApJ...806...59T,vard20} can be turned into a $L_{1.4} \sim 10^{25} \rm ~W/Hz$ luminosity, i.e. across the standard FR-radio dicotomy. Therefore, our two runs can be seen as a first exploration on the propagation of fossil electrons ejected by FRI (Run 1) and FRII (Run 2) radio galaxies, respectively in a poor group and in a more massive and evolved version of the same system.

A more extensive survey of the effect of sensible variations of feedback parameters (e.g. $\alpha_{Bondi}$, $\epsilon_{B,jet}$, $\epsilon_r$ etc), as well as on the impact of multiple or restarted jets within the same galaxy group will be subject of a follow-up campaign of simulations. 

In summary, the scenarios we explored in this work are: 
\begin{itemize}
    \item {\it Run0}: No feedback from SMBH is activated. This simulation is essentially a standard non-radiative MHD simulation, in which the ICM magnetic fields are the result of compression and, partly, of small-scale dynamo amplification of the initial $B_0=0.1 \rm ~nG$ seed field {\footnote{In comparison with our most resolved simulations in \citet{va18mhd} and \citet{dom19}, the fraction of volume where the Alfv\'en scale is well-resolved is smaller, owing to the less aggressive AMR strategy employed here. Hence, small-scale dynamo amplification  can develop only in a smaller volume}}.
    
     \item {\it Run1}: In this case the SMBH at the centre of the most massive halo in the high-resolution region is activated at redshift $z=1$, with an initial mass of $M_{\rm BH}=10^{7} M_{\odot}$ and remains active for $\approx 200$~Myr, releasing its thermal and magnetic feedback into the ICM. We set $\alpha_{\rm Bondi}=10^4$ to boost the accretion rate. The initial magnetic seed field is the same as in Run0. Hence, the magnetisation of the ICM is the combined result of seed field amplification and AGN feedback.
     
      \item {\it Run2}: as in Run0 until $z=0.5$, then a $M_{\rm BH}= 10^{9} M_{\odot}$ SMBH is placed at the centre of the most massive halo in the high-resolution region (which is inside an already formed galaxy group). The SMBH  releases thermal/magnetic feedback for $\approx 10$ Myr.  As in Run1, the  magnetisation of the ICM stems from the  combination of seed field amplification and AGN feedback. In this case, we use $\alpha_{\rm Bondi}=10$ to  boost the (unresolved) mass accretion rate onto the SMBH.  The initial magnetic seed field is the same as in Run0.
   
\end{itemize}
Table 1 gives more details of the list of parameters describing the jet launching in all runs. These are measured within the jet volume, after a time $t_{\rm jet}$. Possible caveats of our method are discussed in Sec.4.

 \subsection{\CRaTer-simulations of Lagrangian tracer particles}
 We use the Lagrangian code \CRaTer \ to follow the spatial evolution of the cosmic-ray electrons in our runs. In previous work, \CRaTer \ has been used to study cosmic rays and turbulence in galaxy clusters \citep[e.g.][]{wi16,wi17,wi17b}. For details of the implementation, we point to these references.\\

In post-processing, we injected $\sim 2 \cdot 10^6$ particles in all runs, starting from the highest-density peaks of the simulation at  $z=1$ in Run1 (and Run0) and $z=0.5$ in Run2 (and Run0), with a mass resolution $m_{\rm trac}=5\cdot 10^5 ~M_{\odot}$, and evolved them using all snapshots of the simulation ($\sim 120$ for Run1 and $\sim 100$ for Run2). The various grid quantities, e.g. density, velocity, are assigned to the tracers using a cloud-in-cell (CIC) interpolation method. Further details on the full procedure implemented to advect tracers in our simulations, in order to properly sample the advection of gas matter performed by the {\enzo} (Eulerian) calculation, we refer to \citet{wiPHD} and \citet{wi16} .

Fig.~\ref{fig:tracer_denis} shows the spatial evolution of all tracers in runs Run0 and Run1, where tracers were initialised at $z=1$. The maps are projected along the jet axis to show the dispersal of tracers in directions perpendicular to the jets. Although the large-scale dynamics of the cluster is the same in both runs, the impulsive feedback from the radio galaxy, even with a single event, appears to increase the 
filling factor of tracer particles as a function of time (see a longer discussion in Sec. 3.3). 

While all tracers (also in Fig.~\ref{fig:scatter}) were injected at the highest density peaks, only $\sim 10^5$ tracers were used to compute the radio spectra. Those tracers were selected based on the values of magnetic field strength ($\geq 10 \rm ~\mu G$) and the magnitude of the velocity with respect to the SMBH ($|v| \geq 600 ~\rm km/s$) at the moment of the first injection of AGN jets, so that only the gas directly entrained by jets is initially enriched with cosmic rays.

After their injection, we used a temperature-jump based shock finder to detect when tracer particles traverse a shock \citep[Sec. 2.2 in][]{wi17}. For each detected shock they compute the Mach number according to:
  \begin{align}
  M = \sqrt{\frac{4}{5} \frac{T_{\mathrm{new}}}{T_{\mathrm{old}}} \frac{\rho_{\mathrm{new}}}{\rho_{\mathrm{old}}} + \frac{1}{5}}.
 \end{align}
 
 Our tracer particles also keep track of the local fluid divergence, $\nabla \cdot \vec{v}$, and of the fluid vorticity, $\nabla \times \vec{v}$, which serve as proxies for the local turbulence experienced by the tracer particles. Hence, we can estimate the effects of compression/rarefaction as well as turbulent re-acceleration on the electron energy distribution (Sec.~\ref{subsec:fokker}). In particular, we used the gas vorticity to estimate the solenoidal turbulence experienced by particles, $\sigma_v= |\nabla \times \vec{v}| l_{\rm scale}$, where for simplicity we used the same fixed reference scale of $l_{\rm scale} = 27$~kpc used to compute vorticity via finite differences (i.e. 3 cells on the high-resolution mesh). The vorticity is used to compute the turbulent re-acceleration model outlined below, as well as the (solenoidal) turbulent kinetic energy flux. In our model for turbulent re-acceleration (Sec. 2.3) the kinetic energy flux is the key input. We assume that this is {\it constant} across the Kolmogorov turbulent cascade, provided the turbulence injection scale is always $\geq l_{\rm scale}$. The assumption of a turbulent cascade close to a Kolmogorov model, and of an injection scale $\geq 27$ kpc is supported by our previous simulations of turbulence in the ICM \citep[e.g.][]{va17turb,wi17b,2020MNRAS.495..864A}. 
 For more recent applications of passive tracers to study the evolution of cosmic rays in galaxy clusters or turbulence in the ICM we refer the reader to \citet{wittor20} and \citet{2020MNRAS.498.4983W}.\\

\begin{figure*}
    \centering
    \includegraphics[width=0.995\textwidth]{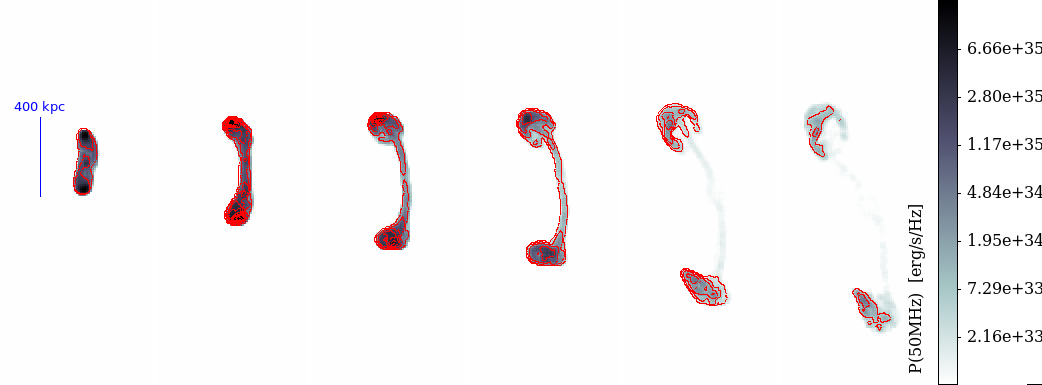}
    \caption{Evolution from $z=0.49$ to $z=0.39$ of the radio emission (50 MHz) from jets in our Run2. The red contours (spaced in square root of the emission) are drawn to better visualise the location of emission peaks in the jet structure. The colorbar gives the emitted power per pixel ($=8.86 \rm ~kpc$ comoving).}
    \label{fig:map_jet_Run2}
\end{figure*}

\begin{figure*}
    \centering
    \includegraphics[width=0.995\textwidth]{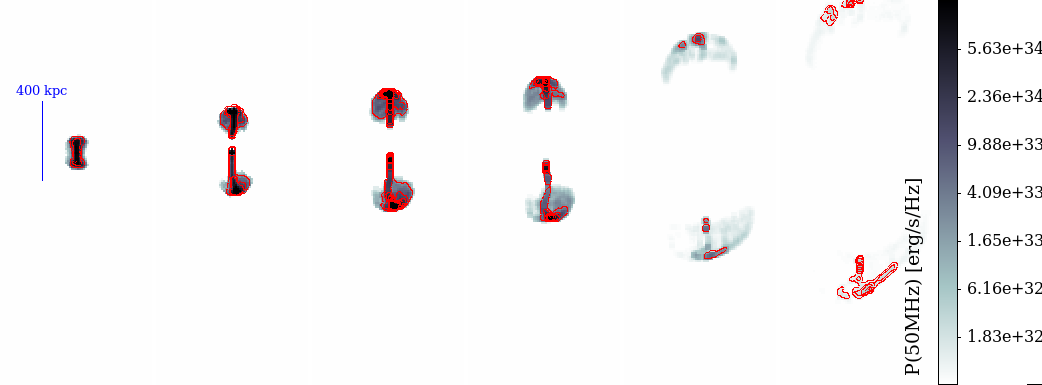}
    \caption{Evolution from $z=0.99$ to $z=0.89$ of the radio emission (50 MHz) from jets in our Run1. The red contours (spaced in square root of the emission) are drawn to better visualise the location of emission peaks in the jet structure. The colorbar gives the emitted power per pixel ($=8.86 \rm ~kpc$ comoving).  }
    \label{fig:map_jet_Run1}
\end{figure*}

\subsection{Simulating the evolution of electron spectra}
\label{subsec:fokker}

We solve the time-dependent diffusion-loss equation of relativistic electrons represented by tracer particles, using the standard \citet{1970JCoPh...6....1C} finite-difference scheme, implemented in the programming language Julia (https://julialang.org). We typically used $N_b=4000$ equal energy bins in the $p_{\rm min} \leq p \leq p_{\rm max}$ momentum range (with $P=\gamma m_e v $ and $p=P/(m_e c)$ is the normalised momentum of electrons and  $p_{\rm min}=1$ and $p_{\rm max}=2 \cdot 10^4$ (hence $dp=5$).

Considering the reduced Fokker-Planck equation without injection and escape terms (i.e. Liouville equation), the evolution of the number density of relativistic electrons as a function of  momentum $N(p)$, can be computed separately for each tracer particle:

\begin{equation}
    {{\partial N}\over{\partial t}}
    =
    {{\partial}\over{\partial p}} \left[
    N \left( \left|{{p}\over{\tau_{\rm rad}}}\right| + \left|{{p}\over{\tau_{\rm c}}}\right| +
    {{p}\over{\tau_{\rm adv}}} - \left|{{p}\over{\tau_{\rm acc}}}\right| \right) \right]
    + {{\partial^2}\over{\partial p}} \left( D_{pp} N \right) ,
    \label{eq11}
\end{equation}
where $D_{\rm pp}$ is the particle diffusion coefficient in momentum space. 
We can thus define 

\begin{equation}
    \dot{p} \approx  \left|{{p}\over{\tau_{\rm rad}}}\right| + \left|{{p}\over{\tau_{\rm c}}}\right| +
    {{p}\over{\tau_{\rm adv}}} - \left|{{p}\over{\tau_{\rm acc}}}\right| .
    \label{eq12}
\end{equation}

\noindent

In the following, we neglect the stochastic term, $D_{\rm pp}$, in Eq. \ref{eq11}, the numerical solution is obtained with the \citet{1970JCoPh...6....1C} finite difference scheme: 

\begin{equation}
N(p,t+dt)=\frac{{{N(p,t)}/{dt}} + N(p+dp,t+dt){{p}}} 
{1/dt + {{p}}/dp} + Q_{\rm inj} ,
    \label{eq13}
\end{equation}
where in the splitting-scheme for finite differences we assumed $N(p +dp/2)=N(p+dp)$ and $N(p-dp/2)=N(p)$, and $Q_{\rm inj}$ accounts for the injection by radio galaxies or shocks, whenever present, regarded as an instantaneous process owing to the timescales much shorter than the time step of our integration (see Eq. \ref{eq:tDSA} below). 

With this simple approach to solve Fokker-Planck equations, we can only model the {\it systematic} acceleration by turbulent re-acceleration. In order to model the {\it stochastic} acceleration, which can lead to higher electron energies, one has to resort to more complex solvers \citep[e.g.][for an application to numerical simulations]{2014MNRAS.443.3564D}.

In order to quantify the effects of different mechanisms on the observed properties of radio emission,  we will compare the outcome of three models applied to our simulated electron evolution:
\begin{itemize}
    \item a model in which relativistic electrons get injected by radio jets (assuming an initially an unbroken power-law energy distribution with slope $\delta_r=-2.0$) and are subsequently only subject to energy losses as they are advected in the ICM;
    \item a model in which, beside the injection of relativistic electrons by jets and their continuous cooling during advection, new relativistic electrons can be injected at shocks waves (via the $Q_{\rm inj}$ term in Eq. \ref{eq13}), as well as by considering the re-acceleration by weak shocks ($\mathcal{M} \leq 3$);
    \item a model in which, beside the two acceleration mechanisms considered above and the continuous cooling, electrons can also be re-accelerated by the turbulent re-acceleration via second-order acceleration (Eq. \ref{eq:ASA}).
\end{itemize}

By comparing the radio properties of cosmic-ray electrons in these three models, we will infer the most likely distribution of fossil electrons, injected by radio galaxies.

The main routines to solve for the evolution of relativistic electrons are written and parallelized in the Julia (v1.4.0) language. This allows us to simultaneously evolve three different realisations of the momentum spectra over $N_b=4000$ bins for $\sim 10^5$ tracers with $\sim 100 ~\rm s/step$, using 16 threads on 8 Intel I9 cores.{\footnote{
A serial version of the code, containing all most important features and a sample sequence of tracer particle data is publicly available here: \url{https://github.com/FrancoVazza/JULIA/tree/master/CR_solver\_pub}.}}
\subsubsection{Loss Terms}

The loss timescales for the radiative, Coulomb and expansion (compression) processes are given by the following formulae, adapted from \citet{bj14}: 

\begin{equation}
    \tau_{\rm rad} =\frac
    {7720 \rm ~Myr} {(\gamma/{300})\left[\left(\frac{B}{3.25 \rm \mu G}\right)^2 + (1+z)^4\right]} , 
    \label{eq:ic}
  \end{equation}

\begin{equation}
    \tau_{\rm c} =
    7934 \rm ~Myr \left\{ {{n/10^{-3} \over{{\gamma/300}}}}
    \left( 1.168 + {1 \over {75}}ln \left( {{\gamma/300 }\over{ n/10^{-3} }} \right) \right) 
\right\}^{-1}    
\label{eq:coulomb}
\end{equation}

and

\begin{equation}
    \tau_{\rm adv} = \frac{951 \rm ~Myr}{ 
    \nabla \cdot \vec{v}/10^{-16}} ,
\label{eq:adv}
\end{equation}

\noindent
where $n$ measured in [$\rm cm^{-3}$], $B$ in [$\rm \mu G$] and $\nabla \cdot \vec{v}$ in [$1/s$].  
We neglect bremsstrahlung losses since their timescale is significantly larger than the ones of all other loss channels for the ICM physical condition considered here. 

\subsubsection{Gain Terms - Shock acceleration}

Concerning energy gains for relativistic electrons, we considered the contribution from
Fermi-I type acceleration from shocks (i.e. via Diffusive Shock Acceleration) and from Fermi-II type acceleration (i.e. from  adiabatic-stochastic-acceleration).
The shock kinetic energy flux that is converted into the acceleration of cosmic rays \citep[e.g.][]{ry03}, is given by 

\begin{equation}
\Psi_{\rm CR} = \xi_e ~\eta(\mathcal{M}) \frac{\rho_u V_s^3 dx_t^2}{2} ,
\end{equation}
where $\rho_u$ is the pre-shock gas density,  $V_s$ is the shock velocity and the combination $\xi_e ~\eta({\mathcal{M}})$ gives the cosmic-ray acceleration efficiency, which comprises a prescription for the energy going into cosmic rays, $\eta(\mathcal{M})$, and the ratio of acceleration rates of electrons to that of protons, $\xi_e$.  For convenience of implementation,  we use the polynomial approximation given by \citet{kj07} for $\eta(\mathcal{M})$, which includes the effects of finite Alfv\'{e}n wave drift and wave energy dissipation in the shock precursor. $dx_t^2$ is the surface associated to each tracers, which we can compute considering that $dx_t^3 = dx^3/n_{\rm tracers}$ is the initial volume associated to every tracer at the epoch of their injection ($n_{\rm tracer}$ being the number of tracers in every cell) and $dx_t(z)^3=dx_t^3 \cdot \rho_t/\rho(z)$ is the relative change of the volume associated to each tracer as a function of $z$, based on the ratio between the density at injection, $\rho$ and the density of cells where each tracer sits as a function of redshift, $\rho(z)$.

For the electron-to-proton ratio, $\xi_e \sim 10^{-2}$ would be commonly used to model 
strong supernova remnant shocks  \citep[e.g.][]{2007Natur.449..576U}. 
However, the exact ratio is extremely uncertain for weak
shocks, and several sophisticated particle-in-cell simulations have been performed over the last years to investigate this \citep[e.g.][]{2011ApJ...733...63R,guo14a,guo14b,2020ApJ...897L..41X}. In order to compare to previous work on fossil electrons, we strictly follow the previous work by \citet{2013MNRAS.435.1061P} and link the injection of fresh electrons to the possible one protons injected by Diffusive Shock Acceleration (DSA),  by requiring an equal number density of cosmic-ray electrons and protons above a fixed injection momentum, which yields  $\xi_e=(m_p/m_e)^{(1-\delta_{\rm inj})/2}$. This approach gives $\xi_e \sim 10^{-2}$ for an injection spectral index of $\delta_{\rm inj} \approx 2.3$, in line with the injection spectral index of local Galactic supernova remnants. 
For simplicity, we also neglect a possible dependence of the shock acceleration efficiency on the local magnetic field topology, which could change the cosmic ray content of galaxy clusters \citep[e.g.][]{wittor20,Banfi20,guo14a,Xie2020}. 

Following the DSA thermal leakage model, we assume that the injection momentum  $P_{\rm inj}$ is a multiple of the thermal momentum of particles, i.e. $P_{\rm inj}= \xi P_{\rm th}$ ($P_{\rm th}=\sqrt{2 k_b T_d m_p}$). Here, we obtain 
$\xi$ based on the fit formula provided by \citet{kr11} for their 1-dimensional convection-diffusion simulations:  

\begin{equation}
\xi=1.17 \frac{m_p v_d}{p_{\rm th}} \cdot \left(1+\frac{1.07}{\epsilon_B}\right)\left(\frac{\mathcal{M}}{3}\right)^{0.1}    
\end{equation}
in which $v_d$ is the downstream shock velocity and $\epsilon_B$ is the ratio between the downstream magnetic field strength ($B_0$) generated by the shock, and is the large scale magnetic field perpendicular to the shock normal ($B_\perp$).
We fixed  $\epsilon_B=0.23$, consistent with \citet{2013MNRAS.435.1061P},  which gives an injection parameter $\xi \sim 2.5-3.5$ for the shocks in our simulation.

We inject relativistic electrons with a momentum distribution that follows a power-law \citep[e.g.][]{1962SvA.....6..317K,sa99}:
\begin{equation}
    Q_{\rm inj}(p) = K_{\rm inj} ~p^{-\delta_{\rm inj}} \left(1-\frac{p}{p_{\rm cut}}\right)^{\delta_{\rm inj}-2} ,
\end{equation}
in which the initial slope of the input momentum spectrum, $\delta_{\rm inj}$, follows from the standard Diffuse Shock Acceleration prediction, $\delta_{\rm inj} = 2 (\mathcal{M}^2+1)/(\mathcal{M}^2-1)$.  
$p_{\rm cut}$ is the cut-off momentum, which is the defined for every shocked tracer as the maximum momentum, beyond which the radiative cooling timescale gets shorter than the acceleration timescale, $\tau_{\rm DSA}$:

\begin{equation}
\tau_{\rm DSA} = \frac{3~D(E)}{V_s^2} \cdot \frac{r(r+1)}{r-1} ,
\label{eq:tDSA}
\end{equation}
where $r$ is the shock compression factor and $D(E)$ is the electron diffusion coefficient as a function of energy \citep[e.g.][]{gb03}.
The specific energy-dependent value of $D(E)$ is poorly constrained and it depends on the local turbulent conditions of the plasma undergoing shocks. It is however critical to set the maximum energy that can reached by shock acceleration \citep[e.g.][]{ka12}. Nevertheless, this is not an issue for our work, as all plausible choices of $D(E)$ in Eq.~\ref{eq:tDSA} give an acceleration timescale which is many orders of magnitude smaller than the typical cooling time of radio emitting electrons, whose momentum distribution at injection can be assumed to follow a power law within our momentum range of interest.

This also motivates the fact that we can model shock injection by DSA by adding the newly created population of particles across timesteps (see Eq.~\ref{eq13} below), without integrating a source term as needed for the much slower re-acceleration by turbulence (see below).  

The expression for the normalisation factor, $K$, can thus be derived by equating the cosmic ray energy flux crossing each tracer volume element, and the product between the total energy of cosmic rays advected with a post-shock velocity  ($v_d$): 

\begin{equation}
\Psi_{\rm CR} ~dx_t = v_d E_{\rm CR}  ,
\end{equation}
where $v_d$ is the downstream (post-shock) velocity 

\begin{equation}
E_{\rm CR} = \int_{p_{\rm inj}}^{p_{\rm cut}} Q_{\rm inj}(p) T(p) dp ,
\end{equation}
with $Q_{\rm inj}(p)$ defined as above and $T(p) = (\sqrt{1+p^2}-1)m_e c^2$. The integration yields \citep[e.g.][]{2013MNRAS.435.1061P}

\begin{equation}
E_{\rm CR} = \frac{K_{\rm inj} m_e c^2}{\delta_{\rm inj}-1} \left[\frac{B_x}{2} \left( \frac{\delta_{\rm inj}-2}{2},\frac{3-\delta_{\rm inj}}{2}\right) + p_{\rm cut}^{1-\delta_{\rm inj}} \left(\sqrt{1+p_{\rm cut}^2}-1  \right)  \right]  
\end{equation}
where $B_x(a,b)$ is the incomplete Bessel function and $\rm x=1/(1+p_{\rm cut}^2)$. 

\subsubsection{Gain Terms - Shock re-acceleration}
Beside the {\it direct} injection of relativistic electrons by shocks, we also include {\it re}-acceleration by shocks waves \citep[e.g.][]{2005ApJ...627..733M,kr11,ka12}.
According to DSA, the input particle spectrum, $N_0(x)$, becomes

\begin{equation}
N(p)=(\delta_{\rm inj}+2) \cdot p^{-\delta_{\rm inj}} \int_{p_{min}}^p N_0(x) x^{\delta+1} dx ,
\end{equation}
where $\delta_{\rm inj}$ is the local slope in momentum space and $\delta$ is the slope corresponding to the new shock, according to DSA.

\begin{figure*}
    \centering
     \includegraphics[width=0.99\textwidth]{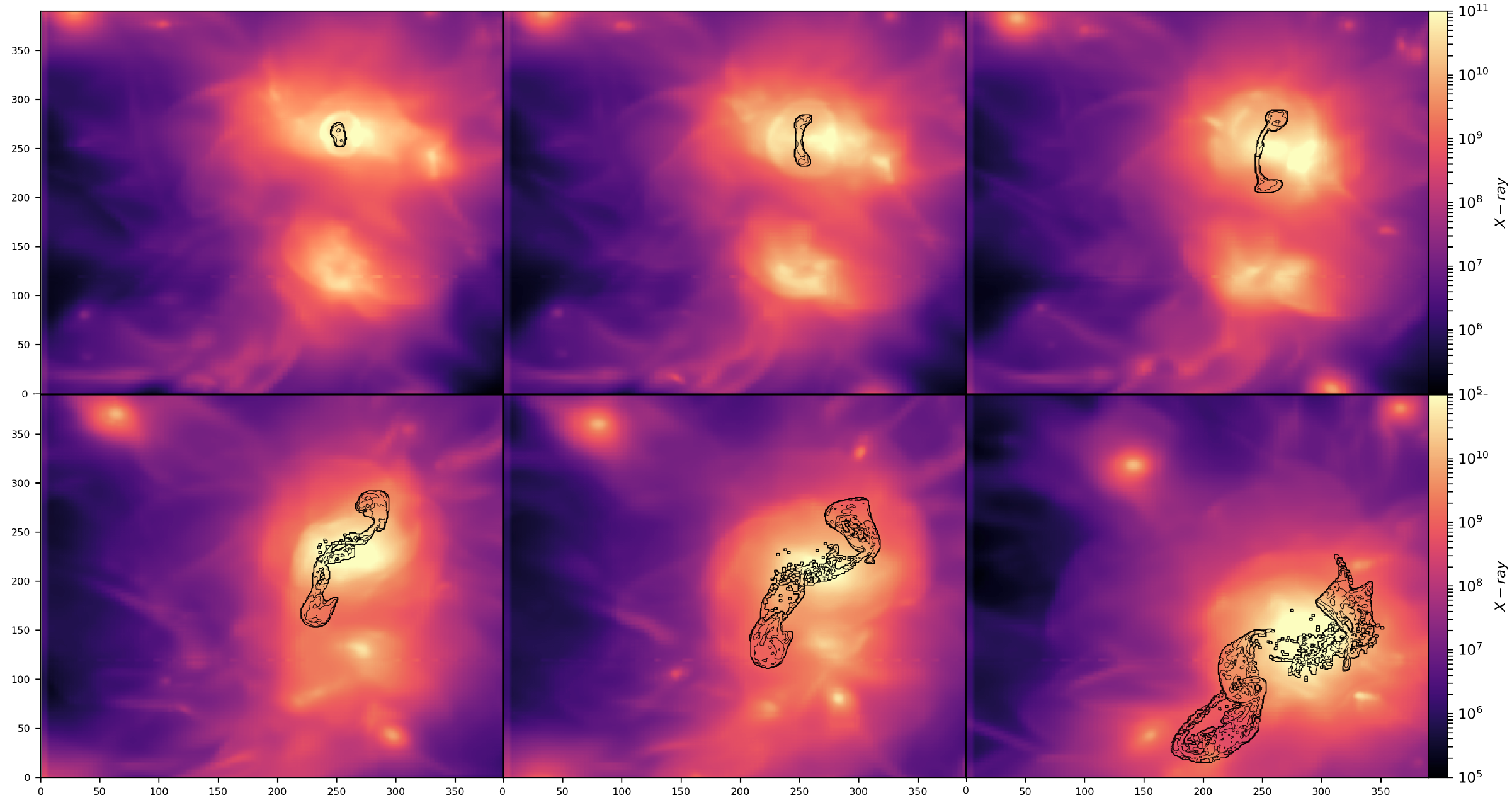}
    \caption{Evolution of the X-ray emission and of the radio emission ($\log_{\rm 10}$ contours) for the Run2 model, in which a radio galaxy is activated at $z=0.5$. The epochs in the panels are $z=0.49$, $z=0.46$, $z=0.45$, $z=0.37$, $z=0.31$ and $z=0.16$. No observational cut is applied to the maps, and the radio data have been smoothed to a $\times 3$ coarser resolution to allow a better visualisation of contours. The coordinates of each axis are in units of cells ($\delta x =8.86$~kpc).}
    \label{fig:map0}
\end{figure*}

\begin{figure*}
    \centering
       \includegraphics[width=0.99\textwidth]{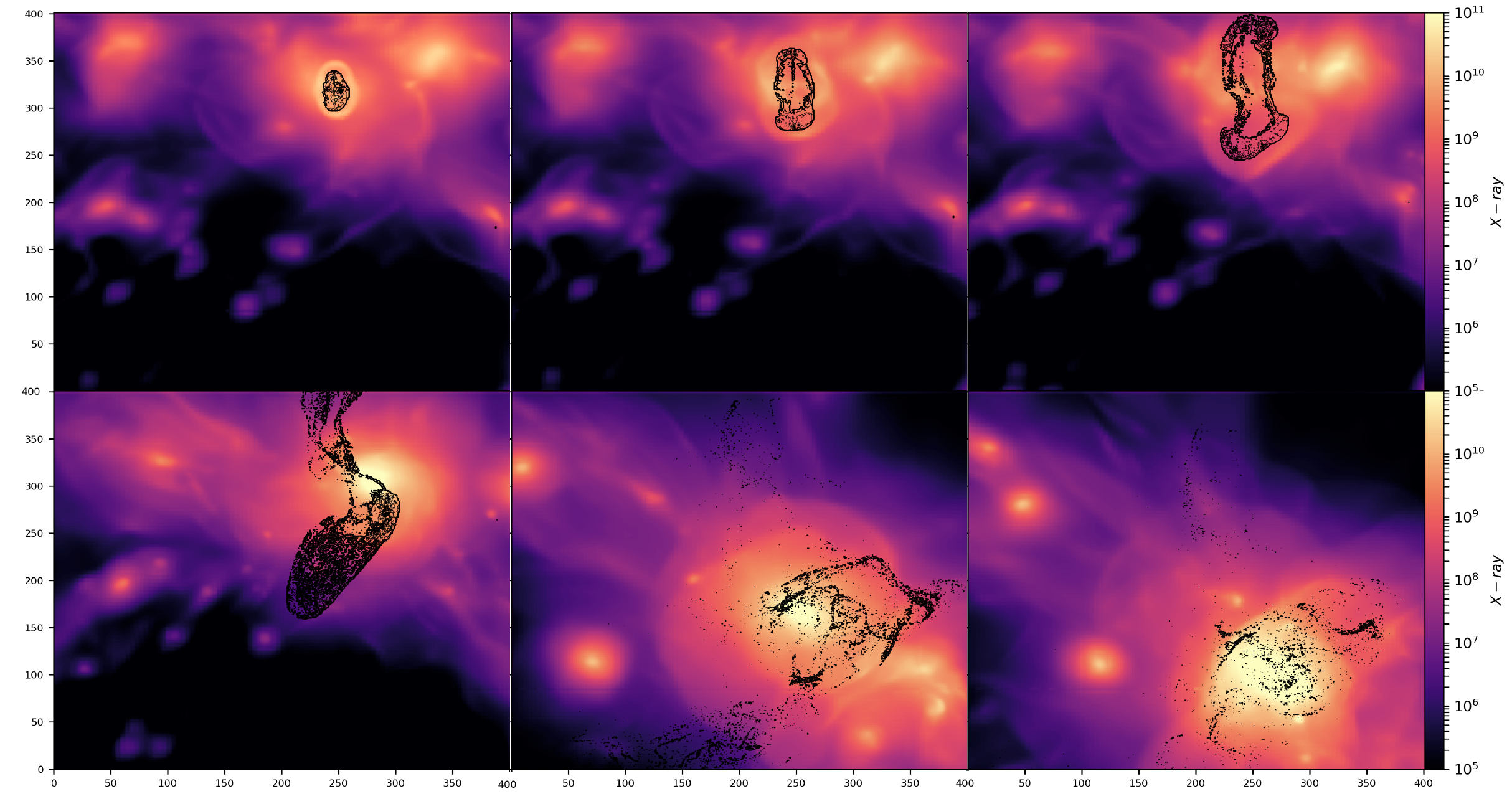}
    \caption{Evolution of the X-ray emission and of the radio emission ($\log_{\rm 10}$ contours) for the Run1 model, in which a radio galaxy is activated at $z=1.0$. The epochs in the panels are $z=0.96$, $z=0.93$, $z=0.89$, $z=0.76$, $z=0.37$ and $z=0.19$. No observational cut is applied to the maps, and the radio data have been smoothed to a $\times 3$ coarser resolution to allow a better visualisation of contours. The coordinates of each axis are in units of cells ($\delta x =8.86$~kpc).}
    \label{fig:map1}
\end{figure*}

\subsubsection{Gain terms - Turbulent re-acceleration} 

\noindent
Turbulence is thought to play a major role in producing giant radio halos in galaxy clusters, in connection with mergers \citep[e.g.][and references therein]{bj14}. Some residual level of turbulence is expected to be present at all times \citep[e.g.][]{va11turbo,2020MNRAS.495..864A}, caused by minor mergers \citep[e.g.][]{bn99,cassano05,in08,wi17}, AGN feedback \citep[e.g.][]{gaspari11b,2020arXiv200812784B} and cool core sloshing \citep[e.g.][]{2010ApJ...717..908Z}. The solenoidal component of turbulence is prominent in clusters \citep[e.g.][]{miniati14,va17turb}. According to \citet{2016MNRAS.458.2584B}, solenoidal turbulence can re-accelerate particles via stochastic interaction with diffusing magnetic field lines in super-Alfvenic turbulence developed in the ICM. In addition to radio halos, this mechanism has also been used in models of Gamma-Ray Bursts \citep[][]{2017ApJ...846L..28X} and radio bridges between galaxy clusters \citep[][]{bv20,bonafede21}. The implied acceleration timescale is thus: 

\begin{equation}
    \tau_{\rm ASA} = \frac{P^2}{4~D_{\rm pp}  } = 1.25 \times 10^5 \rm ~Myr \rm  \frac{L/(0.5) ~ B}{\sqrt{n/10^{-3}} ({\delta V/10^7})^3} ,
    \label{eq:ASA}
\end{equation}
where $L$ is measured in Mpc and $\delta V$ is the local gas turbulent velocity within the scale $L$, measured in cm/s. Based on this formula, it can be seen that for typical ICM conditions, only for (solenoidal) turbulent rms velocities of $\delta V \geq  300 ~\ rm km/s$ can this mechanism produce acceleration on timescales of $\leq \rm Gyr$ (e.g. Sec.3.3.1).

The interaction between particles and turbulence is a stochastic process that can be thought of as the combination of two effects: First, electrons are systematically accelerated on a timescale given by Eq. \ref{eq:ASA} and, second, their energy is changed stochastically (Eq. \ref{eq11}). Starting from a monoenergetic initial distribution of electrons, these two effects result in the progressive widening of the momentum distribution of electrons, with a median value determined by the systematic acceleration (captured by our solver). 

\subsubsection{Injection of electron by jets}
Finally, the injection of relativistic electrons from radio galaxies assumes that a new population of electrons is generated at each tracer located in a jet launching cell, as 

\begin{equation}
K_r \int_{p_{\rm min}}^{p_{\rm max}} p^{-\delta_r} dp = \phi_e m_{\rm trac} ,
\end{equation}
where we assume an injection spectrum of $\delta_r=2$ and we set $\phi_e$ so that the number of all injected relativistic electrons is $10^{-3}$ of the total number of thermal electrons for each tracer,  i.e. a few $\sim 10^{-7} \rm ~part/cm^3$  \citep[e.g.][]{2012ApJ...750..166M}.

\section{Results}\label{sec::res}

Using cosmological, MHD simulations, we addressed the following questions:

\begin{itemize}
    \item What is the impact of jets from radio galaxies on the thermodynamic, kinematic and magnetic properties of the ICM?
    \item How much do these effects influence the distribution of relativistic electrons in the ICM?
    \item How do the properties of relativistic electrons change depending on loss and (re)acceleration mechanisms arising from the interaction with the ICM?
\end{itemize}

In the following sections, we will discuss these question in the light of our results in turn.

\begin{figure*}
    \centering
    \includegraphics[width=0.95\textwidth]{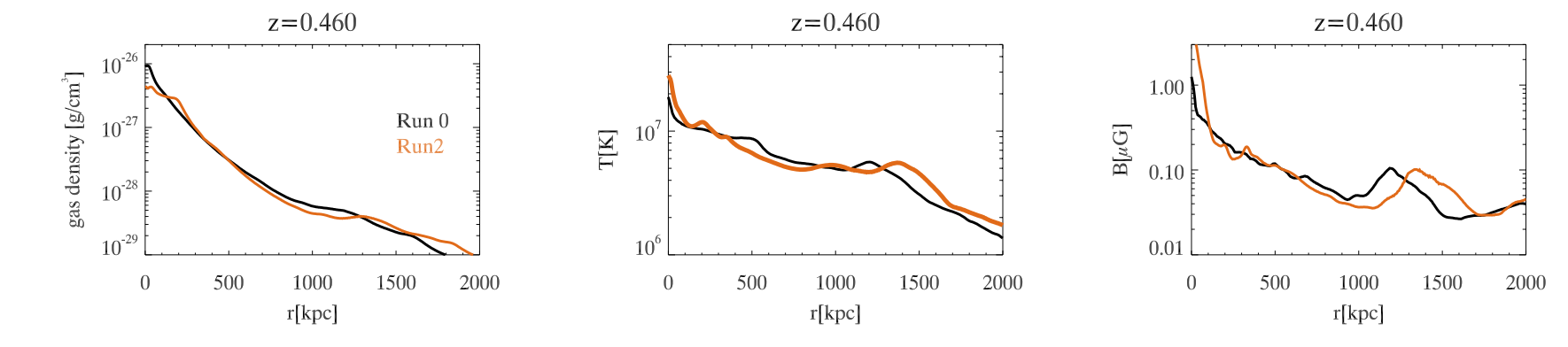}
    \includegraphics[width=0.95\textwidth]{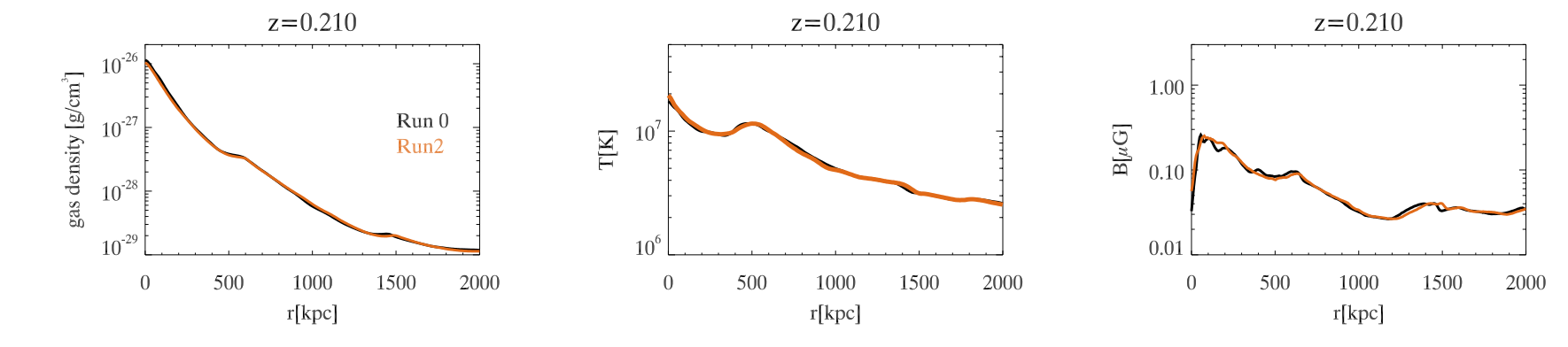}
    \includegraphics[width=0.95\textwidth]{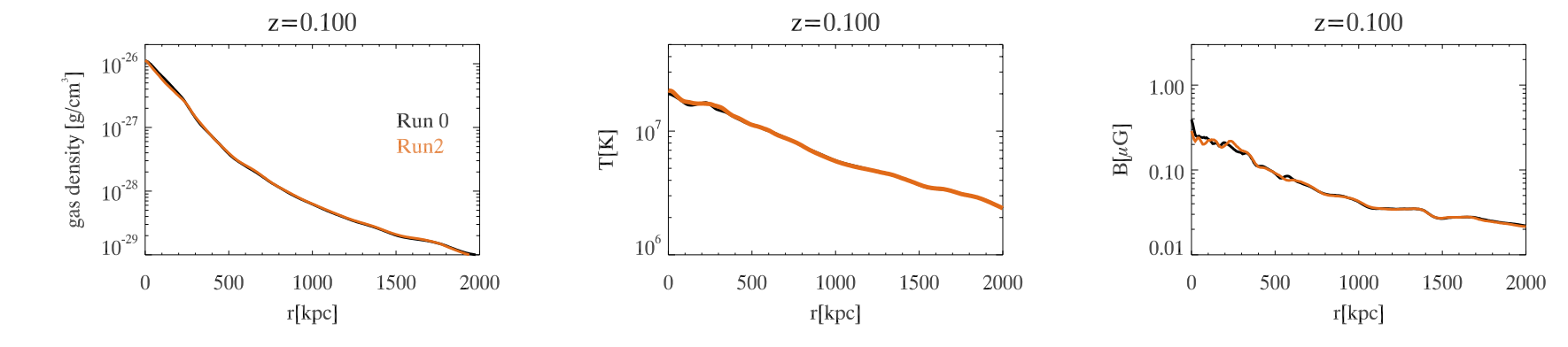}
    \caption{Radial (mass weighted) profiles of gas density, gas temperature and proper magnetic field strength for runs Run0 and Run2 at three different epochs.}
    \label{fig:prof1}
\end{figure*}

\subsection{Radio source evolution}

Radio galaxies have been historically separated into two classes, based on the radio surface brightness of their cocoons \citep[][]{FR74}. While Fanaroff \& Riley type II (FRII) sources show  well-defined jet termination shocks around edge-brightened cocoons, type I sources (FRI) have peaks of surface brightness closer to the core. 
Both class of FR sources are  thought to initially expand supersonically, with the jet-environment interaction likely to have a decisive role in determining when FRII sources evolve into FRI systems, following the disruption of their jets \citep[e.g.][]{1997MNRAS.286..215K,2015ApJ...806...59T}. Moreover, many radio sources associated with low-z galaxies and detected by recent surveys are compact and small in size, and can be regarded as one of the extremes (FR0) of a continuous population of radio galaxies, with a broad distribution of sizes and luminosities (e.g. \citealt[][]{2019A&A...631A.176C}, see however \citealt[][]{hardcastlecroston} for a different take on FR0 sources).

Our simulated radio sources appear to have broad morphological similarity with FRII-type sources, but in both cases show a dynamic evolution. Figures~\ref{fig:map_jet_Run2}-\ref{fig:map_jet_Run1} show the evolution of radio emission (at 50 MHz) in the first evolutionary stages ($\leq 400$ Myr) of radio galaxies in the two runs, where we numerically integrated the synchrotron emission from each tracer assuming the Jaffe Perola ageing model \citep[][]{1973A&A....26..423J}. The frequencies used to compute the emission in this work are the "LOFAR" frequencies $50$ and $140$ MHz and the "JVLA" frequency of $1400$ MHz. The typical total emission power at $140$ MHz of each radio jet is of several $\sim 10^{35} \rm ~erg/s/Hz$  $100$ Myr after their injection, and it fades to $\leq 10^{33}-10^{34} \rm ~erg/s/Hz$ $200$ Myr after injection, due to the combination of radiative cooling and adiabatic losses experienced by lobes.

Jets in Run2 are quickly distorted and bent by the interaction with the surrounding turbulent ICM, while they remain more collimated in Run1, owing to the lower and narrower amplitude of the pressure profile of the gas atmosphere in the cluster at  $z \sim 1$. Additional renderings for the entire evolution of Run1 and Run2 are shown in the Appendix.  

 Based on their appearance during $\sim 100-200$ Myr since the injection of jets, both sources resemble FRII-type galaxies. However, an analysis of the distance of the radio emission peak from the jet core \citep[e.g.][]{vard19,mingo19} suggests that Run1 source is an FRI galaxy in a fraction of its evolution. However, in the absence of a  dominant relativistic components in our simulated jets, our model cannot reproduce hot spots of real radio sources.

\begin{figure}
    \centering
    \includegraphics[width=0.499\textwidth]{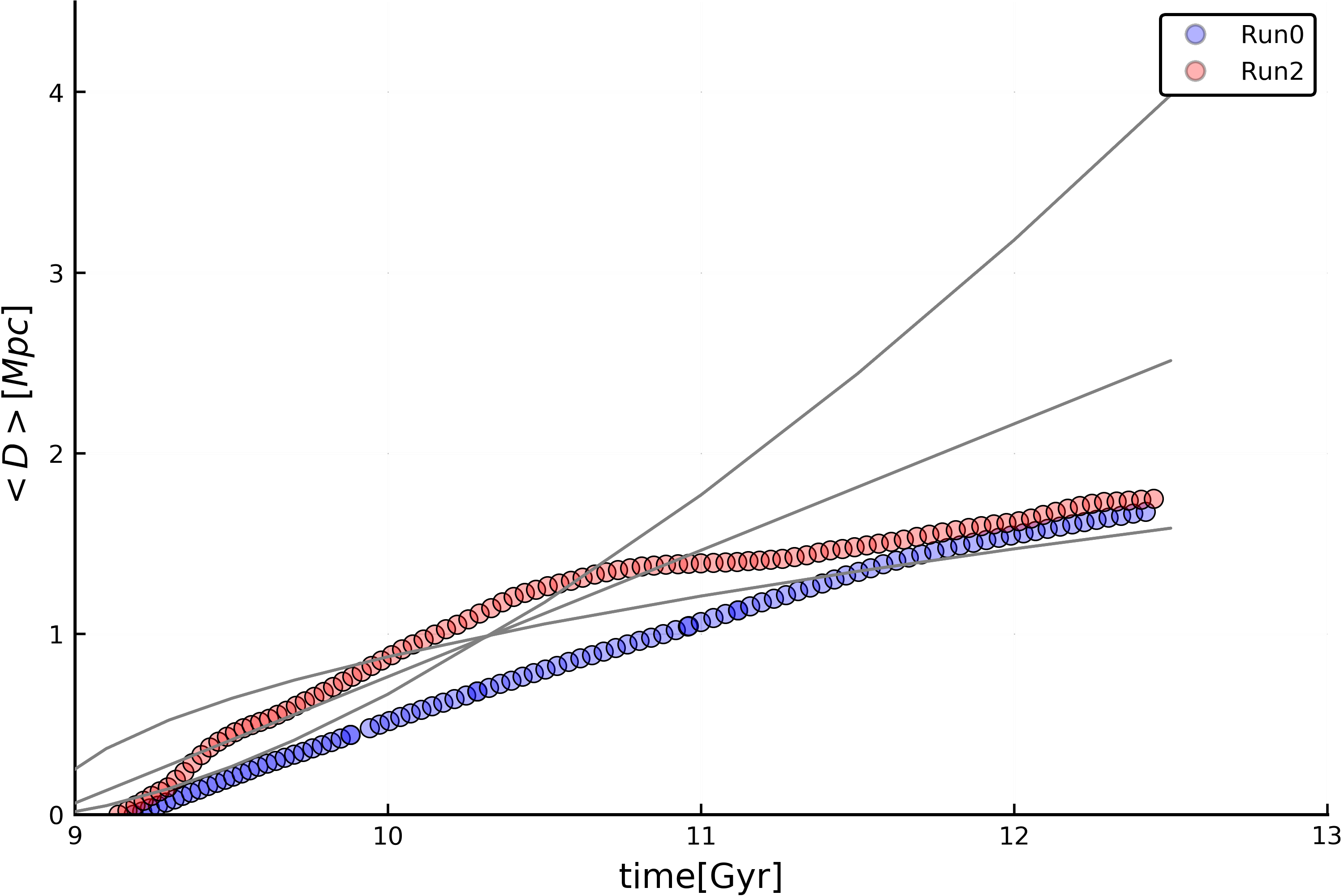}
    \includegraphics[width=0.499\textwidth]{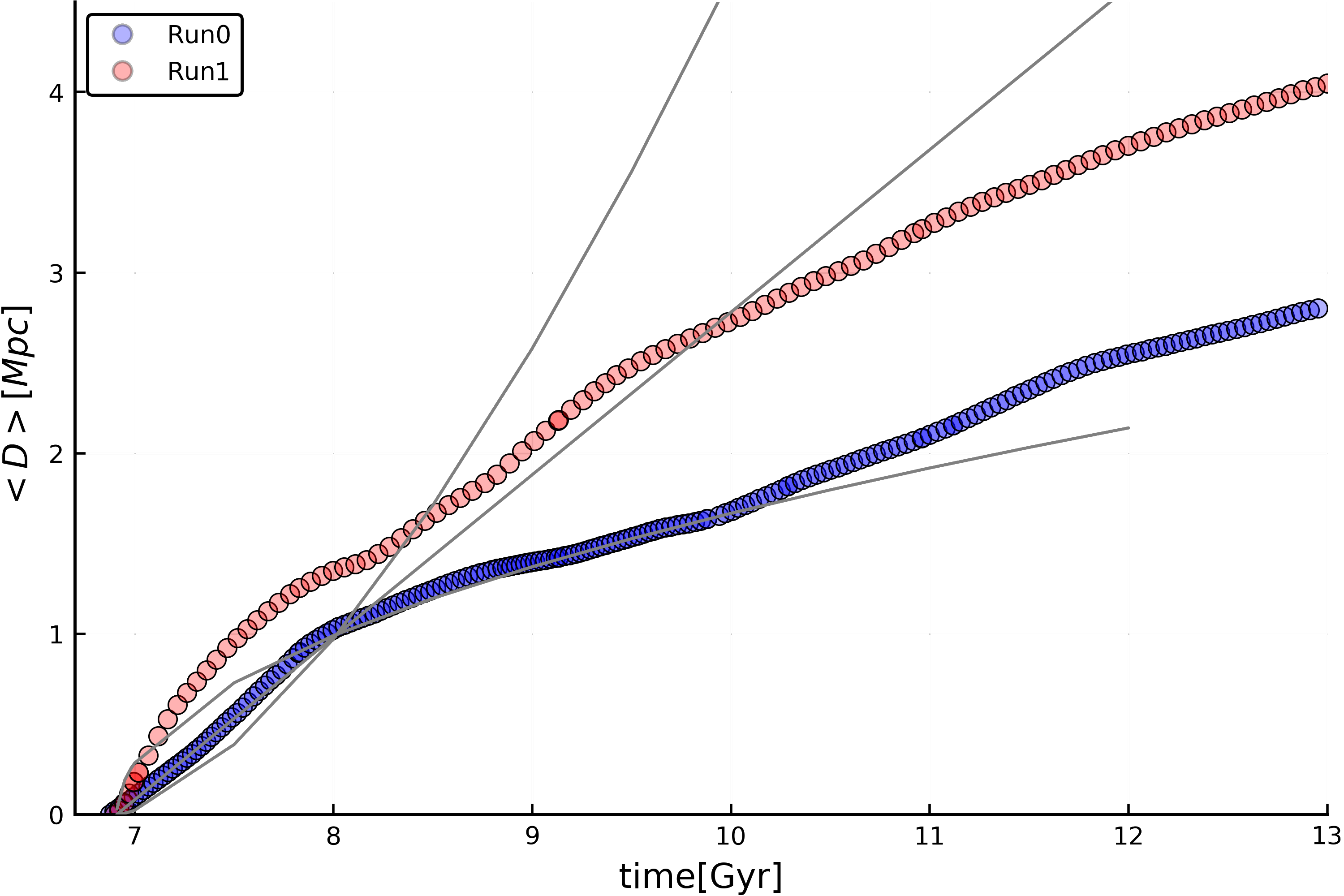}
    
    \caption{Evolution of the median distance (in comoving Mpc) covered by tracers since their initial injection in our runs, as a function of time, for Run2 vs Run0 (top, injection at $z=0.5$) and for Run1 vs Run0 (bottom, injection at $z=1$). The additional grey lines give the $D \propto t^{3/2}$, $\propto t$ and $\propto t^{1/2}$ trends, to guide the eye.}
    \label{fig:distance}
\end{figure}

\begin{figure*}
    \centering
    \includegraphics[width=0.495\textwidth]{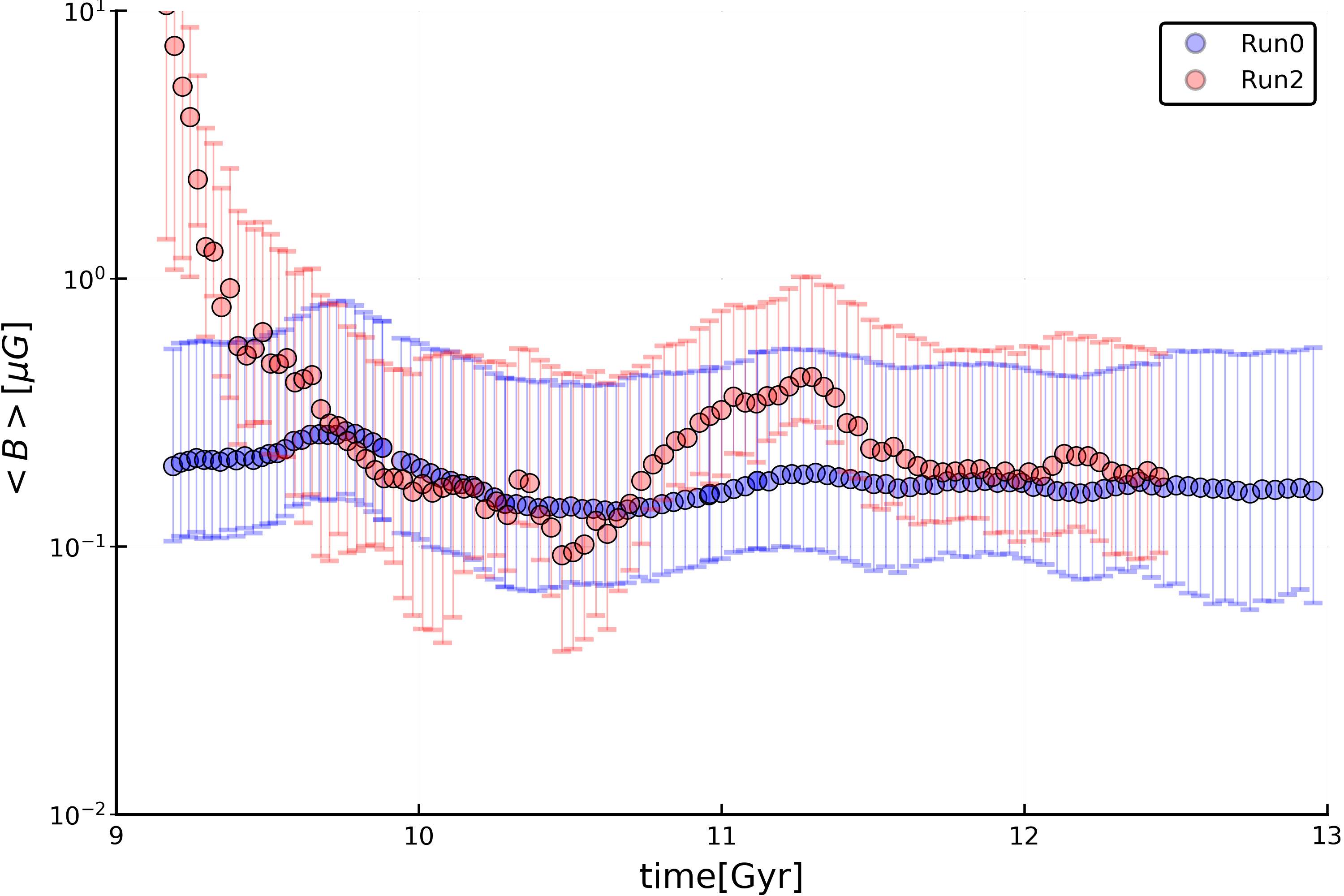}
     \includegraphics[width=0.495\textwidth]{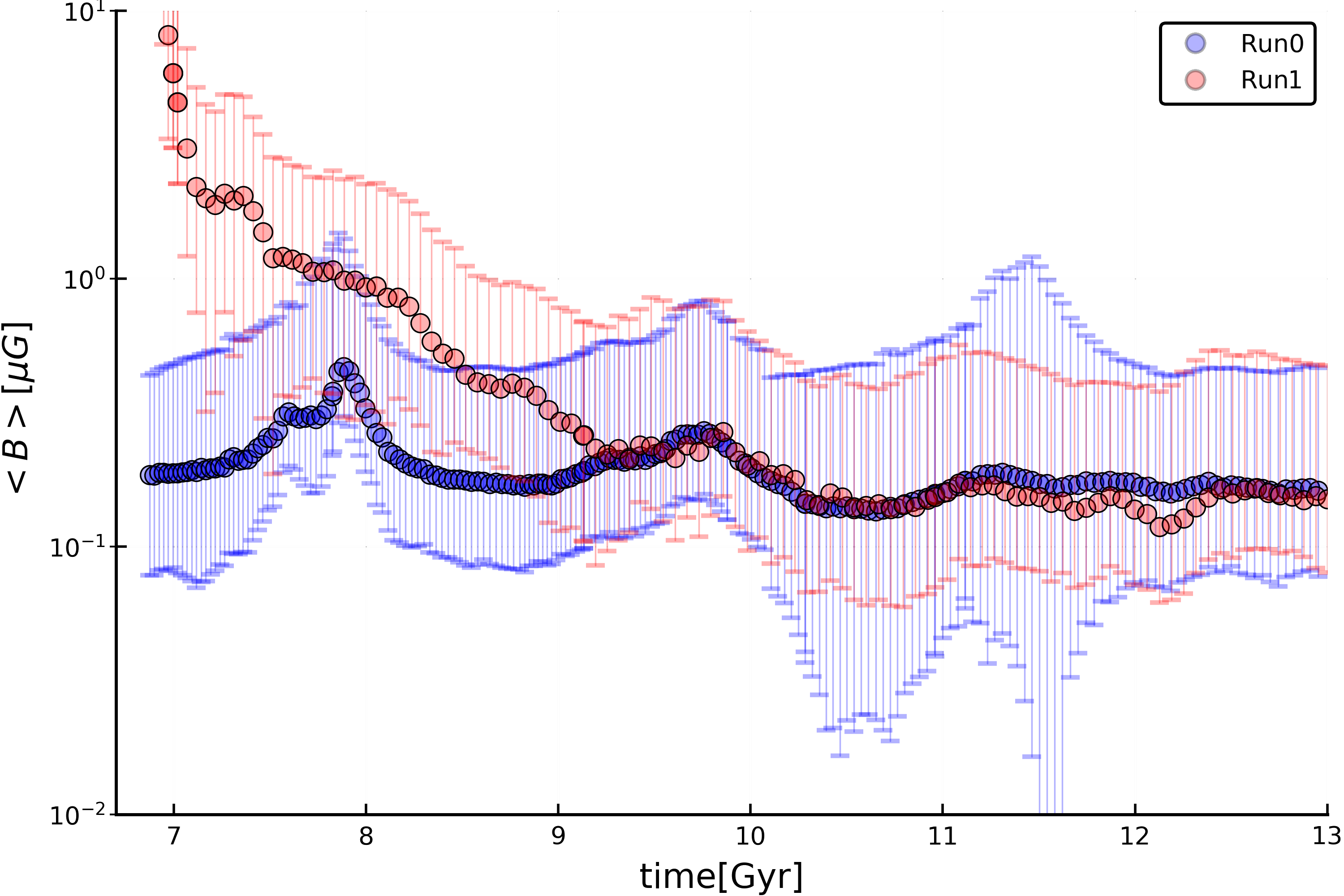}
     
    \includegraphics[width=0.495\textwidth]{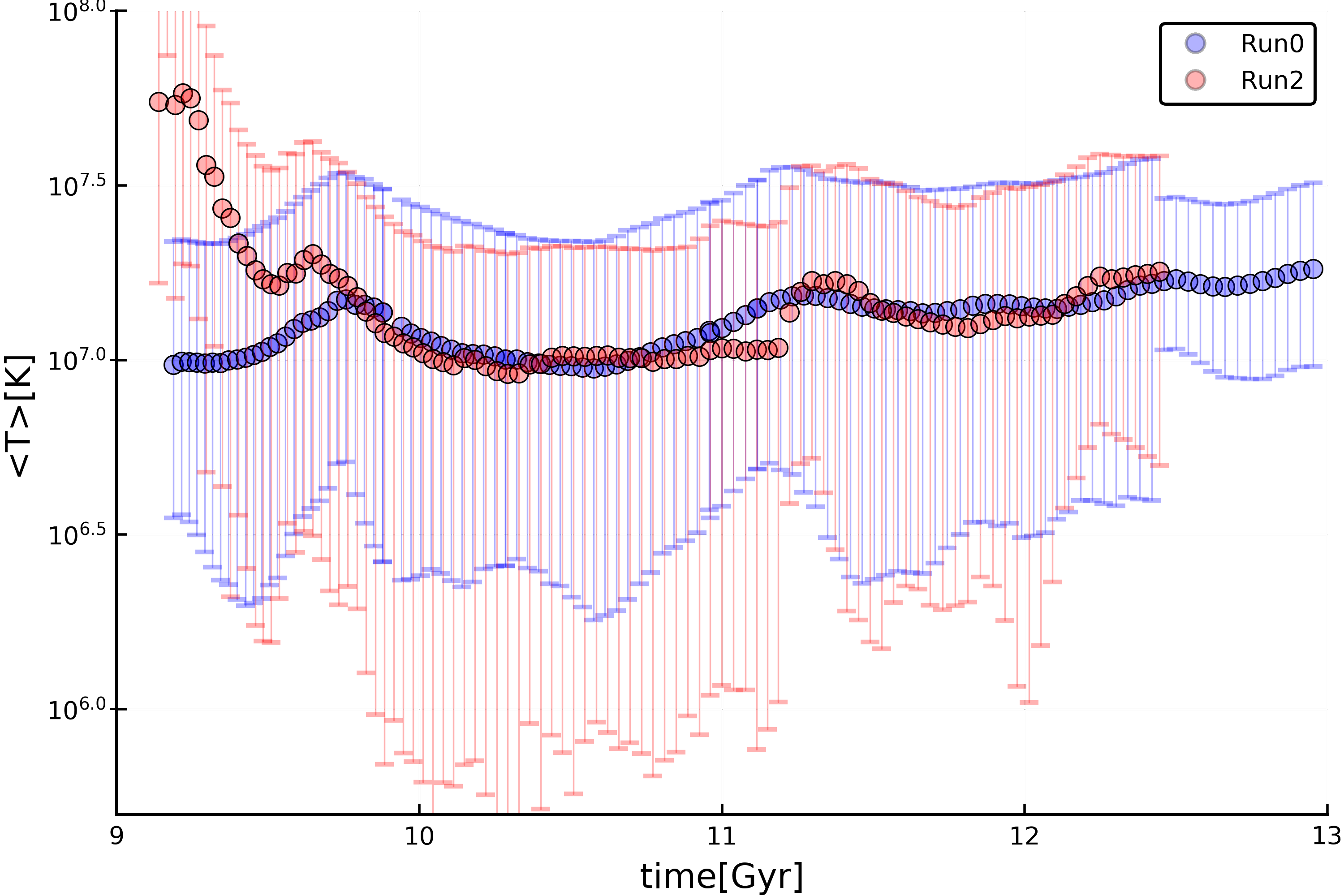}
    \includegraphics[width=0.495\textwidth]{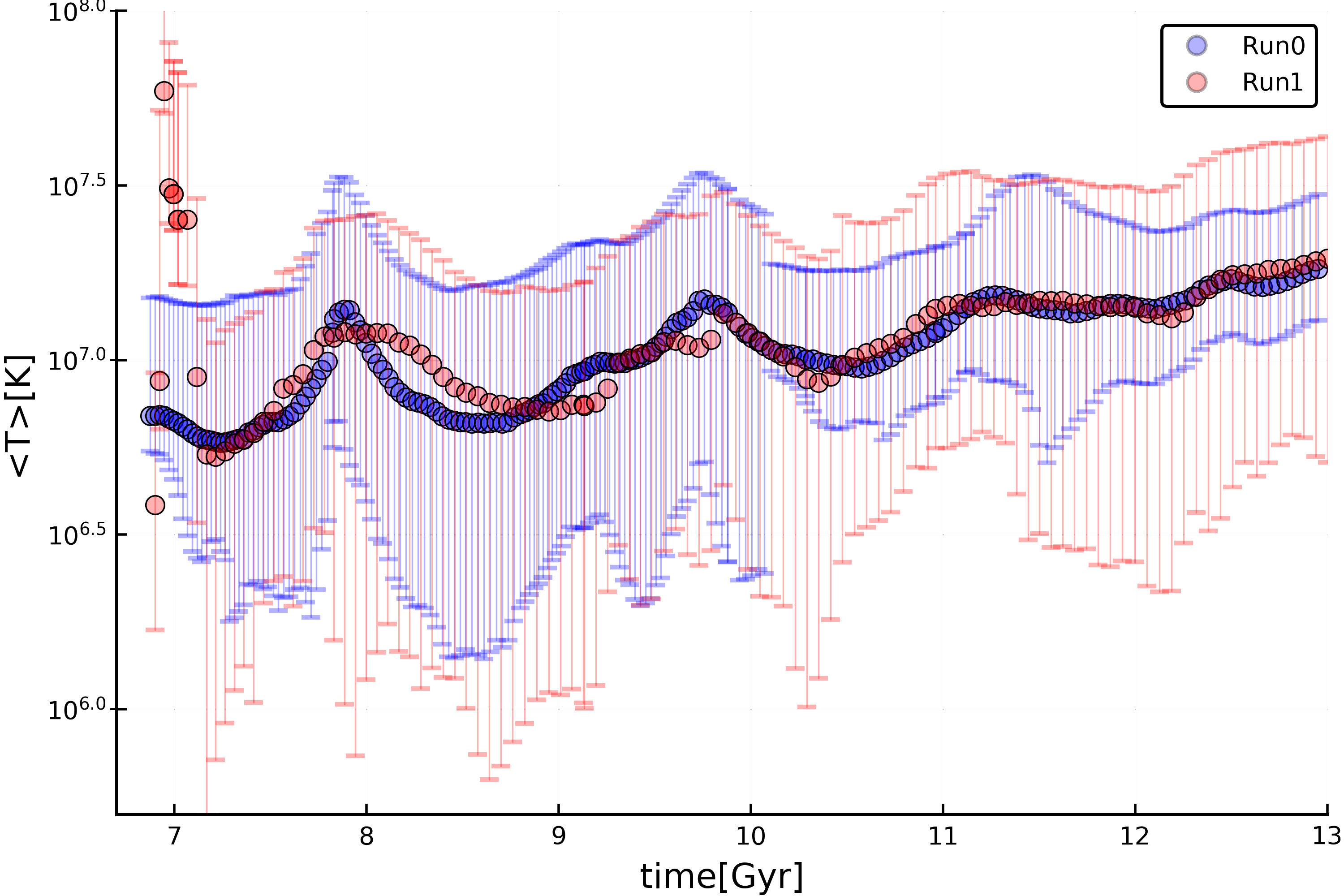}
     \includegraphics[width=0.495\textwidth]{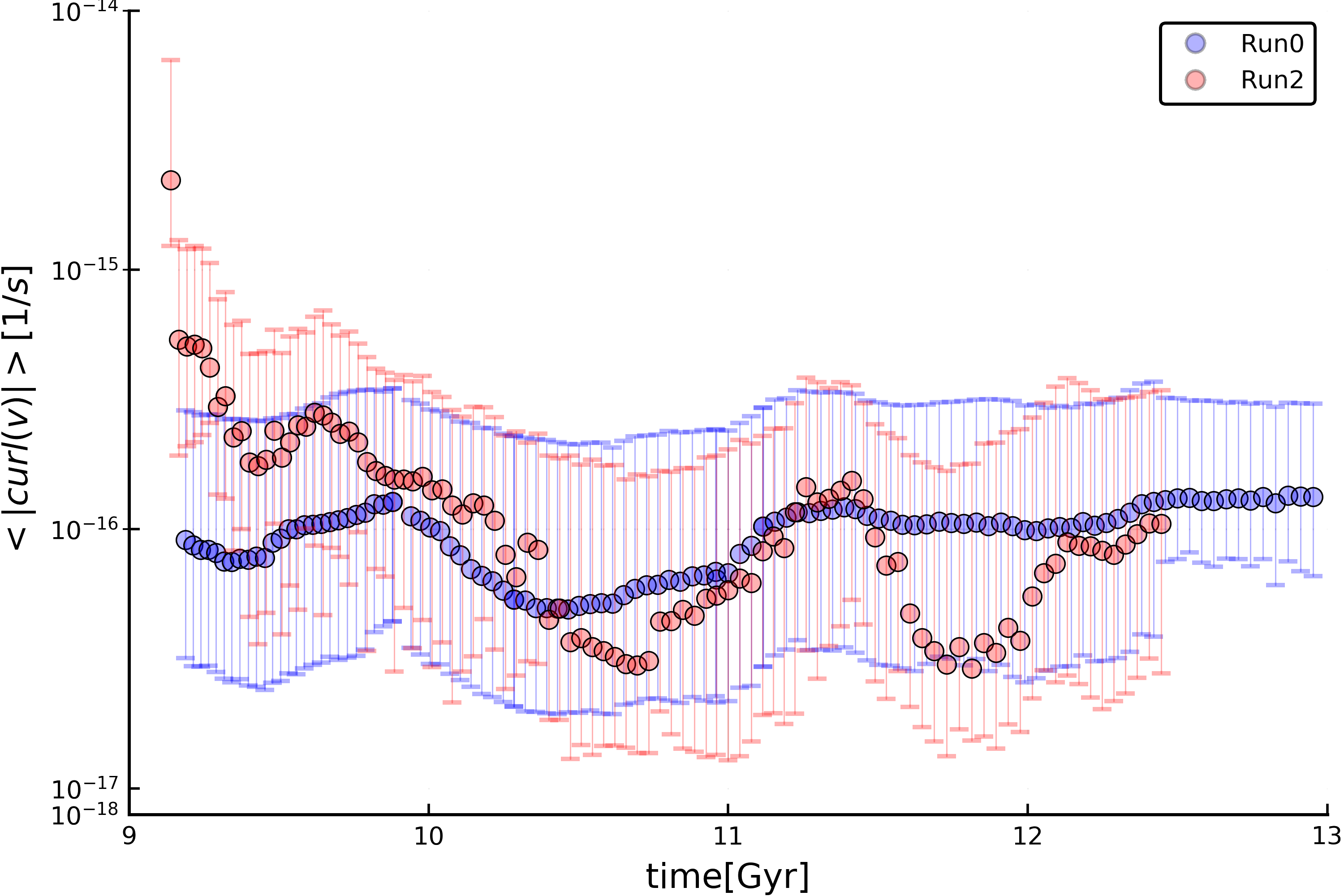}
      \includegraphics[width=0.495\textwidth]{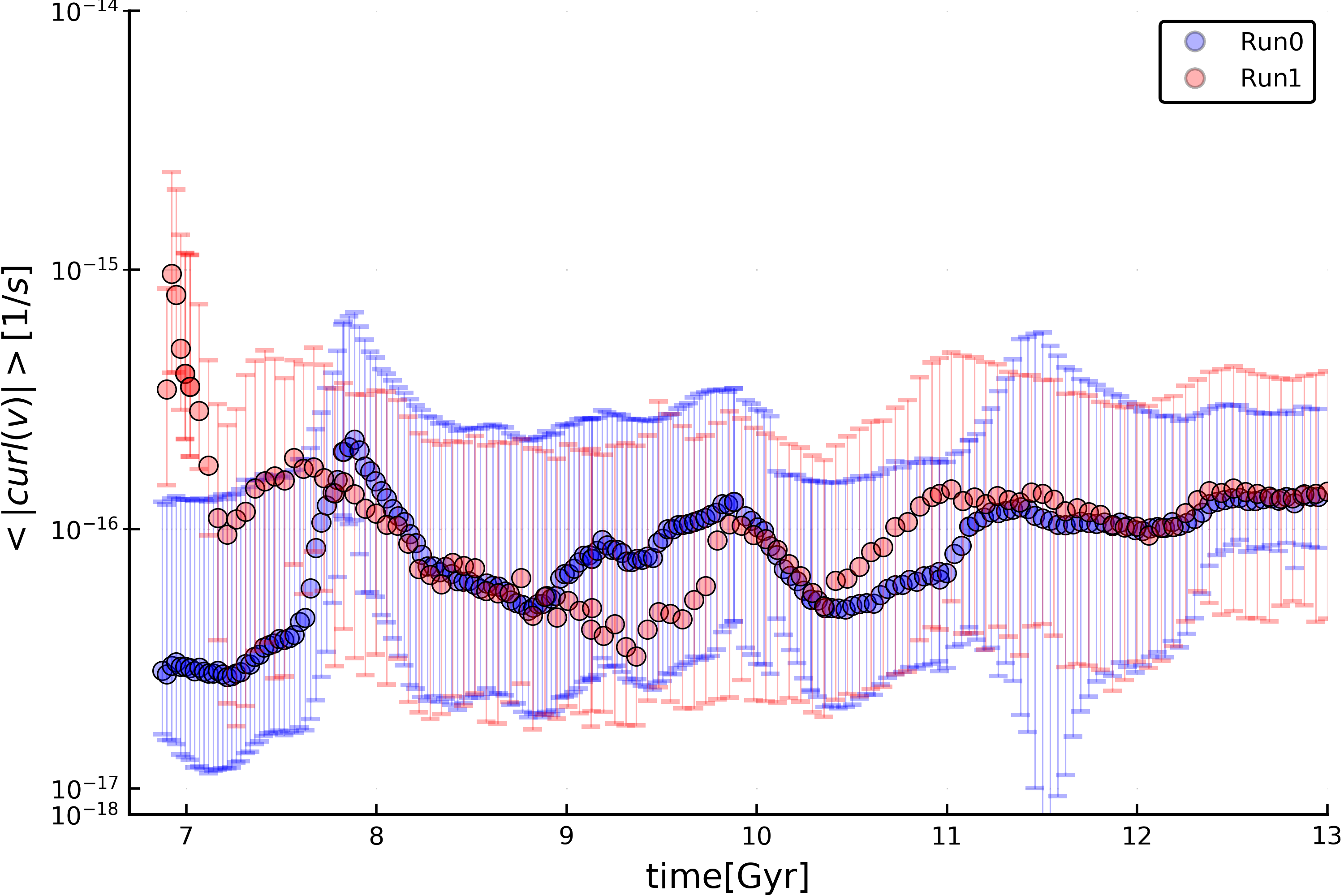}
      
    \caption{Evolution of the median magnetic field strength, temperature and absolute value of vorticity for tracers in our runs (the error bars show the 16-84th percentiles of the distributions at each redshift) comparing Run2 vs Run0 (left panels, injection at $z=0.5$) and Run1 vs Run0 (right, injection at $z=1$). The properties of tracers are measured at each redshift for particles at the radial distance from the cluster centre $\leq 0.5$~Mpc. }
    \label{fig:scatter}
\end{figure*}

\begin{figure*}
    \centering
       \includegraphics[width=0.99\textwidth]{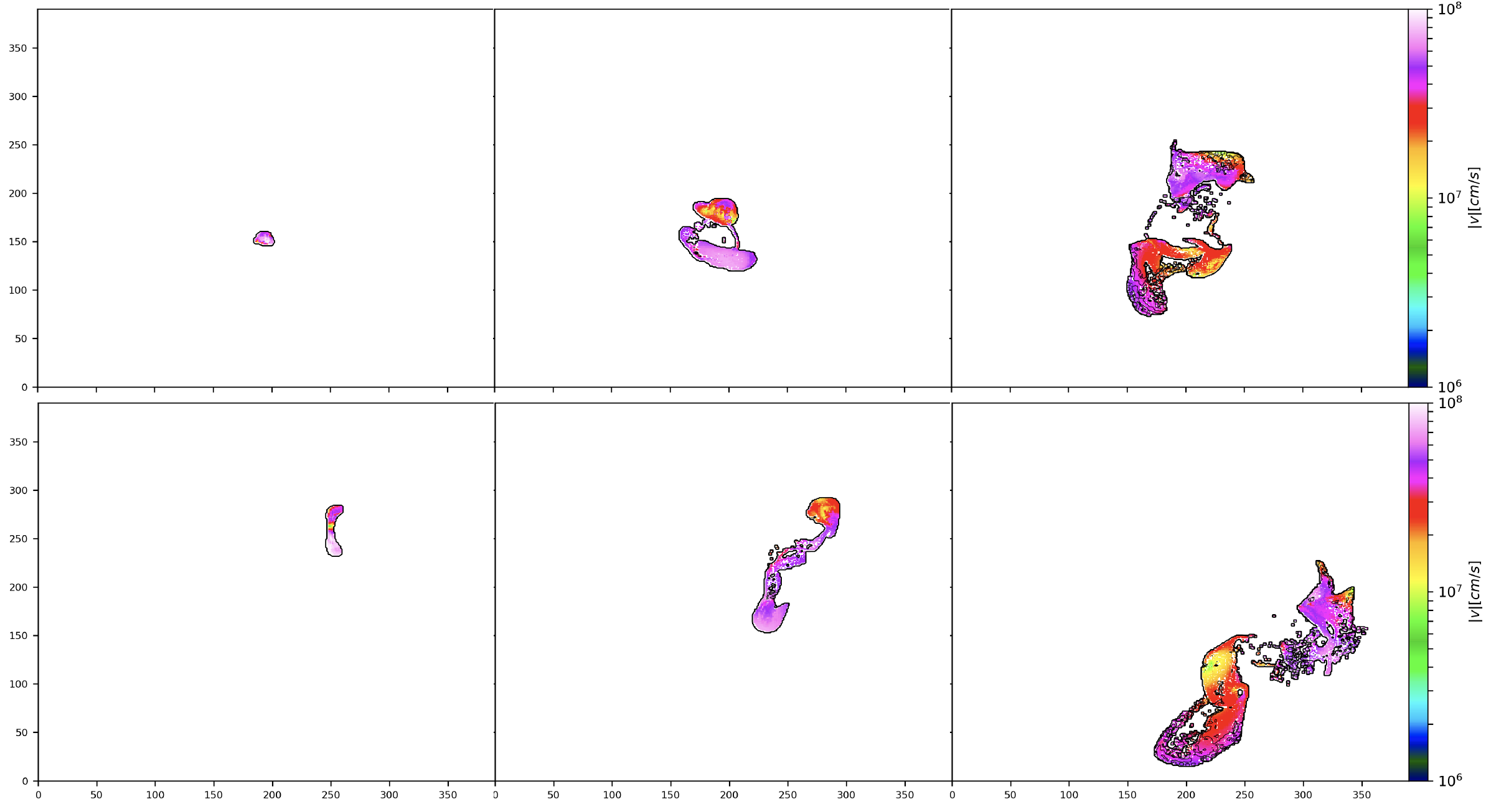}
         
    \caption{Velocity magnitude for tracers released by jets in the Run2, at three different epochs ($z=0.46$, $z=0.37$ and  $z=0.16$ from left to right, respectively), and averaged along the two different line of sights. The top row show the evolution of the lobes seen along the line of sight parallel to the jet axis, while the second row shows their evolution seen along a perpendicular line of sight (the same of Fig. \ref{fig:map0}). The axes are in units of cells of the simulation ($\delta x=8.86$ $\rm kpc$ comoving).}
    \label{fig:velocity}
\end{figure*}

\begin{figure*}
    \centering
       \includegraphics[width=0.99\textwidth]{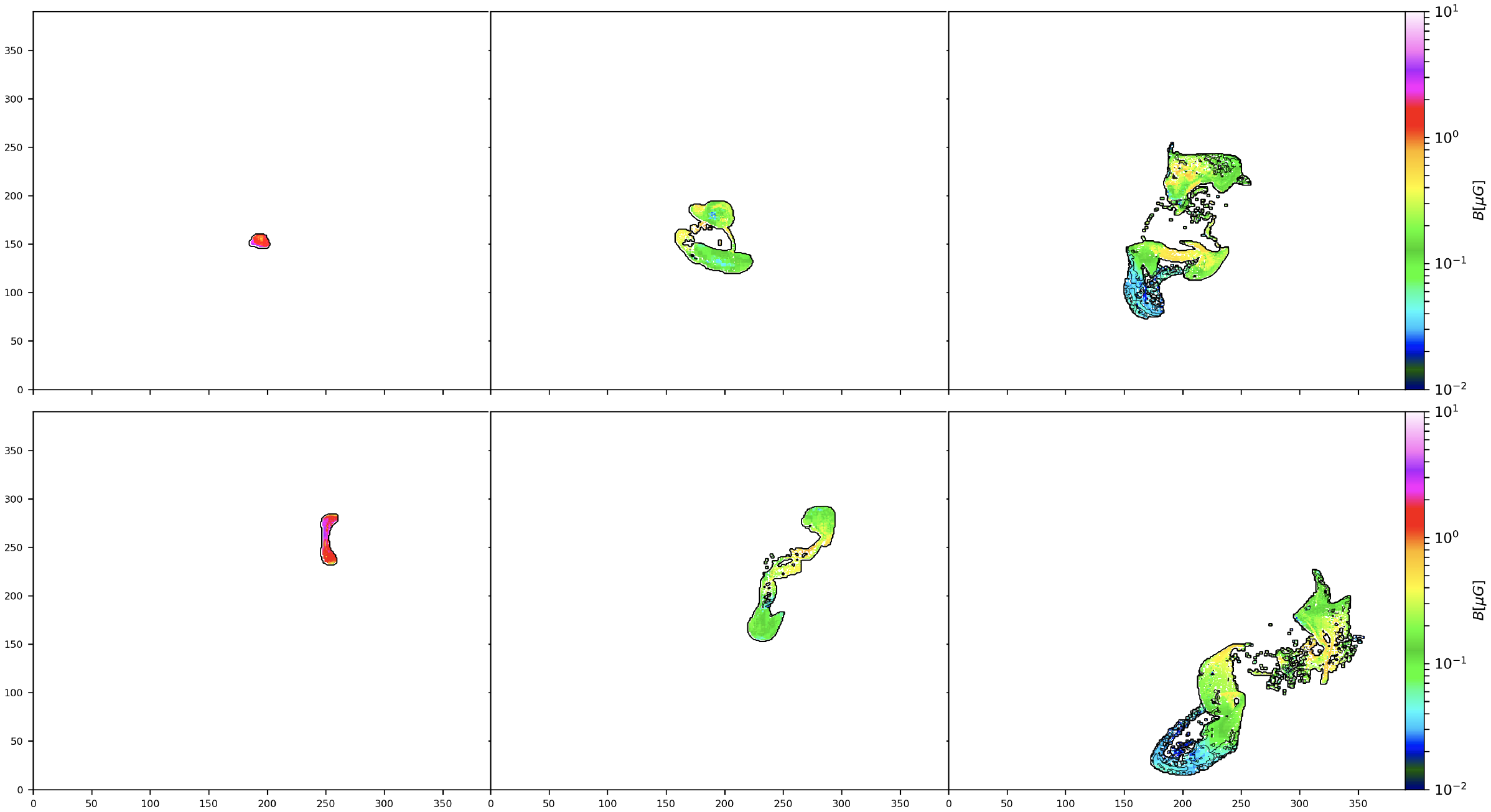}
    \caption{Magnetic field strength for tracers released by jets in Run2, averaged along the same lines of sight and epochs of Fig.\ref{fig:velocity}.}
    \label{fig:Bfield}
\end{figure*}

\begin{figure}
    \centering
    \includegraphics[width=0.495\textwidth]{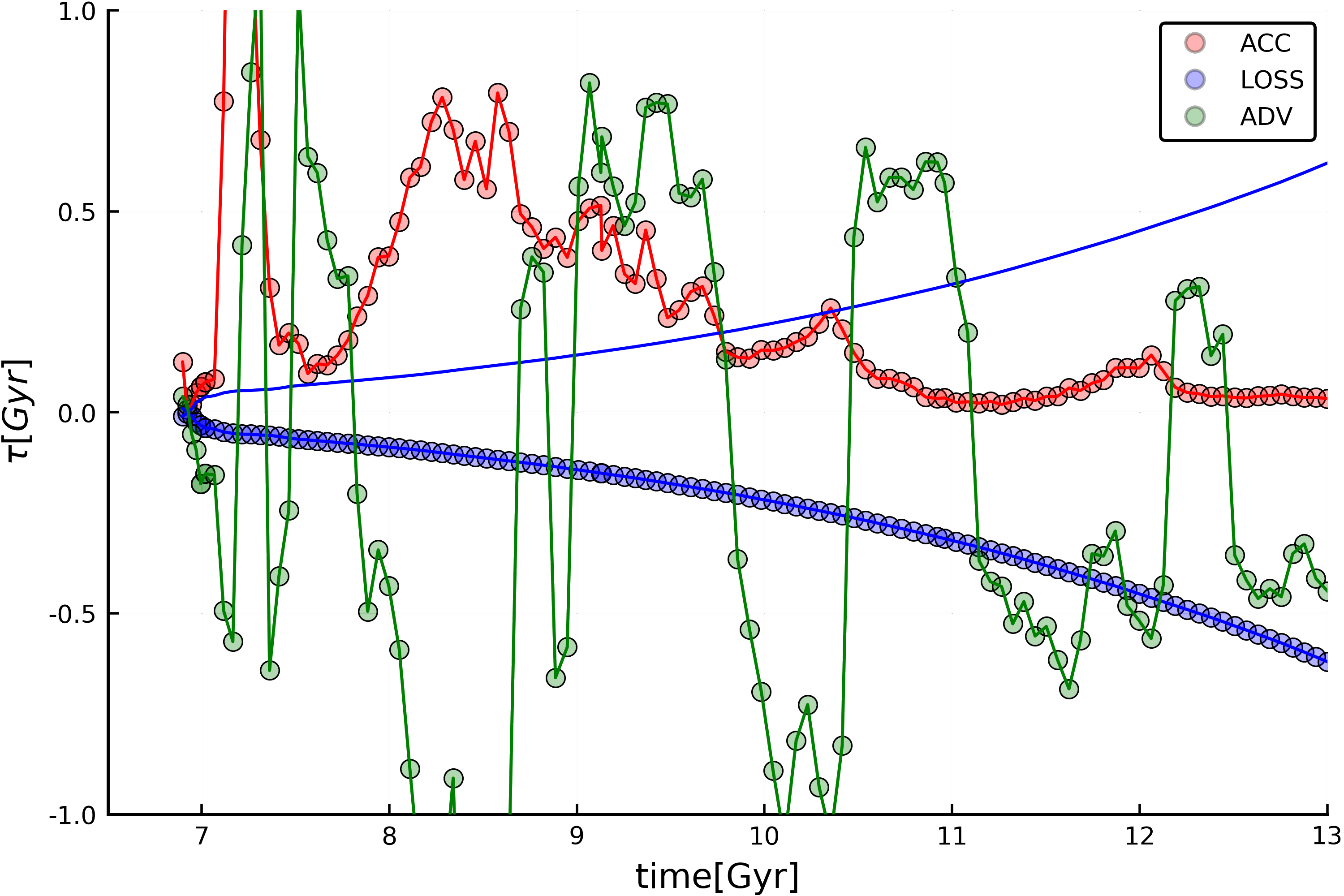}
    \includegraphics[width=0.495\textwidth]{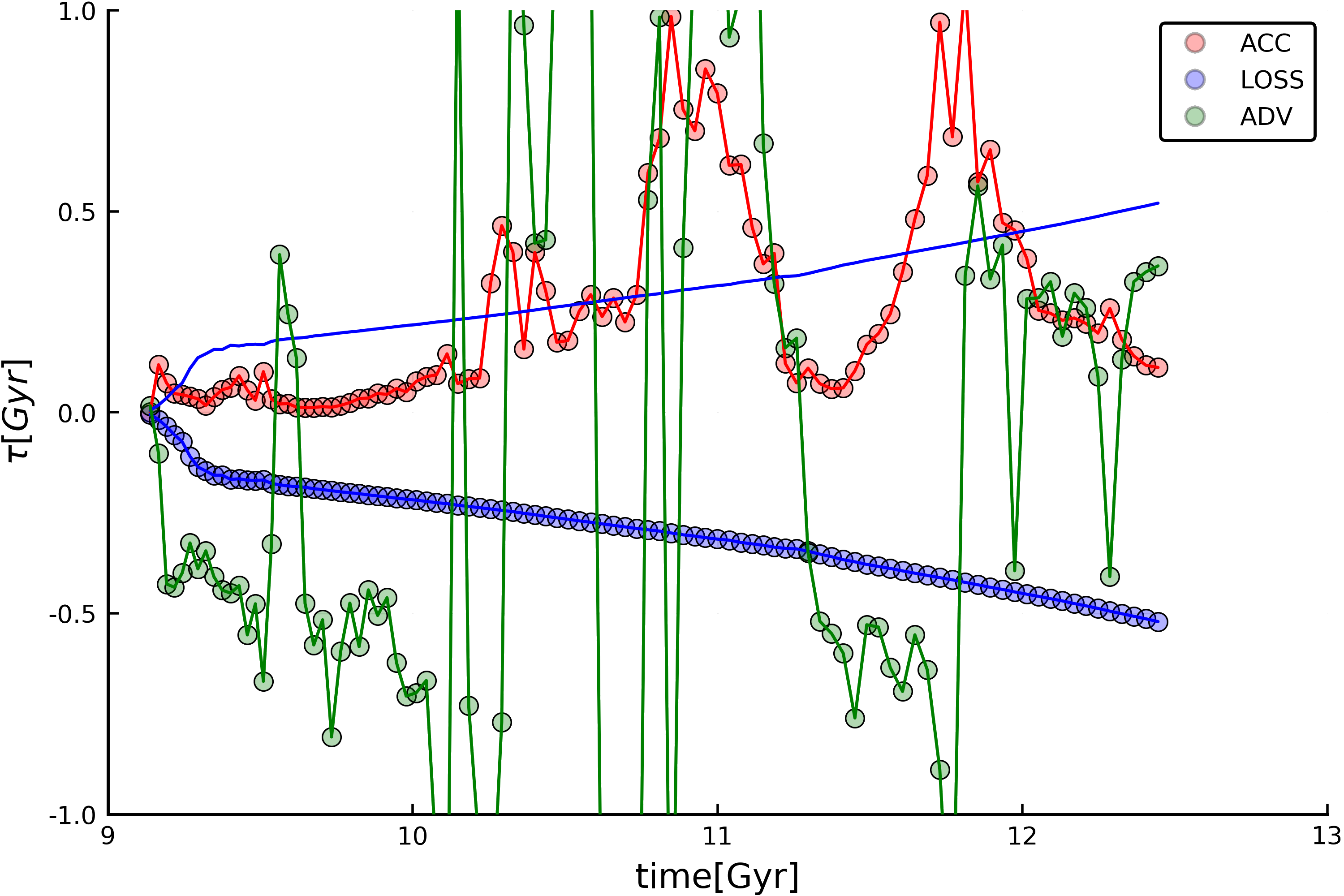}
         \caption{Evolution of the median timescales for acceleration or losses (see Sec. 2.3) experienced by electrons (here we consider $\gamma=10^3$ as a reference) released by radio galaxies in Run1 (top) and Run2 (bottom). Timescales for loss terms and shown as negative, while timescales for gain terms are shown as positive (to ease the comparison with radiative losses, the latter timescale is also mirrored in the upper panel). The timescales for shock acceleration ($\tau_{\rm DSA}$) and for Coulomb losses ($\tau_{\rm C}$) are not visible as they are much smaller and much bigger, respectively, than the time range covered by vertical axis in the plots.}
    \label{fig:spectra_tau}
\end{figure}

\begin{figure}
    \centering
    \includegraphics[width=0.49\textwidth,height=0.187\textheight]{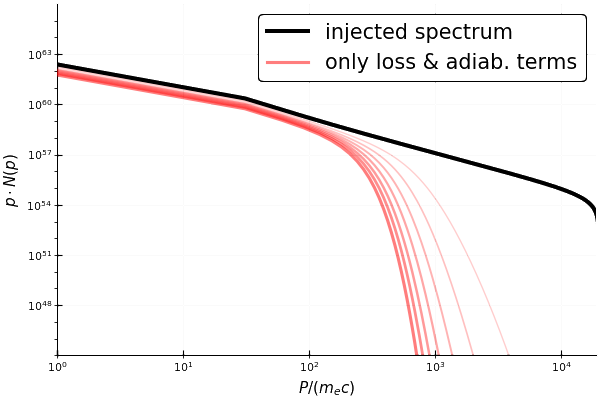}
    \includegraphics[width=0.49\textwidth,height=0.187\textheight]{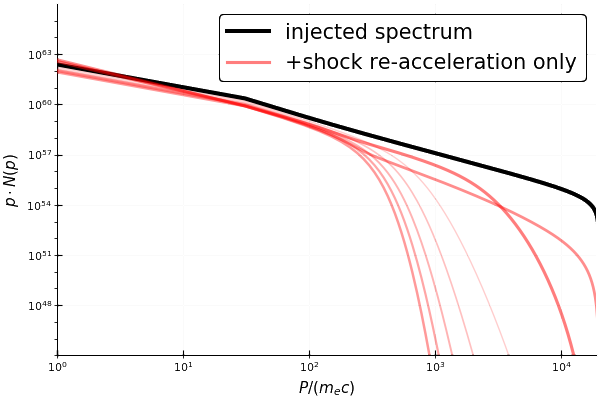}
     \includegraphics[width=0.49\textwidth,height=0.187\textheight]{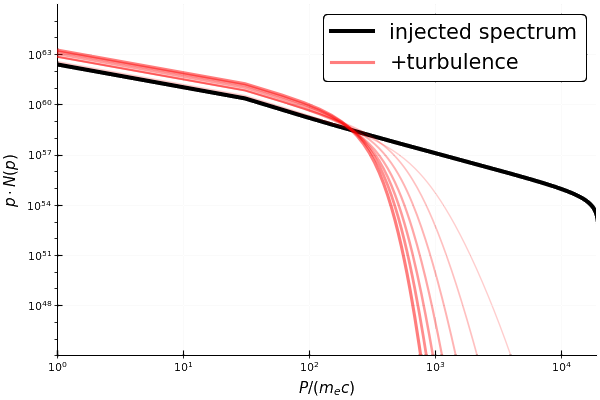}
     \includegraphics[width=0.49\textwidth,height=0.187\textheight]{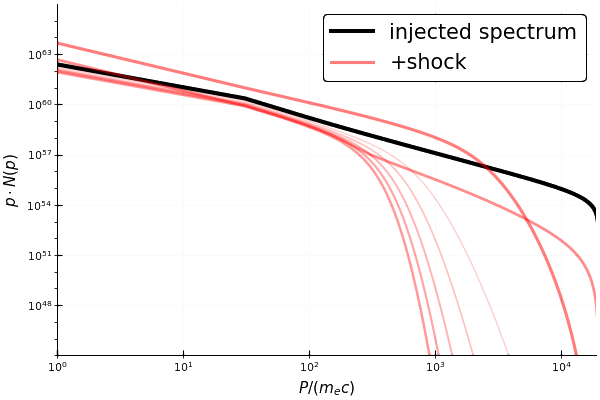}
       \includegraphics[width=0.49\textwidth,height=0.187\textheight]{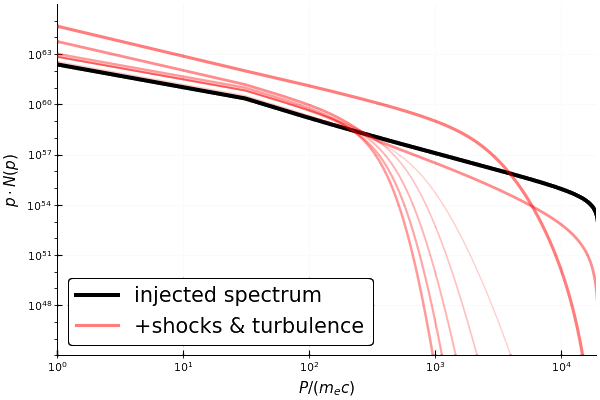}
    \caption{Evolution (from $z=0.5$ to $z=0.1$ of spectral energy distribution of a sample of $100$ electron tracers  initially located in the most magnetised regions of jets in the  Run2 model, for several combinations of re-acceleration scenarios (see labels).  The initial particle spectrum is given by the solid black line and the evolution of spectra (equally spaced every $\approx 0.5 \rm Gyr$) goes from the most to the least transparent lines.}
    \label{fig:spectra_evol}
\end{figure}

\begin{figure}
    \centering
    \includegraphics[width=0.49\textwidth,height=0.2\textheight]{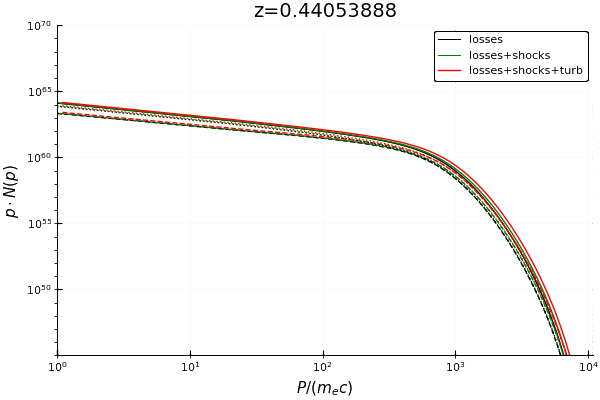}
    \includegraphics[width=0.49\textwidth,height=0.2\textheight]{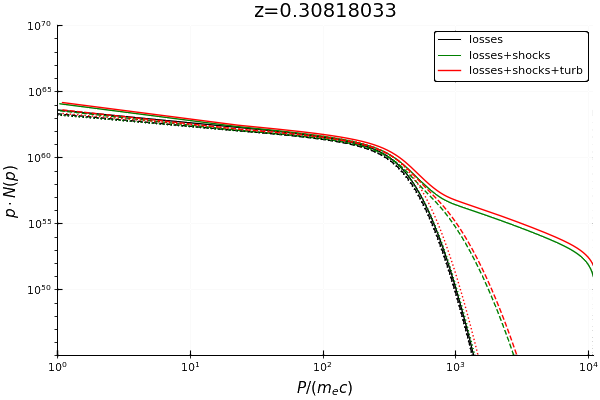}
     \includegraphics[width=0.49\textwidth,height=0.2\textheight]{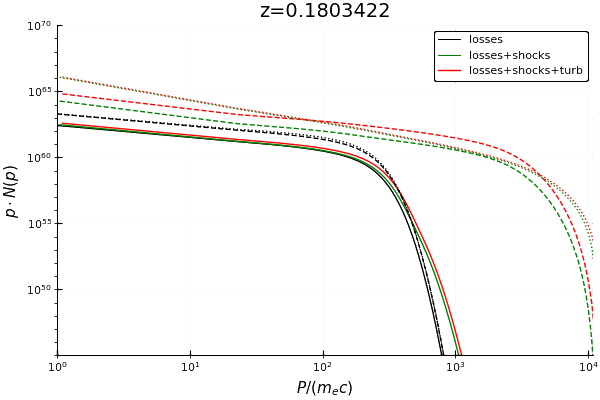}
     \includegraphics[width=0.49\textwidth,height=0.2\textheight]{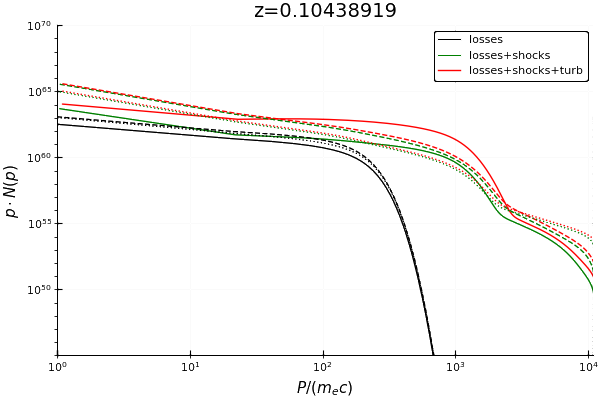}
     
    \caption{Evolution of the spectral energy distributions of electrons emitted by the radio galaxy in Run2, for three different acceleration/cooling models (colours) and for three different radial ranges:$r \leq 300$~kpc (solid),  300 kpc $< r \leq 600$~kpc (dashed) and $r > 600$~kpc (dotted).}
    \label{fig:spectra1}
\end{figure}

\begin{figure}
    \centering
    \includegraphics[width=0.49\textwidth,height=0.2\textheight]{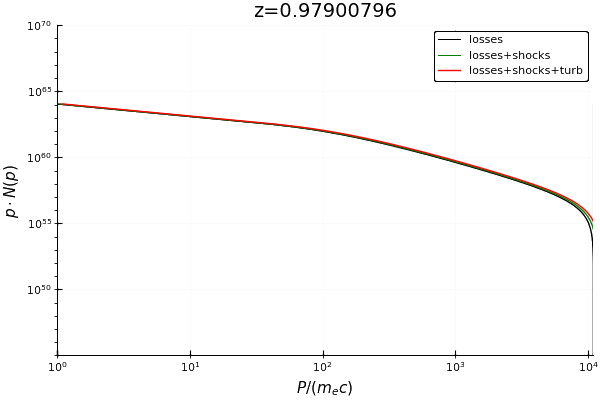}
    \includegraphics[width=0.49\textwidth,height=0.2\textheight]{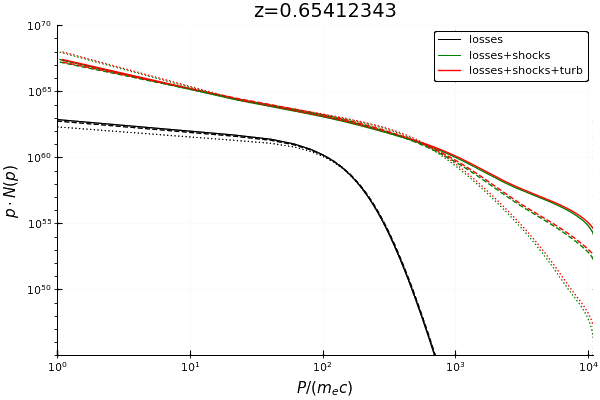}
     \includegraphics[width=0.49\textwidth,height=0.2\textheight]{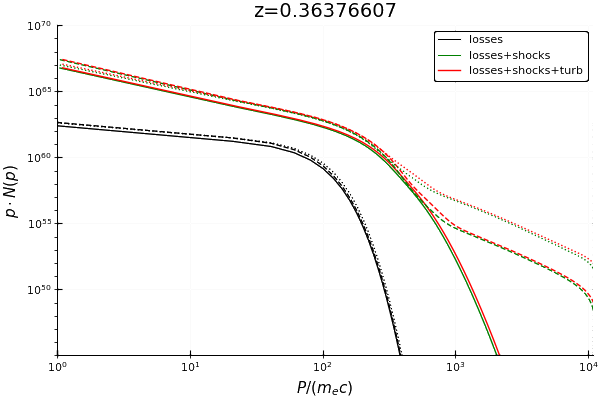}
     \includegraphics[width=0.49\textwidth,height=0.2\textheight]{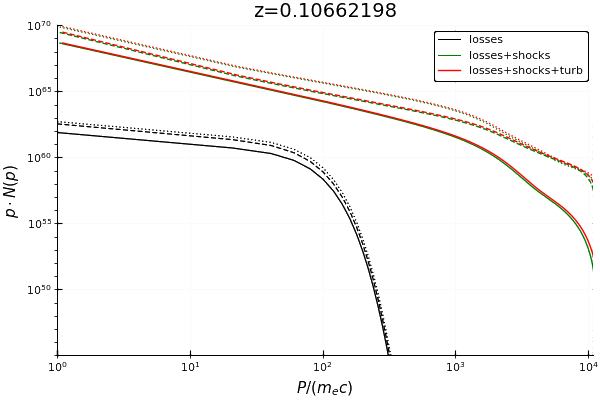}
    \caption{Evolution of the spectral energy distributions of electrons emitted by the radio galaxy in Run1, for three different acceleration/cooling models (colours) and for three different radial ranges: $r \leq 300$~kpc (solid),  300~kpc $< r \leq 600$~kpc (dashed) and $r > 600$~kpc (dotted).}
    \label{fig:spectra2}
\end{figure}

\begin{figure}
    \centering
        \includegraphics[width=0.49\textwidth,height=0.203\textheight]{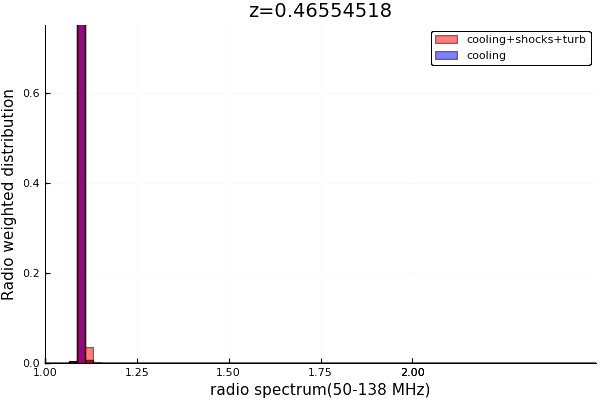}
     \includegraphics[width=0.49\textwidth,height=0.203\textheight]{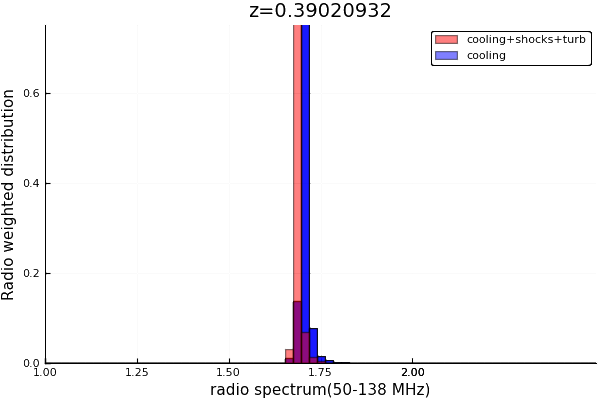}
      \includegraphics[width=0.49\textwidth,height=0.203\textheight]{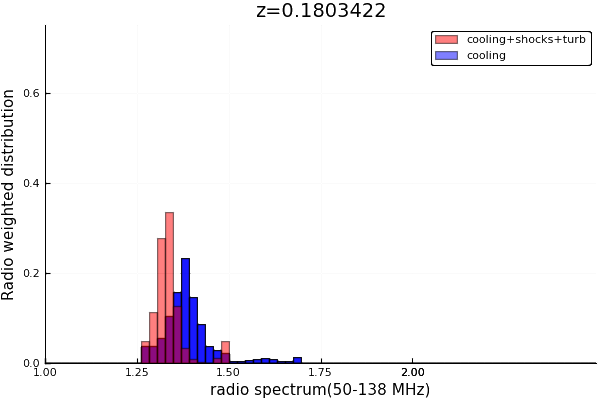}
       \includegraphics[width=0.49\textwidth,height=0.203\textheight]{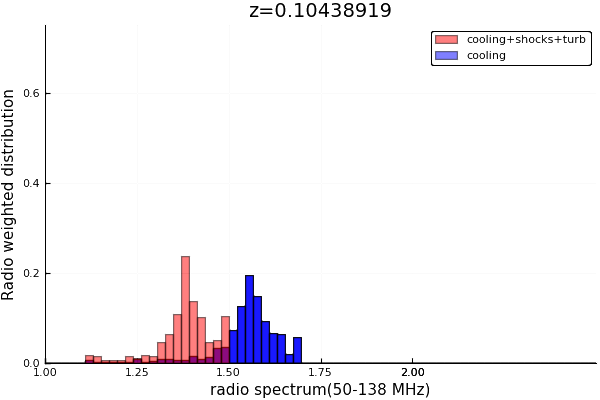}
    \caption{Evolution of the (radio power weighted) distribution of radio spectral index  between 50 and 138 MHz, for four epochs in the Run2 and only including radio detectable pixels at 50MHz (assuming a LOFAR LBA observation). Only the cooling and cooling+shocks+turbulence cases are shown for clarity.}
    \label{fig:spec_pdf}
\end{figure}

\subsection{The impact of radio galaxies on the ICM} 
\label{subsec:icm}
We start by focusing on the dynamics of the ICM as a results of radio galaxies launched at 
two different times. 

In Fig.~\ref{fig:map0} and Fig.~\ref{fig:map1}, we show the evolution of the X-ray emission (colours) overlaid with the projected distribution of radio emission at $50 ~\rm MHz$ from our tracers (contour) for the two runs. In both cases, no observational cuts were applied in drawing the radio contours. Altogether, we show six snapshots to 
 display the most salient features of AGN feedback: 
\begin{itemize}
\item i) the release of powerful shocks ahead of jets (with $\mathcal{M} \sim 3.0$ in Run2 and $\mathcal{M} \sim 4.0$ in Run1), marked by discontinuities in the X-ray surface brightness in the first panel of each figure; 

\item ii) the formation of pairs of cavities, partially void of X-ray emitting gas in the region inflated by the two radio jets (second and third panel of each figure);
\item iii) in the Run2 case, the development of two $\sim 100-250$~kpc long jets, which remain straight at least $\sim 400$~Myr since their ejection; in the Run1 case, jets are dissolved already $\sim 200$~Myr after their ejection;
\item iv) the subsequent evolution of jet-inflated radio lobes, which expand laterally (i.e. perpendicular to the jet axis) up to $\sim 300-500$~kpc;
\item v) the progressive mixing of the radio-emitting plasma with the ICM, which covers an increasingly larger number of cells in the innermost cluster regions. While in Run2  the distribution of electrons has a large covering factor in the innermost cluster regions down to $z=0.1$ (or below), electrons released in Run1 have a much lower covering factor by the end of the simulation. The reason is that they have been dispersed in the ICM earlier (last panel).  
\end{itemize}

Next, we investigate the impact of radio galaxies on the radial profiles of gas density, temperature and magnetic field strength of the ICM.
Figure~\ref{fig:prof1} shows the 
sequence of radial profiles of gas density, temperature and magnetic field strength comparing Run2 and Run0. The impact of the radio galaxy on the overall properties of the ICM is far from dramatic in these runs. The difference to Run0 models are limited to quite localised excesses of gas heating due to the thermal feedback by AGN-driven shock waves, and to slightly more pronounced excesses of magnetic fields due to radio jets. However, while the differences are clear close to the epoch of jet launching ($z \geq 0.46$), they tend to decrease with time at most radii. However, some excess magnetisation in Run2 (of order $\sim 3-10$) remains visible in the radial profile of the cluster even at later redshift, within $\leq 250$~kpc.  Very similar trends are found when comparing the radial profiles of Run1 and Run0 (not shown).

As a caveat, we should stress that our limited (and typically decreasing with radius) spatial resolution prevents the Reynolds number from getting large enough for an efficient amplification of magnetic fields seeded by radio galaxies, which can proceed only when the spatial resolution of the simulation is enough to resolve the Alfven length-scale 
\citep[e.g.][and references therein]{review_dynamo,bv20}.  Hence, these simulations likely provide a {\it lower limit} to the ICM magnetic field strength after the jet activity has ceased, and future higher resolution (and more expensive) simulations will be needed to assess the maximal possible amplification here.  However, recent simulations by \citet{2020arXiv201113964E} support the fact that magnetic fields injected by jets only affect the close proximity of AGN, while the cluster-wide magnetic field distribution is dominated by large-scale turbulence.

\begin{figure*}
    \centering
    
    \includegraphics[width=0.99\textwidth]{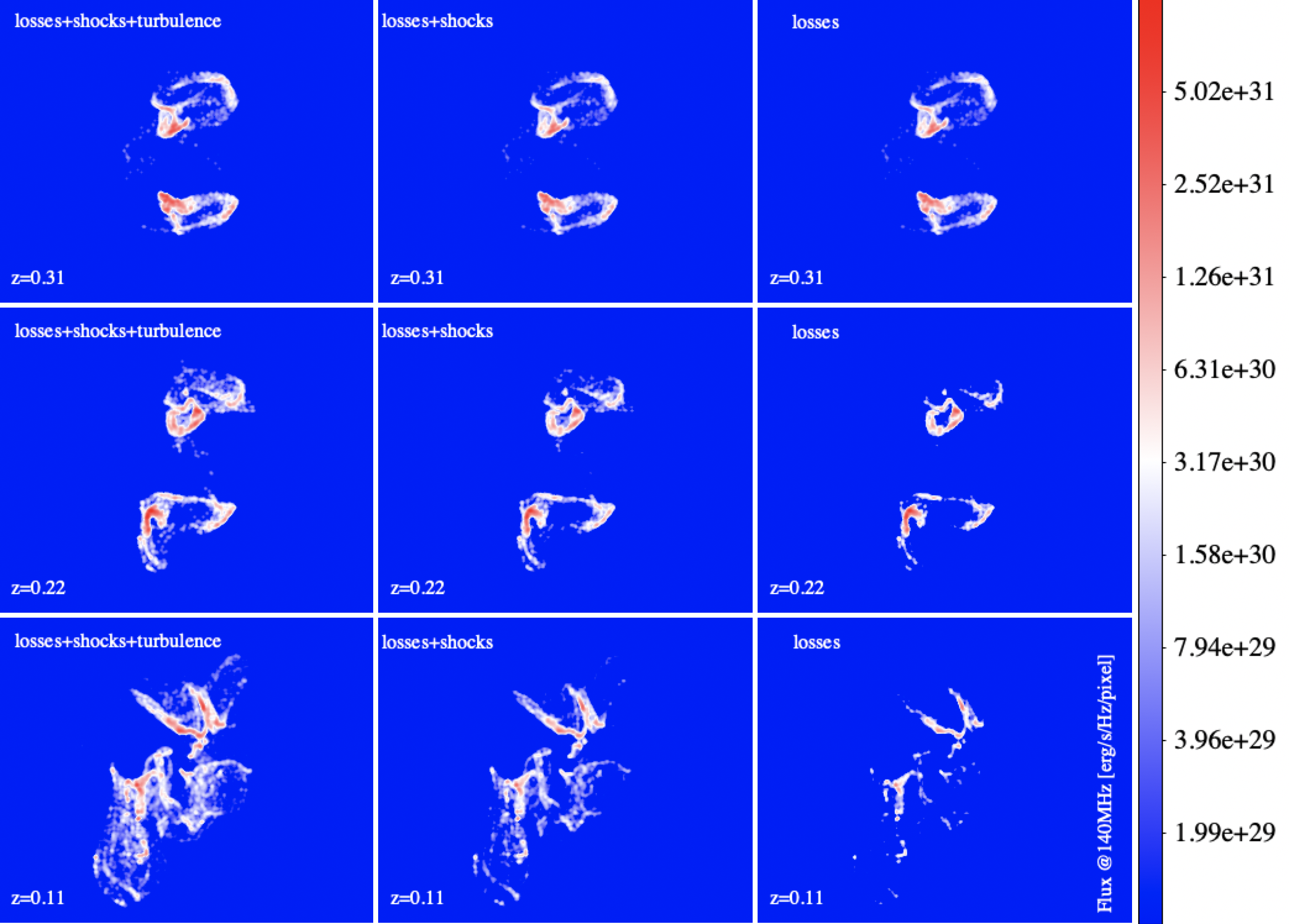}
    \caption{Evolution of the radio emission for Run2, for three epochs and the three re-acceleration models investigated in the paper. Only the emission above the putative $3 ~\sigma$ detectability using LOFAR-HBA  (140 MHz) is shown.}
    \label{fig:map_radio1}
\end{figure*}

\begin{figure*}
    \centering
    
    \includegraphics[width=0.99\textwidth]{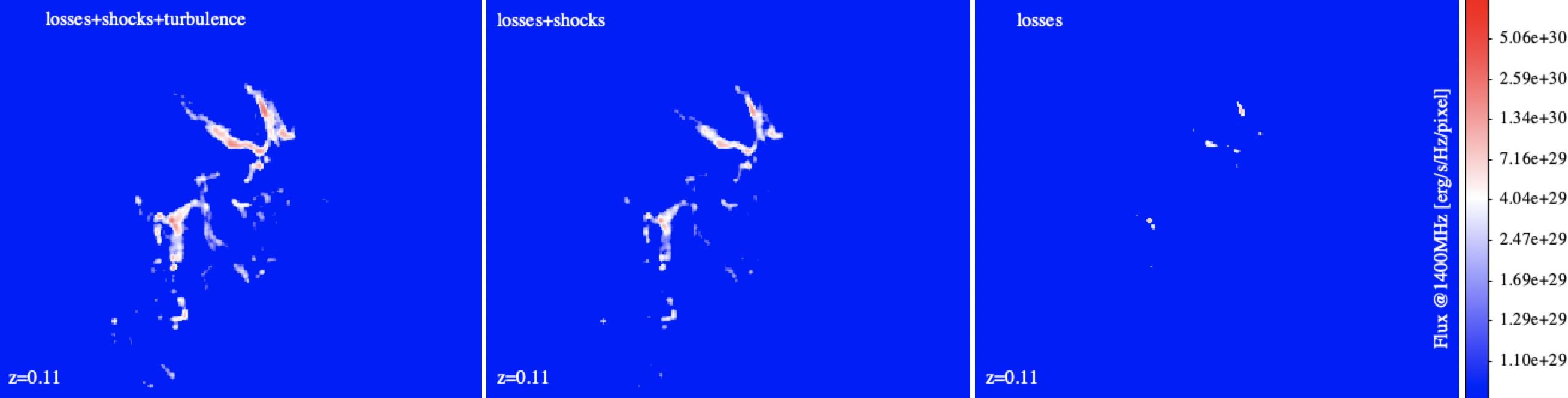}
    \caption{Radio emission for Run2 at 1.4 GHz, for the three re-acceleration models investigated in the paper. Only the emission above the putative $3 ~\sigma$ detectability using JVLA is shown.}
    \label{fig:map_radio2}
\end{figure*}

\begin{figure*}
    \centering
      \includegraphics[width=0.99\textwidth]{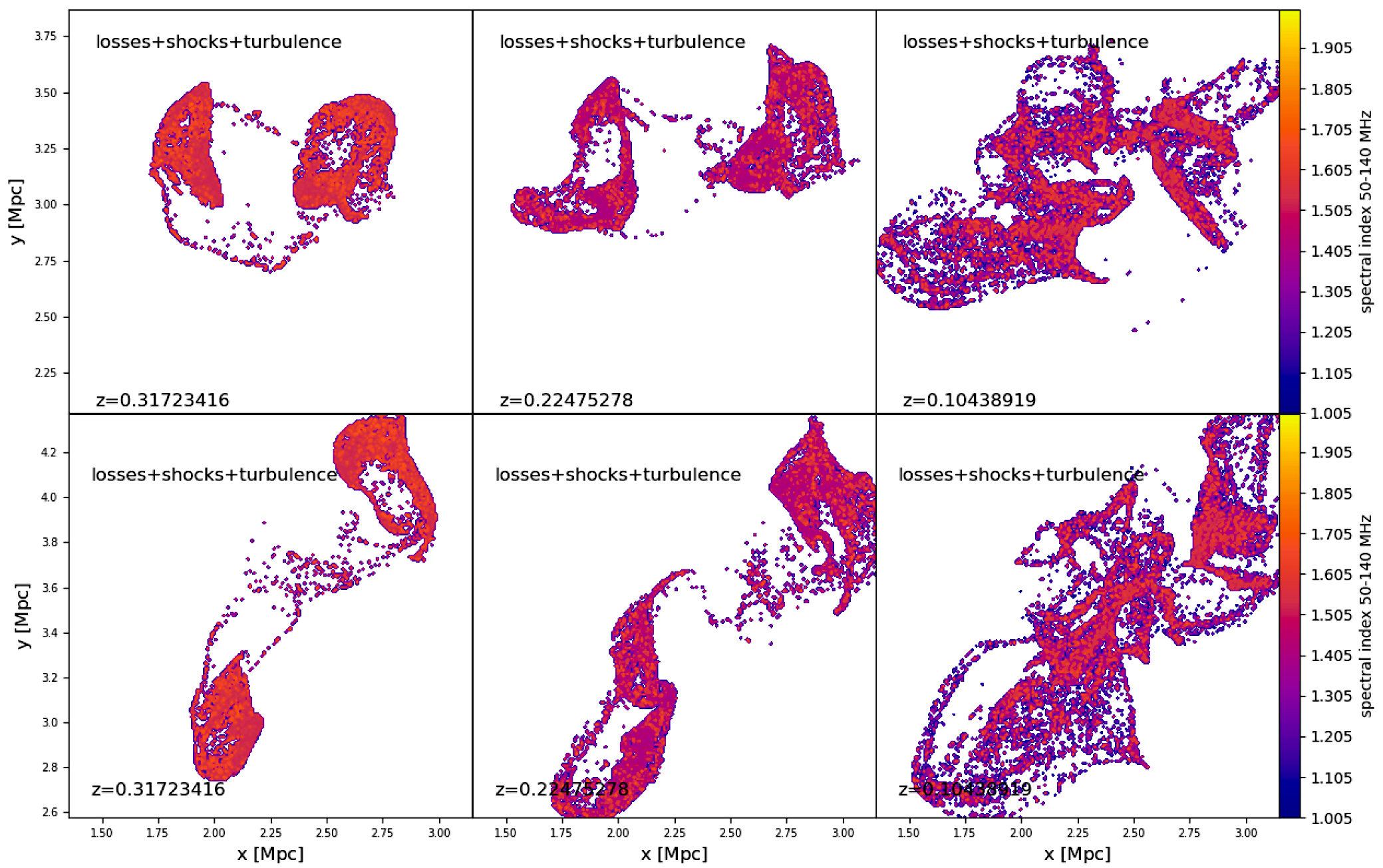}
    \caption{Radio spectral index maps for the 50-140 MHz range, for three epochs and two lines of sight, for the Run2 simulation. Only regions detectable with LOFAR-HBA (as in Fig.\ref{fig:map_radio1}) are shown.}
    \label{fig:map_radio3}
\end{figure*}

\subsection{The long-term evolution of relativistic electrons injected by radio galaxies}

We now focus on the properties of the ICM as viewed by the tracer particles as they move through the ICM.  
 Fig.~\ref{fig:distance} gives the evolution of the distance travelled by tracers, in which we compared the median distance covered by tracer particles initially placed in jets (or in corresponding cells, in Run0 models). Tracers ejected by radio jets always travel a larger distance compared to their corresponding particles in Run0 models. This follows from the combined effect of outflows, as well as of large-scale mixing by turbulent motions, which are typically found to be significant at large radii \citep[e.g.][]{2020MNRAS.495..864A}.
The effect is more significant in Run1, where the jet power is large enough to propel a fraction of the lobe to beyond the cluster virial radius, and in general is sufficient to spread the jets material over the entire cluster volume. 
In Run2, the jets are not powerful enough to escape through the denser and hotter ICM, and the jet material settles to a similar distance as the Run0 case over time.
In both jet simulations, being dominated by the fast bipolar transport, the average spatial separation of tracers is faster than a Richardson ($\propto t^{3/2}$) diffusion, which applies to the pair dispersion statistics of passive tracers in Kolmogorov-type turbulence \citep[e.g.][]{va10tracers}. The latter trend is instead overall measured initially in the pair dispersion statistics of tracers in  Run0. Later on ($\geq 1-2$  $ \rm Gyr$) the average separation progresses slower with time, following normal diffusion ($\propto t^{1/2}$),  approximately after tracers have travelled a distance larger than the typical size of turbulent eddies. Interestingly, tracers ejected at $z=1$ as in Run1 continue to spread in the volume at a relatively fast rate, compared to Run2, owing to the fact that their initially faster expansion has placed them (on average) on outer cluster regions, where the mixing due to turbulence and bulk motions is larger  \citep[][]{lau09,2020MNRAS.495..864A}.

The various panels in Fig.~\ref{fig:scatter} show the evolution of magnetic field strength, temperature and velocity curl for tracers in the various runs.
Since tracers in Run2/Run1 typically spread to a larger radius (and a lower density) compared to the corresponding tracers in the control run (Run0), in this case we always measure the quantities for tracers found within the same radius from the cluster centre ($\leq 500$ kpc) at all times. The changes of the ambient medium that tracer particles are subjected to, are typically violent and involve velocity fluctuations with amplitude several $\sim 100 \rm ~km/s$, even during the later stages of lobe evolution.  Figures ~\ref{fig:velocity} and ~\ref{fig:Bfield} show the evolution of the absolute velocity and the magnetic field strength averaged along the line of sight(s), for different epochs in Run2. The sequence shows how the interaction with the environment, as well as the variations of local conditions experienced by the lobes as they expand, cause variations in the velocity and magnetic field experienced by relativistic particles.

Nevertheless, the activity of radio galaxies in both runs adds additional fluctuations to the dynamical evolution of Lagrangian tracers, especially in the first hundreds of Myr since the AGN bursts.
In particular, the magnetisation, temperature and vorticity of particles tracing jets are found to be significantly higher (up to an order of magnitude) up to $\sim 200-400$ Myr after the release of jets.  Interestingly, the magnetic field strengths carried by injected tracers show the largest difference compared to Run0. In the case of Run1, the injected particles carry a significantly higher magnetic field strength even $\sim 2$ Gyr after their release by jets. For comparison, the typical temperature of the same particles are indistinguishable from the tracers in Run0 already after $\sim \rm ~400$ Myr since their injection.  Also in Run2, particles released by jets have a significantly higher magnetic field than their counterparts in Run0, up to $\sim 1 \rm ~Gyr$ after their release. 
These differences are intriguing, in the sense that the non-thermal energy content of  particles tracking jets can carry a  long dynamical memory of previous AGN seeding events, unlike their corresponding thermal properties, which seem to carry a much shorter (a factor $\sim 2-4$ here) memory of AGN events. This is consistent with the  thermodynamic radial profiles of the ICM previously studied in Sec.~\ref{subsec:icm}: the violent mixing of gas phases typically found in galaxy clusters at all times quickly entangles the past energy output from AGN with the cluster atmosphere (which is, after all, the most salient feature of AGN feedback). The fact that the magnetic field properties of a fraction of the ICM gas (directly affected by jets) carries a long lasting memory of seeding events offers a powerful probe into the past activity of AGN, which can be used by radio observations to assess the energetic of past feedback events in the lifetime of a cluster, to a larger extent compared to what standard X-ray analysis can do \citep[e.g.][]{2012A&A...547A..56D,2012MNRAS.427.3468B,2018MNRAS.476.1614C}.

\subsubsection{The evolution of electron and radio spectra}

In our runs, relativistic electrons are always first seeded in the ICM by radio jets, and are later subject to energy losses and/or re-acceleration processes,  depending on the local gas conditions they encounter during their propagation.

Most electrons released by radio galaxies suffer from significant adiabatic losses as the radio lobes expand into the ICM, and hence their radio synchrotron power decreases quickly. However, a fraction of the electrons can undergo further (re)acceleration by shocks and turbulence. 

Fig.~\ref{fig:spectra_tau} shows the electron acceleration and losses in typical conditions of a dynamical ICM. We give the median values of the timescales for energy losses $\tau_{\rm loss}$, acceleration  $\tau_{\rm acc}$ and advection $\tau_{\rm adv}$, using $p=10^3$ electrons as a reference.

The timescales for shock acceleration ($\tau_{\rm DSA}$) are much smaller because shocks only cover a tiny fraction of the cluster volume. Hence, we show the turbulent re-acceleration timescale, $\tau_{ASA}$ (Sec. 2.3). Likewise, the timescale for Coulomb losses ($\tau_{\rm c}$) is much larger than the radiative timescale for IC and synchrotron losses ($\tau_{\rm rad}$) for $p = 10^3$, hence only the second is displayed in the plot. 

We should point out some interesting trends: the radiative loss time for electrons steadily increases with time, following the decrease of the  $\propto (1+z)^4$ term related to Inverse Compton losses (Eq. \ref{eq:ic}). The latter dominates the radiative loss terms considering that the typical magnetic field strength probed by most tracers in cells gets smaller than the Cosmic Microwave Background (CMB)-equivalent field strength most of the time.
The contributions of adiabatic compression/expansion often  switches roles in their impact on the evolution of particles, i.e. from enhancing to decreasing the particle energies, on timescales of $\geq 100-200$ Myr. In both runs, the particles are first subjected to strong adiabatic losses, followed after $5 \times 10^8$ yr by adiabatic gains due to strong compression by the external ICM. Later on, particles follow more erratic sequences of expansion/compression as shocks waves cross the ICM. Moreover, particles are dispersed in a less dense ICM, particularly in Run1. 
The acceleration timescale related to the turbulence also significantly varies during the evolution and depending on the local ICM conditions; in general turbulent re-acceleration is less efficient to balance cooling losses for $p=10^3$ electrons in the Run1 case, at least for half of the simulated evolution, because particles diffuse to a larger cluster radius, where the relative impact of solenoidal turbulence is generally smaller \citep[e.g.][]{va17turb}, with the exception of a boost of turbulent acceleration in at least one peripheral sector of the group ($\sim 11-13$ $\rm Gyr$) following the last major merger experienced by the cluster (see Fig.1).

Conversely, electrons ejected by the radio galaxy in Run2 remains more confined in the innermost ICM, where solenoidal turbulence is dominant, and the median $\tau_{ASA}$ is measured to be smaller (or of the same order) of the median synchrotron cooling time for $p=10^3$ electrons for most of the investigated evolution. The complex evolution of the amplitude of different loss/acceleration terms once more stresses the importance of full numerical simulations for a detailed simulation of the full evolution of radio emitting electrons in a realistic environment. 

The panels in Fig.~\ref{fig:spectra_evol} show an example of the spectral energy evolution of $100$ electron tracers injected by the radio galaxy in Run2,  initially placed in the cells with the highest magnetic field strength in the jets, i.e. roughly corresponding to the location of the brightest emission spots in the southern lobe of Fig.\ref{fig:map_jet_Run2}.
While they are advected at a close distance between each other, these electron tracers are  subject to several possible combinations of re-acceleration events (or absence of thereof): only losses, losses and injection and re-acceleration by shocks, losses and only shock re-acceleration, losses and only turbulent re-acceleration, or all mechanisms together, from top to bottom, respectively. 

In the presence of cooling and adiabatic losses alone, this family of particles steadily accumulates at a momentum of $p \sim 10^2$ by the end of the run, consistent with the literature \citep[][]{gb01,br07,2013MNRAS.435.1061P}.
The re-acceleration by shocks and turbulence instead largely increases the energy budget of  particles in the $p \leq 10^3$ range, while the  direct injection of "fresh" electrons by shocks (here seeded by a $\mathcal{M} \sim 3.0$ shock at $\sim 2 \rm ~Gyr$ after the jet injection), replenishes the high energy tail of the momentum distribution with a power-law.
When all the above mechanisms are considered, we can see that the interaction of electrons with the ICM dynamics produce a complex spectrum with several time-dependent features, which overall allows a steep population of radio emitting electrons to survive several $\rm Gyr$ after the release by radio jets, and it overall generates a more prominent reservoir of low energy relativisic electrons, for a significant fraction of the lifetime of this galaxy group. 

The complete view of the the radial distribution in the spectra for different re-acceleration scenarios is given in Fig.~\ref{fig:spectra1}-\ref{fig:spectra2}. In this figure we use different line styles to identify the total spectra of all simulated tracers, binned in three different radial shells ($r \leq 300$~kpc with a  solid line,  300~kpc $< r \leq 600$~kpc with a dashed line and $r > 600$~kpc with a dotted line).
We  observe a similar evolution at all radii of a complex spectrum (albeit smoothed by the combination of the many more spectra considered here), with tails of  $p \geq 10^3$ electrons. This is a consequence of the crossing of shock waves, and the steady build-up of a fossil electron reservoir for $p \leq 10^2$ due to the combination of shocks and turbulent re-acceleration. The $p\sim 10^3$ bump caused by turbulent re-acceleration is more evident in Run2, where the electrons remain more confined in the cluster innermost regions. In those core regions, $\sim 30-50\%$ of the volume is typically filled by solenoidal turbulent motions \citep[][]{va17turb}, while only $\leq 0.1 \%$ of cells in clusters are shocked \citep[e.g.][]{va09shocks}. However, we should point out that our model probably underestimates the highest energy tail of the energy distribution, by a factor $\sim 2$, owing to the missing term of stochastic turbulent acceleration,  whose computation would requires more advanced solvers \citep[e.g.][]{2014MNRAS.443.3564D}. Moreover, turbulent reacceleration is predicted to be more efficient at larger cluster masses \citep[e.g.][]{cassano05,cassano10}, hence shorter acceleration times can be expected for more massive objects. 

On the other hand, few electrons can exist with momenta larger than a few $\sim 10^2$ if they are only subject to radiative cooling and adiabatic expansion. As seen already above, re-accelerated electrons can reach a higher energy in Run2, while in Run1 the bulk of the energy distribution of electrons spreads out to larger cluster radii due to their earlier ejection  from their source. 

Since only in Run2, enough electrons are confined in the ICM at late redshift to give rise to potentially detectable radio emission, we will focus on those in the following, and work out the observable radio properties of this run .

The above differences in the energy distribution driven by re-acceleration in the ICM obviously have a direct impact on the observable radio emission which depends on the slope of available CR electrons. 
The distribution of the radio spectral index of the detectable emission in Run2 at different epochs is shown in  Fig.~\ref{fig:spec_pdf}. For clarity, only the extreme cases with  all re-acceleration terms included, or only with losses, are shown. The radio-weighted distributions are generated only for pixels above the putative detection threshold of a LOFAR LBA observation. As expected,  the radio spectral index distribution of models where re-acceleration terms are included shows the marked tendency of having flatter radio spectra, compared to the pure cooling scenario, at each epoch and more increasingly in more evolved epochs. For most of their evolution after $z \leq 0.4$ in Run2, most of the detectable radio emission from tracers undergoing shock and turbulent re-acceleration has a radio spectrum $\alpha \geq 1.25$ (measured from $50$ to $138$ MHz), while the residual emission in case only cooling and adiabatic losses are included is typically a factor $\Delta \alpha \sim 0.5$ steeper at equal epochs.  In general, the steep spectra of fossil electrons in scenarios without re-acceleration from the ICM make them difficult to observe at low frequencies ($\leq 140 \rm ~MHz$), and almost invisible at high frequencies ($\geq 1.4 \rm ~GHz$). 

Figure~\ref{fig:map_radio1} gives an overview of the radio emission at 140 MHz for electrons in Run2. We compare the three re-acceleration scenarios and only sow the emission above the approximate $\geq 3 \sigma$ detection level of LOFAR-LBA $\sim 8$ hr observations (with $\sigma \approx 0.2 \rm mJy/beam$ for a $30" \times 30"$ resolution beam, e.g. \citealt{bonafede21}), placing in all cases our source at a reference distance of $z=0.1$. Figure~ \ref{fig:map_radio2} also shows the emission at 1.4 GHz, limited to the last snapshot of the above panel, only for the detectable emission above a realistic approximate sensitivity of a $\sim 5$ hr observation with the JVLA ($\sigma \approx 7 \rm ~\mu Jy/beam$ for a $5" \times 5"$ resolution beam, e.g. \citealt{rajp20}). 
Clearly, only when all sources of re-acceleration from the ICM are included, the diffuse emission from fossil electrons injected by radio galaxies several billions years ago can become again detectable in the radio band. 
In the absence of turbulent and shock re-acceleration, the brightest emission patches of fossil electrons can hardly be seen as connected on $\sim \rm ~Mpc$ scales at high frequencies,

Hence, the late radio morphologies ($\geq 2$ Gyr since first injection) are entirely determined by the passive advection of fossil electrons in the perturbed ICM. Such structures do not have any clear symmetry, appear entirely different when observed from a different viewing angle and in general do not show simple distributions of radio spectral indices, as shown in Fig.~\ref{fig:map_radio3}. 

Interestingly, some of the observed radio morphologies are qualitatively similar to recent deep LOFAR observations of complex radio structures in nearby galaxy clusters \citep[e.g.][]{2017SciA....3E1634D,2020A&A...643A.172I,2020ApJ...897...93B,2020A&A...634A...4M}.
In particular, we would like to point to the rich, filamentary structures that are found in and around radio galaxies. For example, \cite{2020A&A...636L...1R} showed collimated synchrotron threads linking the radio lobes of ESO 137-006. These thin structures are most likely magnetically dominated and resemble the morphology seen in Fig.~15.

A careful comparison to observations is beyond the goal of this paper. However, our results regarding morphology suggest that it will be challenging to relate the age of such radio sources to their original radio galaxy. This is because their evolution appears to be mostly related to their interaction with cluster weather. On the other hand, the extensive theoretical study of all possible re-acceleration scenarios that can take place in the realistic ICM, also for a larger variety of objects and merger scenarios, will have the potential to make such remnants of radio galaxy activity a probe of ICM physics \citep[][]{2017PhPl...24d1402J}.

\section{Caveats}

This paper is only a first step towards a more realistic model of the complex interplay between radio galaxies and their environment, and it is admittedly very limited in several important aspects.  

First, our runs do not include radiative gas cooling, hence there is no self-consistent coupling between the feeding of gas undergoing overcooling and the duty cycle of SMBH jets. This would introduce a number of numerical and physical complications in the modelling of the ICM, which are not entirely solved in cosmological simulations \citep[e.g.][]{2008A&A...482L..13D,co11,ruszkowski11,2017MNRAS.470.1121T} 
but are nevertheless crucial for any realistic modelling of ICM atmospheres \citep[e.g.][]{2020ApJ...899...70V}. 
While this choice was motivated to best focus on the dynamical effect of the feedback from radio galaxies (or absence thereof) on the amplification of magnetic fields in the ICM and on the evolution of relativistic electrons, removed by the additional effects of compression from gas cooling,  future 
developments will have to couple the launching of radio jets with the duty cycle of AGN feedback \citep[e.g.][]{2020arXiv200812784B,2020arXiv201011225T}.
 While our choice of having single and well defined episodes of jets release by radio galaxies is convenient to study the dynamics of ejected electrons in a clean way,  recent results from radio surveys have instead suggested that at least the most massive radio galaxies can be "on" for most of their lifetime \citep[e.g.][]{2019A&A...622A..17S}, unlike the discontinuous activity scenario explored in this work. Only through a more consistent coupling of cooling/feedback in our simulations we can approach the complexity of real duty cycles of radio galaxies. The extension of this study with clusters in a different dynamical state will also allow us to explore different combinations of shocks and turbulent re-acceleration events, naturally triggered by different merger configurations. 

Secondly, our simulations cannot reproduce the actual interaction between the relativistic (either electron or proton dominated) content of jets and the thermal ICM, as our jets are filled entirely with magnetic fields and hot thermal gas, as commonly done by simulations of this kind \citep[e.g.][]{xu09,gaspari11b,teyssier11}.  
In order to directly test the impact of a significant relativistic energy content in protons or electrons, future development will have to resort to two-fluid modelling of the ICM \citep[e.g.][]{2018MNRAS.481.2878E,2019ApJ...871....6Y} or similar \citep[e.g.][]{va13feedback}. 

The use of passive tracer particles to track the propagation of relativistic electrons introduces an element of discreteness in our modelling, in the sense that
compared to a fluid approach (i.e. via two-fluid modelling, e.g. \citealt{2020MNRAS.491..993G}, \citealt{2020arXiv200906941O}) it cannot fully map the spatial diffusion of jet ejecta over time. In this respect, the small-scale morphology of our radio jets may change if the spatial diffusion of relativistic electrons is taken into account, even if the global dynamics of lobes is not expected to change significantly.

Our Fokker-Planck model for particles (Sec. 2.3) is only concerned with specific choices for the acceleration of relativistic electrons via shock waves. It also includes a simplified treatment of the turbulent re-acceleration, in which only a systematic acceleration effect on the global distribution can be captured. Significant effects on the highest energy tail of the electron distributions can be expected for different choices in acceleration scenarios \citep[e.g.][]{wittor20}, and will be explored in future work with more sophisticated numerical schemes \citep[e.g.][]{2014MNRAS.443.3564D}.

Finally, our simulations only evolved a $P_M=R_{\rm M}/R_{\rm e}=\nu/\eta \approx 1$ plasma, i.e. a plasma in which the resistivity and viscosity are of the same order and are numerical by nature. 
This is clearly a strong and probably unrealistic assumption for the ICM  \citep[e.g.][]{2004ApJ...612..276S,bl11b,2016ApJ...817..127B}, but more expensive, fully kinetic simulations will be required to address the small-scale dynamo in a more realistic way \citep[e.g.][]{2016PNAS..113.3950R}.

\section{Conclusions}

In this paper, we simulated the evolution of relativistic electrons injected in the ICM by active radio galaxies. Using a Fokker-Planck method on Lagrangian tracer particles, we tracked the cooling and re-acceleration of electrons as a function of time.
These new simulations allow us to start exploring in detail, the idea that diffuse radio emission in the intracluster medium requires a volume-filling population of old relativistic electrons, seeded by radio galaxies \citep[e.g.][]{1977ApJ...212....1J, 1987A&A...182...21S,sa99,gb01}.  Understanding how  radio jets disperse their relativistic and magnetic content on  $\sim ~ \rm Mpc$ scales has been shown to be critical to explain how radio halos, radio relics, mini radio halos and radio bridges power their large-scale emission: current models need to postulate the existence of an ubiquitous, volume-filling reservoir of  "fossil" , $\gamma \sim 10^2-10^3$ electrons to account for the normalisation of the observed emission \cite[e.g.][for reviews]{bj14,2019SSRv..215...16V}. However the existence of such a reservoir is a mere hypothesis, and whether or not radio galaxies could produce them in their lifetime remains unproven.  

Our analysis shows that relativistic electrons from radio galaxies efficiently spread across the intracluster volume, as a result of cluster-wide turbulence and the momentum of powerful jets, even considering a single episode of jet injection in the lifetime of a radio galaxy.

Close to the injection epoch, we find that the magnetic output of jets dominates close to the radio galaxies ($\leq 500$~kpc). Yet, the large-scale distribution of magnetic fields seems to be dominated by the growth of the galaxy cluster. Hence, the input from jets is subdominant with respect to the overall cluster magnetic field - with the exception of the cavities inflated by jets, which preserve a higher and distinct magnetic field up to $\sim 2 \rm ~Gyr$ since the jet release. 

Meanwhile, the cosmic-ray electrons deposited by jets into the ICM steadily build (at least  part of) the reservoir of suprathermal particles, which can be further re-accelerated 
by shock waves and turbulence.
However, the final energy (and radio) distributions critically depend on the details of (re)acceleration events caused by shock waves and turbulence.

Spatially and spectrally resolved radio observations (especially in the low-frequency regime explored by LOFAR, MWA, and in the future by the SKA-LOW) will be able to retrace the sequence of radio-mode feedback and probe a number of plasma processes \citep[e.g.][]{Hodgson21a}.

In conclusion, electrons and magnetic fields seeded by radio galaxies, or AGN in general, are becoming an additional and required ingredient for cosmological simulations to properly model the complexity of non-thermal emission in groups and clusters of galaxies, and this work represents a first, already very informative step into this direction.

\section*{Acknowledgments}
  We are grateful to the anonymous referee for their helpful comments on the first draft of this work.
  The authors gratefully acknowledge the Gauss Centre for Supercomputing e.V. (www.gauss-centre.eu) for supporting this project by providing computing time through the John von Neumann Institute for Computing (NIC) on the GCS Supercomputer JUWELS at J\"ulich Supercomputing Centre (JSC), under projects "stressicm" and projects no. 11823, 10755 and 9016 as well as hhh44. \\ 
  F.V. acknowledges financial support from the European Union's Horizon 2020 program under the ERC Starting Grant "MAGCOW", no. 714196. D.W. is funded by the Deutsche Forschungsgemeinschaft (DFG, German Research Foundation) - 441694982. MB acknowledges support from the Deutsche Forschungsgemeinschaft under Germany's Excellence Strategy - EXC 2121 "Quantum Universe" - 390833306.  We also acknowledge the usage of online storage tools kindly provided by the Inaf Astronomica Archive (IA2) initiave (http://www.ia2.inaf.it) and of the Cosmological Calculator by N. Wrigth  (http://www.astro.ucla.edu/~wright/CosmoCalc.html). 
  We gratefully acknowledge helpful scientific discussions with M. Brienza and E. Vardoulaki, concerning details of radio observations of radio galaxies. 
 \bibliographystyle{aa}
 \bibliography{franco3} 

\begin{thebibliography}{149}
\expandafter\ifx\csname natexlab\endcsname\relax\def\natexlab#1{#1}\fi

\bibitem[{{Abdulla} {et~al.}(2019){Abdulla}, {Carlstrom}, {Mantz}, {Marrone},
  {Greer}, {Lamb}, {Leitch}, {Muchovej}, {O'Donnell}, {Plagge}, \&
  {Woody}}]{2019ApJ...871..195A}
{Abdulla}, Z., {Carlstrom}, J.~E., {Mantz}, A.~B., {et~al.} 2019, \apj, 871,
  195

\bibitem[{{Ackermann} {et~al.}(2014){Ackermann}, {Ajello}, {Albert},
  {Allafort}, {Atwood}, {Baldini}, {Ballet}, {Barbiellini}, {Bastieri},
  {Bechtol}, {Bellazzini}, {Bloom}, {Bonamente}, {Bottacini}, {Brandt},
  {Bregeon}, {Brigida}, {Bruel}, {Buehler}, {Buson}, {Caliandro}, {Cameron},
  {Caraveo}, {Cavazzuti}, {Chaves}, {Chiang}, {Chiaro}, {Ciprini}, {Claus},
  {Cohen-Tanugi}, {Conrad}, {D'Ammando}, {de Angelis}, {de Palma}, {Dermer},
  {Digel}, {Drell}, {Drlica-Wagner}, {Favuzzi}, {Franckowiak}, {Funk}, {Fusco},
  {Gargano}, {Gasparrini}, {Germani}, {Giglietto}, {Giordano}, {Giroletti},
  {Godfrey}, {Gomez-Vargas}, {Grenier}, {Guiriec}, {Gustafsson}, {Hadasch},
  {Hayashida}, {Hewitt}, {Hughes}, {Jeltema}, {J{\'o}hannesson}, {Johnson},
  {Kamae}, {Kataoka}, {Kn{\"o}dlseder}, {Kuss}, {Lande}, {Larsson},
  {Latronico}, {Llena Garde}, {Longo}, {Loparco}, {Lovellette}, {Lubrano},
  {Mayer}, {Mazziotta}, {McEnery}, {Michelson}, {Mitthumsiri}, {Mizuno},
  {Monzani}, {Morselli}, {Moskalenko}, {Murgia}, {Nemmen}, {Nuss}, {Ohsugi},
  {Orienti}, {Orlando}, {Ormes}, {Perkins}, {Pesce-Rollins}, {Piron}, {Pivato},
  {Rain{\`o}}, {Rando}, {Razzano}, {Razzaque}, {Reimer}, {Reimer}, {Ruan},
  {S{\'a}nchez-Conde}, {Schulz}, {Sgr{\`o}}, {Siskind}, {Spandre}, {Spinelli},
  {Storm}, {Strong}, {Suson}, {Takahashi}, {Thayer}, {Thayer}, {Thompson},
  {Tibaldo}, {Tinivella}, {Torres}, {Troja}, {Uchiyama}, {Usher},
  {Vandenbroucke}, {Vianello}, {Vitale}, {Winer}, {Wood}, {Zimmer}, {Fermi-LAT
  Collaboration}, {Pinzke}, \& {Pfrommer}}]{fermi14}
{Ackermann}, M., {Ajello}, M., {Albert}, A., {et~al.} 2014, \apj, 787, 18

\bibitem[{{Angelinelli} {et~al.}(2020){Angelinelli}, {Vazza}, {Giocoli},
  {Ettori}, {Jones}, {Brunetti}, {Br{\"u}ggen}, \&
  {Eckert}}]{2020MNRAS.495..864A}
{Angelinelli}, M., {Vazza}, F., {Giocoli}, C., {et~al.} 2020, \mnras, 495, 864

\bibitem[{{Banfi} {et~al.}(2020){Banfi}, {Vazza}, \& {Wittor}}]{Banfi20}
{Banfi}, S., {Vazza}, F., \& {Wittor}, D. 2020, arXiv e-prints,
  arXiv:2006.10063

\bibitem[{{Beresnyak} \& {Miniati}(2016)}]{2016ApJ...817..127B}
{Beresnyak}, A. \& {Miniati}, F. 2016, \apj, 817, 127

\bibitem[{{Berezinsky} {et~al.}(1997){Berezinsky}, {Blasi}, \&
  {Ptuskin}}]{bbp97}
{Berezinsky}, V.~S., {Blasi}, P., \& {Ptuskin}, V.~S. 1997, \apj, 487, 529

\bibitem[{{B{\^\i}rzan} {et~al.}(2020){B{\^\i}rzan}, {Rafferty}, {Br{\"u}ggen},
  {Botteon}, {Brunetti}, {Cuciti}, {Edge}, {Morganti}, {R{\"o}ttgering}, \&
  {Shimwell}}]{2020MNRAS.496.2613B}
{B{\^\i}rzan}, L., {Rafferty}, D.~A., {Br{\"u}ggen}, M., {et~al.} 2020, \mnras,
  496, 2613

\bibitem[{{B{\^\i}rzan} {et~al.}(2012){B{\^\i}rzan}, {Rafferty}, {Nulsen},
  {McNamara}, {R{\"o}ttgering}, {Wise}, \& {Mittal}}]{2012MNRAS.427.3468B}
{B{\^\i}rzan}, L., {Rafferty}, D.~A., {Nulsen}, P.~E.~J., {et~al.} 2012,
  \mnras, 427, 3468

\bibitem[{{Bodo} {et~al.}(1998){Bodo}, {Rossi}, {Massaglia}, {Ferrari},
  {Malagoli}, \& {Rosner}}]{1998A&A...333.1117B}
{Bodo}, G., {Rossi}, P., {Massaglia}, S., {et~al.} 1998, \aap, 333, 1117

\bibitem[{{Bonafede} {et~al.}(2021){Bonafede}, {Brunetti}, {Vazza},
  {Simionescu}, {Giovannini}, {Bonnassieux}, {Shimwell}, {Br{\"u}ggen}, {van
  Weeren}, {Botteon}, {Brienza}, {Cassano}, {Drabent}, {Feretti}, {de
  Gasperin}, {Gastaldello}, {di Gennaro}, {Rossetti}, {Rottgering}, {Stuardi},
  \& {Venturi}}]{bonafede21}
{Bonafede}, A., {Brunetti}, G., {Vazza}, F., {et~al.} 2021, \apj, 907, 32

\bibitem[{{Bonafede} {et~al.}(2014){Bonafede}, {Intema}, {Br{\"u}ggen},
  {Girardi}, {Nonino}, {Kantharia}, {van Weeren}, \&
  {R{\"o}ttgering}}]{2014ApJ...785....1B}
{Bonafede}, A., {Intema}, H.~T., {Br{\"u}ggen}, M., {et~al.} 2014, \apj, 785, 1

\bibitem[{{Booth} \& {Schaye}(2009)}]{2009MNRAS.398...53B}
{Booth}, C.~M. \& {Schaye}, J. 2009, \mnras, 398, 53

\bibitem[{{Borse} {et~al.}(2020){Borse}, {Acharya}, {Vaidya}, {Mukherjee},
  {Bodo}, {Rossi}, \& {Mignone}}]{2020arXiv200913540B}
{Borse}, N., {Acharya}, S., {Vaidya}, B., {et~al.} 2020, arXiv e-prints,
  arXiv:2009.13540

\bibitem[{{Botteon} {et~al.}(2020{\natexlab{a}}){Botteon}, {Brunetti}, {Ryu},
  \& {Roh}}]{2020A&A...634A..64B}
{Botteon}, A., {Brunetti}, G., {Ryu}, D., \& {Roh}, S. 2020{\natexlab{a}},
  \aap, 634, A64

\bibitem[{{Botteon} {et~al.}(2020{\natexlab{b}}){Botteon}, {Brunetti}, {van
  Weeren}, {Shimwell}, {Pizzo}, {Cassano}, {Iacobelli}, {Gastaldello},
  {B{\^\i}rzan}, {Bonafede}, {Br{\"u}ggen}, {Cuciti}, {Dallacasa}, {de
  Gasperin}, {Di Gennaro}, {Drabent}, {Hardcastle}, {Hoeft}, {Mandal},
  {R{\"o}ttgering}, \& {Simionescu}}]{2020ApJ...897...93B}
{Botteon}, A., {Brunetti}, G., {van Weeren}, R.~J., {et~al.}
  2020{\natexlab{b}}, \apj, 897, 93

\bibitem[{{Bourne} \& {Sijacki}(2020)}]{2020arXiv200812784B}
{Bourne}, M.~A. \& {Sijacki}, D. 2020, arXiv e-prints, arXiv:2008.12784

\bibitem[{{Brighenti} \& {Mathews}(2000)}]{2000ApJ...535..650B}
{Brighenti}, F. \& {Mathews}, W.~G. 2000, \apj, 535, 650

\bibitem[{{Br{\"u}ggen} \& {Kaiser}(2002)}]{2002Natur.418..301B}
{Br{\"u}ggen}, M. \& {Kaiser}, C.~R. 2002, \nat, 418, 301

\bibitem[{{Br{\"u}ggen} {et~al.}(2020){Br{\"u}ggen}, {Reiprich}, {Bulbul},
  {Koribalski}, {Andernach}, {Rudnick}, {Hoang}, {Wilber}, {Duchesne},
  {Veronica}, {Pacaud}, {Hopkins}, {Norris}, {Johnston-Hollitt}, {Brown},
  {Bonafede}, {Brunetti}, {Collier}, {Sanders}, {Vardoulaki}, {Venturi},
  {Kapinska}, \& {Marvil}}]{mb21}
{Br{\"u}ggen}, M., {Reiprich}, T.~H., {Bulbul}, E., {et~al.} 2020, arXiv
  e-prints, arXiv:2012.08775

\bibitem[{{Brunetti} \& {Jones}(2014)}]{bj14}
{Brunetti}, G. \& {Jones}, T.~W. 2014, International Journal of Modern Physics
  D, 23, 1430007

\bibitem[{{Brunetti} \& {Lazarian}(2011)}]{bl11b}
{Brunetti}, G. \& {Lazarian}, A. 2011, \mnras, 410, 127

\bibitem[{{Brunetti} \& {Lazarian}(2016)}]{2016MNRAS.458.2584B}
{Brunetti}, G. \& {Lazarian}, A. 2016, \mnras, 458, 2584

\bibitem[{{Brunetti} {et~al.}(2001){Brunetti}, {Setti}, {Feretti}, \&
  {Giovannini}}]{gb01}
{Brunetti}, G., {Setti}, G., {Feretti}, L., \& {Giovannini}, G. 2001, \mnras,
  320, 365

\bibitem[{{Brunetti} \& {Vazza}(2020)}]{bv20}
{Brunetti}, G. \& {Vazza}, F. 2020, \prl, 124, 051101

\bibitem[{{Brunetti} {et~al.}(2007){Brunetti}, {Venturi}, {Dallacasa},
  {Cassano}, {Dolag}, {Giacintucci}, \& {Setti}}]{br07}
{Brunetti}, G., {Venturi}, T., {Dallacasa}, D., {et~al.} 2007, \apjl, 670, L5

\bibitem[{{Bryan} {et~al.}(2014){Bryan}, {Norman}, {O'Shea}, {Abel}, {Wise},
  {Turk}, {Reynolds}, {Collins}, {Wang}, {Skillman}, {Smith}, {Harkness},
  {Bordner}, {Kim}, {Kuhlen}, {Xu}, {Goldbaum}, {Hummels}, {Kritsuk}, {Tasker},
  {Skory}, {Simpson}, {Hahn}, {Oishi}, {So}, {Zhao}, {Cen}, {Li}, \& {Enzo
  Collaboration}}]{enzo14}
{Bryan}, G.~L., {Norman}, M.~L., {O'Shea}, B.~W., {et~al.} 2014, \apjs, 211, 19

\bibitem[{{Bykov} {et~al.}(2019){Bykov}, {Vazza}, {Kropotina}, {Levenfish}, \&
  {Paerels}}]{Bykov19}
{Bykov}, A.~M., {Vazza}, F., {Kropotina}, J.~A., {Levenfish}, K.~P., \&
  {Paerels}, F.~B.~S. 2019, \ssr, 215, 14

\bibitem[{{Candelaresi} \& {Del Sordo}(2020)}]{2020ApJ...896...86C}
{Candelaresi}, S. \& {Del Sordo}, F. 2020, \apj, 896, 86

\bibitem[{{Capetti} {et~al.}(2019){Capetti}, {Baldi}, {Brienza}, {Morganti}, \&
  {Giovannini}}]{2019A&A...631A.176C}
{Capetti}, A., {Baldi}, R.~D., {Brienza}, M., {Morganti}, R., \& {Giovannini},
  G. 2019, \aap, 631, A176

\bibitem[{{Cassano} \& {Brunetti}(2005)}]{cassano05}
{Cassano}, R. \& {Brunetti}, G. 2005, \mnras, 357, 1313

\bibitem[{{Cassano} {et~al.}(2010){Cassano}, {Ettori}, {Giacintucci},
  {Brunetti}, {Markevitch}, {Venturi}, \& {Gitti}}]{cassano10}
{Cassano}, R., {Ettori}, S., {Giacintucci}, S., {et~al.} 2010, \apjl, 721, L82

\bibitem[{{Chang} \& {Cooper}(1970)}]{1970JCoPh...6....1C}
{Chang}, J.~S. \& {Cooper}, G. 1970, Journal of Computational Physics, 6, 1

\bibitem[{{Churazov} {et~al.}(2001){Churazov}, {Br{\"u}ggen}, {Kaiser},
  {B{\"o}hringer}, \& {Forman}}]{2001ApJ...554..261C}
{Churazov}, E., {Br{\"u}ggen}, M., {Kaiser}, C.~R., {B{\"o}hringer}, H., \&
  {Forman}, W. 2001, \apj, 554, 261

\bibitem[{{Collins} {et~al.}(2010){Collins}, {Xu}, {Norman}, {Li}, \&
  {Li}}]{co11}
{Collins}, D.~C., {Xu}, H., {Norman}, M.~L., {Li}, H., \& {Li}, S. 2010, \apjs,
  186, 308

\bibitem[{{Croston} {et~al.}(2018){Croston}, {Ineson}, \&
  {Hardcastle}}]{2018MNRAS.476.1614C}
{Croston}, J.~H., {Ineson}, J., \& {Hardcastle}, M.~J. 2018, \mnras, 476, 1614

\bibitem[{{de Gasperin} {et~al.}(2017){de Gasperin}, {Intema}, {Shimwell},
  {Brunetti}, {Br{\"u}ggen}, {En{\ss}lin}, {van Weeren}, {Bonafede}, \&
  {R{\"o}ttgering}}]{2017SciA....3E1634D}
{de Gasperin}, F., {Intema}, H.~T., {Shimwell}, T.~W., {et~al.} 2017, Science
  Advances, 3, e1701634

\bibitem[{{de Gasperin} {et~al.}(2012){de Gasperin}, {Orr{\'u}}, {Murgia},
  {Merloni}, {Falcke}, {Beck}, {Beswick}, {B{\^i}rzan}, {Bonafede},
  {Br{\"u}ggen}, {Brunetti}, {Chy{\.z}y}, {Conway}, {Croston}, {En{\ss}lin},
  {Ferrari}, {Heald}, {Heidenreich}, {Jackson}, {Macario}, {McKean}, {Miley},
  {Morganti}, {Offringa}, {Pizzo}, {Rafferty}, {R{\"o}ttgering}, {Shulevski},
  {Steinmetz}, {Tasse}, {van der Tol}, {van Driel}, {van Weeren}, {van
  Zwieten}, {Alexov}, {Anderson}, {Asgekar}, {Avruch}, {Bell}, {Bell},
  {Bentum}, {Bernardi}, {Best}, {Breitling}, {Broderick}, {Butcher}, {Ciardi},
  {Dettmar}, {Eisloeffel}, {Frieswijk}, {Gankema}, {Garrett}, {Gerbers},
  {Griessmeier}, {Gunst}, {Hassall}, {Hessels}, {Hoeft}, {Horneffer},
  {Karastergiou}, {K{\"o}hler}, {Koopman}, {Kuniyoshi}, {Kuper}, {Maat},
  {Mann}, {Mevius}, {Mulcahy}, {Munk}, {Nijboer}, {Noordam}, {Paas}, {Pandey},
  {Pandey}, {Polatidis}, {Reich}, {Schoenmakers}, {Sluman}, {Smirnov}, {Sobey},
  {Stappers}, {Swinbank}, {Tagger}, {Tang}, {van Bemmel}, {van Cappellen}, {van
  Duin}, {van Haarlem}, {van Leeuwen}, {Vermeulen}, {Vocks}, {White}, {Wise},
  {Wucknitz}, \& {Zarka}}]{2012A&A...547A..56D}
{de Gasperin}, F., {Orr{\'u}}, E., {Murgia}, M., {et~al.} 2012, \aap, 547, A56

\bibitem[{{Dom{\'\i}nguez-Fern{\'a}ndez}
  {et~al.}(2019){Dom{\'\i}nguez-Fern{\'a}ndez}, {Vazza}, {Br{\"u}ggen}, \&
  {Brunetti}}]{dom19}
{Dom{\'\i}nguez-Fern{\'a}ndez}, P., {Vazza}, F., {Br{\"u}ggen}, M., \&
  {Brunetti}, G. 2019, \mnras, 486, 623

\bibitem[{{Donnert} \& {Brunetti}(2014)}]{2014MNRAS.443.3564D}
{Donnert}, J. \& {Brunetti}, G. 2014, \mnras, 443, 3564

\bibitem[{{Donnert} {et~al.}(2018){Donnert}, {Vazza}, {Br{\"u}ggen}, \&
  {ZuHone}}]{review_dynamo}
{Donnert}, J., {Vazza}, F., {Br{\"u}ggen}, M., \& {ZuHone}, J. 2018, ArXiv
  e-prints [\eprint[arXiv]{1810.09783}]

\bibitem[{{Dubois} {et~al.}(2010){Dubois}, {Devriendt}, {Slyz}, \&
  {Teyssier}}]{dubois10}
{Dubois}, Y., {Devriendt}, J., {Slyz}, A., \& {Teyssier}, R. 2010, \mnras, 409,
  985

\bibitem[{{Dubois} \& {Teyssier}(2008)}]{2008A&A...482L..13D}
{Dubois}, Y. \& {Teyssier}, R. 2008, \aap, 482, L13

\bibitem[{{Ehlert} {et~al.}(2018){Ehlert}, {Weinberger}, {Pfrommer}, {Pakmor},
  \& {Springel}}]{2018MNRAS.481.2878E}
{Ehlert}, K., {Weinberger}, R., {Pfrommer}, C., {Pakmor}, R., \& {Springel}, V.
  2018, \mnras, 481, 2878

\bibitem[{{Ehlert} {et~al.}(2020){Ehlert}, {Weinberger}, {Pfrommer}, \&
  {Springel}}]{2020arXiv201113964E}
{Ehlert}, K., {Weinberger}, R., {Pfrommer}, C., \& {Springel}, V. 2020, arXiv
  e-prints, arXiv:2011.13964

\bibitem[{{Fabjan} {et~al.}(2010){Fabjan}, {Borgani}, {Tornatore}, {Saro},
  {Murante}, \& {Dolag}}]{2010MNRAS.401.1670F}
{Fabjan}, D., {Borgani}, S., {Tornatore}, L., {et~al.} 2010, \mnras, 401, 1670

\bibitem[{{Fanaroff} \& {Riley}(1974)}]{FR74}
{Fanaroff}, B.~L. \& {Riley}, J.~M. 1974, \mnras, 167, 31P

\bibitem[{{Furlanetto} \& {Loeb}(2001)}]{Furlanetto&Loeb..ApJ2001}
{Furlanetto}, S.~R. \& {Loeb}, A. 2001, \apj, 556, 619

\bibitem[{{Gabici} \& {Blasi}(2003)}]{gb03}
{Gabici}, S. \& {Blasi}, P. 2003, \apj, 583, 695

\bibitem[{{Gaspari} {et~al.}(2011){Gaspari}, {Brighenti}, {D'Ercole}, \&
  {Melioli}}]{gaspari11b}
{Gaspari}, M., {Brighenti}, F., {D'Ercole}, A., \& {Melioli}, C. 2011, \mnras,
  415, 1549

\bibitem[{{Gaspari} {et~al.}(2012){Gaspari}, {Ruszkowski}, \&
  {Sharma}}]{gaspari12}
{Gaspari}, M., {Ruszkowski}, M., \& {Sharma}, P. 2012, \apj, 746, 94

\bibitem[{{Gendron-Marsolais} {et~al.}(2021){Gendron-Marsolais}, {Hull},
  {Perley}, {Rudnick}, {Kraft}, {Hlavacek-Larrondo}, {Fabian}, {Roediger}, {van
  Weeren}, {Richard-Laferri{\`e}re}, {Golden-Marx}, {Arakawa}, \&
  {McBride}}]{2021arXiv210105305G}
{Gendron-Marsolais}, M.-L., {Hull}, C. L.~H., {Perley}, R., {et~al.} 2021,
  arXiv e-prints, arXiv:2101.05305

\bibitem[{{Girichidis} {et~al.}(2020){Girichidis}, {Pfrommer}, {Hanasz}, \&
  {Naab}}]{2020MNRAS.491..993G}
{Girichidis}, P., {Pfrommer}, C., {Hanasz}, M., \& {Naab}, T. 2020, \mnras,
  491, 993

\bibitem[{{Guo} {et~al.}(2014{\natexlab{a}}){Guo}, {Sironi}, \&
  {Narayan}}]{guo14a}
{Guo}, X., {Sironi}, L., \& {Narayan}, R. 2014{\natexlab{a}}, \apj, 794, 153

\bibitem[{{Guo} {et~al.}(2014{\natexlab{b}}){Guo}, {Sironi}, \&
  {Narayan}}]{guo14b}
{Guo}, X., {Sironi}, L., \& {Narayan}, R. 2014{\natexlab{b}}, \apj, 797, 47

\bibitem[{{Hahn} \& {Abel}(2011)}]{music}
{Hahn}, O. \& {Abel}, T. 2011, \mnras, 415, 2101

\bibitem[{{Hardcastle} \& {Croston}(2020)}]{hardcastlecroston}
{Hardcastle}, M.~J. \& {Croston}, J.~H. 2020, \nar, 88, 101539

\bibitem[{{Hardcastle} \& {Krause}(2014)}]{2014MNRAS.443.1482H}
{Hardcastle}, M.~J. \& {Krause}, M.~G.~H. 2014, \mnras, 443, 1482

\bibitem[{{Heinz} {et~al.}(2006){Heinz}, {Br{\"u}ggen}, {Young}, \&
  {Levesque}}]{2006MNRAS.373L..65H}
{Heinz}, S., {Br{\"u}ggen}, M., {Young}, A., \& {Levesque}, E. 2006, \mnras,
  373, L65

\bibitem[{{Hodgson} {et~al.}(2021){Hodgson}, {Bartalucci}, {Johnston-Hollitt},
  {McKinley}, {Vazza}, \& {Wittor}}]{Hodgson21a}
{Hodgson}, T., {Bartalucci}, I., {Johnston-Hollitt}, M., {et~al.} 2021, arXiv
  e-prints, arXiv:2103.06462

\bibitem[{{Iapichino} \& {Niemeyer}(2008)}]{in08}
{Iapichino}, L. \& {Niemeyer}, J.~C. 2008, \mnras, 388, 1089

\bibitem[{{Ignesti} {et~al.}(2020){Ignesti}, {Shimwell}, {Brunetti}, {Gitti},
  {Intema}, {van Weeren}, {Hardcastle}, {Clarke}, {Botteon}, {Di Gennaro},
  {Br{\"u}ggen}, {Browne}, {Mandal}, {R{\"o}ttgering}, {Cuciti}, {de Gasperin},
  {Cassano}, \& {Scaife}}]{2020A&A...643A.172I}
{Ignesti}, A., {Shimwell}, T., {Brunetti}, G., {et~al.} 2020, \aap, 643, A172

\bibitem[{{Jaffe}(1977)}]{1977ApJ...212....1J}
{Jaffe}, W.~J. 1977, \apj, 212, 1

\bibitem[{{Jaffe} \& {Perola}(1973)}]{1973A&A....26..423J}
{Jaffe}, W.~J. \& {Perola}, G.~C. 1973, \aap, 26, 423

\bibitem[{{Jones} {et~al.}(2017){Jones}, {Nolting}, {O'Neill}, \&
  {Mendygral}}]{2017PhPl...24d1402J}
{Jones}, T.~W., {Nolting}, C., {O'Neill}, B.~J., \& {Mendygral}, P.~J. 2017,
  Physics of Plasmas, 24, 041402

\bibitem[{{Kaiser} \& {Alexander}(1997)}]{1997MNRAS.286..215K}
{Kaiser}, C.~R. \& {Alexander}, P. 1997, \mnras, 286, 215

\bibitem[{{Kang}(2018)}]{2018JKAS...51..185K}
{Kang}, H. 2018, Journal of Korean Astronomical Society, 51, 185

\bibitem[{{Kang} \& {Jones}(2007)}]{kj07}
{Kang}, H. \& {Jones}, T.~W. 2007, Astroparticle Physics, 28, 232

\bibitem[{{Kang} \& {Ryu}(2011)}]{kr11}
{Kang}, H. \& {Ryu}, D. 2011, \apj, 734, 18

\bibitem[{{Kang} {et~al.}(2012){Kang}, {Ryu}, \& {Jones}}]{ka12}
{Kang}, H., {Ryu}, D., \& {Jones}, T.~W. 2012, \apj, 756, 97

\bibitem[{{Kardashev}(1962)}]{1962SvA.....6..317K}
{Kardashev}, N.~S. 1962, \sovast, 6, 317

\bibitem[{{Kim} {et~al.}(2011){Kim}, {Wise}, {Alvarez}, \&
  {Abel}}]{2011ApJ...738...54K}
{Kim}, J.-h., {Wise}, J.~H., {Alvarez}, M.~A., \& {Abel}, T. 2011, \apj, 738,
  54

\bibitem[{{Kronberg} {et~al.}(1999){Kronberg}, {Lesch}, \&
  {Hopp}}]{Kronberg..1999ApJ}
{Kronberg}, P.~P., {Lesch}, H., \& {Hopp}, U. 1999, \apj, 511, 56

\bibitem[{{Lau} {et~al.}(2009){Lau}, {Kravtsov}, \& {Nagai}}]{lau09}
{Lau}, E.~T., {Kravtsov}, A.~V., \& {Nagai}, D. 2009, \apj, 705, 1129

\bibitem[{{Locatelli} {et~al.}(2018){Locatelli}, {Vazza}, \&
  {Dom{\'\i}nguez-Fern{\'a}ndez}}]{2018Galax...6..128L}
{Locatelli}, N., {Vazza}, F., \& {Dom{\'\i}nguez-Fern{\'a}ndez}, P. 2018,
  Galaxies, 6, 128

\bibitem[{{Mandal} {et~al.}(2020){Mandal}, {Intema}, {van Weeren}, {Shimwell},
  {Botteon}, {Brunetti}, {de Gasperin}, {Br{\"u}ggen}, {Di Gennaro}, {Kraft},
  {R{\"o}ttgering}, {Hardcastle}, \& {Tasse}}]{2020A&A...634A...4M}
{Mandal}, S., {Intema}, H.~T., {van Weeren}, R.~J., {et~al.} 2020, \aap, 634,
  A4

\bibitem[{{Markevitch} {et~al.}(2005){Markevitch}, {Govoni}, {Brunetti}, \&
  {Jerius}}]{2005ApJ...627..733M}
{Markevitch}, M., {Govoni}, F., {Brunetti}, G., \& {Jerius}, D. 2005, \apj,
  627, 733

\bibitem[{{Massaglia} {et~al.}(2016){Massaglia}, {Bodo}, {Rossi}, {Capetti}, \&
  {Mignone}}]{2016A&A...596A..12M}
{Massaglia}, S., {Bodo}, G., {Rossi}, P., {Capetti}, S., \& {Mignone}, A. 2016,
  \aap, 596, A12

\bibitem[{{Massaglia} {et~al.}(2019){Massaglia}, {Bodo}, {Rossi}, {Capetti}, \&
  {Mignone}}]{2019A&A...621A.132M}
{Massaglia}, S., {Bodo}, G., {Rossi}, P., {Capetti}, S., \& {Mignone}, A. 2019,
  \aap, 621, A132

\bibitem[{{Mathews} \& {Brighenti}(2007)}]{mb07}
{Mathews}, W.~G. \& {Brighenti}, F. 2007, \apj, 660, 1137

\bibitem[{{McCarthy} {et~al.}(2010){McCarthy}, {Schaye}, {Ponman}, {Bower},
  {Booth}, {Dalla Vecchia}, {Crain}, {Springel}, {Theuns}, \&
  {Wiersma}}]{mcc2010}
{McCarthy}, I.~G., {Schaye}, J., {Ponman}, T.~J., {et~al.} 2010, \mnras, 406,
  822

\bibitem[{{McCourt} {et~al.}(2011){McCourt}, {Parrish}, {Sharma}, \&
  {Quataert}}]{mcc10}
{McCourt}, M., {Parrish}, I.~J., {Sharma}, P., \& {Quataert}, E. 2011, \mnras,
  413, 1295

\bibitem[{{Mendygral} {et~al.}(2012){Mendygral}, {Jones}, \&
  {Dolag}}]{2012ApJ...750..166M}
{Mendygral}, P.~J., {Jones}, T.~W., \& {Dolag}, K. 2012, \apj, 750, 166

\bibitem[{{Mignone} {et~al.}(2005){Mignone}, {Massaglia}, \&
  {Bodo}}]{2005SSRv..121...21M}
{Mignone}, A., {Massaglia}, S., \& {Bodo}, G. 2005, \ssr, 121, 21

\bibitem[{{Mignone} {et~al.}(2010){Mignone}, {Rossi}, {Bodo}, {Ferrari}, \&
  {Massaglia}}]{2010MNRAS.402....7M}
{Mignone}, A., {Rossi}, P., {Bodo}, G., {Ferrari}, A., \& {Massaglia}, S. 2010,
  \mnras, 402, 7

\bibitem[{{Mingo} {et~al.}(2019){Mingo}, {Croston}, {Hardcastle}, {Best},
  {Duncan}, {Morganti}, {Rottgering}, {Sabater}, {Shimwell}, {Williams},
  {Brienza}, {Gurkan}, {Mahatma}, {Morabito}, {Prandoni}, {Bondi}, {Ineson}, \&
  {Mooney}}]{mingo19}
{Mingo}, B., {Croston}, J.~H., {Hardcastle}, M.~J., {et~al.} 2019, \mnras, 488,
  2701

\bibitem[{{Miniati}(2014)}]{miniati14}
{Miniati}, F. 2014, \apj, 782, 21

\bibitem[{{Nolting} {et~al.}(2019{\natexlab{a}}){Nolting}, {Jones}, {O'Neill},
  \& {Mendygral}}]{nolting19a}
{Nolting}, C., {Jones}, T.~W., {O'Neill}, B.~J., \& {Mendygral}, P.~J.
  2019{\natexlab{a}}, \apj, 876, 154

\bibitem[{{Nolting} {et~al.}(2019{\natexlab{b}}){Nolting}, {Jones}, {O'Neill},
  \& {Mendygral}}]{nolting19b}
{Nolting}, C., {Jones}, T.~W., {O'Neill}, B.~J., \& {Mendygral}, P.~J.
  2019{\natexlab{b}}, \apj, 885, 80

\bibitem[{{Norman} \& {Bryan}(1999)}]{bn99}
{Norman}, M.~L. \& {Bryan}, G.~L. 1999, in Lecture Notes in Physics, Berlin
  Springer Verlag, Vol. 530, The Radio Galaxy Messier 87, ed. {H.-J.~R{\"o}ser
  \& K.~Meisenheimer}, 106--+

\bibitem[{{Norman} {et~al.}(1982){Norman}, {Winkler}, {Smarr}, \&
  {Smith}}]{1982A&A...113..285N}
{Norman}, M.~L., {Winkler}, K. H.~A., {Smarr}, L., \& {Smith}, M.~D. 1982,
  \aap, 113, 285

\bibitem[{{Ogrodnik} {et~al.}(2020){Ogrodnik}, {Hanasz}, \&
  {W{\'o}lta{\'n}ski}}]{2020arXiv200906941O}
{Ogrodnik}, M., {Hanasz}, M., \& {W{\'o}lta{\'n}ski}, D. 2020, arXiv e-prints,
  arXiv:2009.06941

\bibitem[{{O'Neill} \& {Jones}(2010)}]{oj10}
{O'Neill}, S.~M. \& {Jones}, T.~W. 2010, \apj, 710, 180

\bibitem[{{Owen} {et~al.}(2014){Owen}, {Rudnick}, {Eilek}, {Rau}, {Bhatnagar},
  \& {Kogan}}]{2014ApJ...794...24O}
{Owen}, F.~N., {Rudnick}, L., {Eilek}, J., {et~al.} 2014, \apj, 794, 24

\bibitem[{{Perucho} \& {Mart{\'\i}}(2007)}]{2007MNRAS.382..526P}
{Perucho}, M. \& {Mart{\'\i}}, J.~M. 2007, \mnras, 382, 526

\bibitem[{{Pinzke} {et~al.}(2013){Pinzke}, {Oh}, \&
  {Pfrommer}}]{2013MNRAS.435.1061P}
{Pinzke}, A., {Oh}, S.~P., \& {Pfrommer}, C. 2013, \mnras, 435, 1061

\bibitem[{{Planck Collaboration} {et~al.}(2016){Planck Collaboration}, {Ade},
  {Aghanim}, {Arnaud}, {Ashdown}, {Aumont}, {Baccigalupi}, {Banday},
  {Barreiro}, {Bartlett}, \& et~al.}]{2016A&A...594A..13P}
{Planck Collaboration}, {Ade}, P.~A.~R., {Aghanim}, N., {et~al.} 2016, \aap,
  594, A13

\bibitem[{{Puchwein} {et~al.}(2008){Puchwein}, {Sijacki}, \&
  {Springel}}]{2008ApJ...687L..53P}
{Puchwein}, E., {Sijacki}, D., \& {Springel}, V. 2008, \apjl, 687, L53

\bibitem[{{Rajpurohit} {et~al.}(2020){Rajpurohit}, {Hoeft}, {Vazza}, {Rudnick},
  {van Weeren}, {Wittor}, {Drabent}, {Brienza}, {Bonnassieux}, {Locatelli},
  {Kale}, \& {Dumba}}]{rajp20}
{Rajpurohit}, K., {Hoeft}, M., {Vazza}, F., {et~al.} 2020, \aap, 636, A30

\bibitem[{{Ramatsoku} {et~al.}(2020){Ramatsoku}, {Murgia}, {Vacca}, {Serra},
  {Makhathini}, {Govoni}, {Smirnov}, {Andati}, {de Blok}, {J{\'o}zsa},
  {Kamphuis}, {Kleiner}, {Maccagni}, {Moln{\'a}r}, {Ramaila}, {Thorat}, \&
  {White}}]{2020A&A...636L...1R}
{Ramatsoku}, M., {Murgia}, M., {Vacca}, V., {et~al.} 2020, \aap, 636, L1

\bibitem[{{Rasia} {et~al.}(2015){Rasia}, {Borgani}, {Murante}, {Planelles},
  {Beck}, {Biffi}, {Ragone-Figueroa}, {Granato}, {Steinborn}, \&
  {Dolag}}]{2015ApJ...813L..17R}
{Rasia}, E., {Borgani}, S., {Murante}, G., {et~al.} 2015, \apjl, 813, L17

\bibitem[{{Ricarte} {et~al.}(2019){Ricarte}, {Tremmel}, {Natarajan}, \&
  {Quinn}}]{2019MNRAS.489..802R}
{Ricarte}, A., {Tremmel}, M., {Natarajan}, P., \& {Quinn}, T. 2019, \mnras,
  489, 802

\bibitem[{{Rincon} {et~al.}(2016){Rincon}, {Califano}, {Schekochihin}, \&
  {Valentini}}]{2016PNAS..113.3950R}
{Rincon}, F., {Califano}, F., {Schekochihin}, A.~A., \& {Valentini}, F. 2016,
  Proceedings of the National Academy of Science, 113, 3950

\bibitem[{{Riquelme} \& {Spitkovsky}(2011)}]{2011ApJ...733...63R}
{Riquelme}, M.~A. \& {Spitkovsky}, A. 2011, \apj, 733, 63

\bibitem[{{Ruszkowski} {et~al.}(2007){Ruszkowski}, {En{\ss}lin}, {Br{\"u}ggen},
  {Heinz}, \& {Pfrommer}}]{2007MNRAS.378..662R}
{Ruszkowski}, M., {En{\ss}lin}, T.~A., {Br{\"u}ggen}, M., {Heinz}, S., \&
  {Pfrommer}, C. 2007, \mnras, 378, 662

\bibitem[{{Ruszkowski} {et~al.}(2011){Ruszkowski}, {Lee}, {Br{\"u}ggen},
  {Parrish}, \& {Oh}}]{ruszkowski11}
{Ruszkowski}, M., {Lee}, D., {Br{\"u}ggen}, M., {Parrish}, I., \& {Oh}, S.~P.
  2011, \apj, 740, 81

\bibitem[{{Ryu} {et~al.}(2003){Ryu}, {Kang}, {Hallman}, \& {Jones}}]{ry03}
{Ryu}, D., {Kang}, H., {Hallman}, E., \& {Jones}, T.~W. 2003, \apj, 593, 599

\bibitem[{{Sabater} {et~al.}(2019){Sabater}, {Best}, {Hardcastle}, {Shimwell},
  {Tasse}, {Williams}, {Br{\"u}ggen}, {Cochrane}, {Croston}, {de Gasperin},
  {Duncan}, {G{\"u}rkan}, {Mechev}, {Morabito}, {Prandoni}, {R{\"o}ttgering},
  {Smith}, {Harwood}, {Mingo}, {Mooney}, \& {Saxena}}]{2019A&A...622A..17S}
{Sabater}, J., {Best}, P.~N., {Hardcastle}, M.~J., {et~al.} 2019, \aap, 622,
  A17

\bibitem[{{Sarazin}(1999)}]{sa99}
{Sarazin}, C.~L. 1999, \apj, 520, 529

\bibitem[{{Schekochihin} {et~al.}(2004){Schekochihin}, {Cowley}, {Taylor},
  {Maron}, \& {McWilliams}}]{2004ApJ...612..276S}
{Schekochihin}, A.~A., {Cowley}, S.~C., {Taylor}, S.~F., {Maron}, J.~L., \&
  {McWilliams}, J.~C. 2004, \apj, 612, 276

\bibitem[{{Schlickeiser} {et~al.}(1987){Schlickeiser}, {Sievers}, \&
  {Thiemann}}]{1987A&A...182...21S}
{Schlickeiser}, R., {Sievers}, A., \& {Thiemann}, H. 1987, \aap, 182, 21

\bibitem[{{Sijacki} {et~al.}(2007){Sijacki}, {Springel}, {Di Matteo}, \&
  {Hernquist}}]{2007MNRAS.380..877S}
{Sijacki}, D., {Springel}, V., {Di Matteo}, T., \& {Hernquist}, L. 2007,
  \mnras, 380, 877

\bibitem[{{Stuardi} {et~al.}(2019){Stuardi}, {Bonafede}, {Wittor}, {Vazza},
  {Botteon}, {Locatelli}, {Dallacasa}, {Golovich}, {Hoeft}, {van Weeren},
  {Br{\"u}ggen}, \& {de Gasperin}}]{2019MNRAS.489.3905S}
{Stuardi}, C., {Bonafede}, A., {Wittor}, D., {et~al.} 2019, \mnras, 489, 3905

\bibitem[{{Teyssier} {et~al.}(2011){Teyssier}, {Moore}, {Martizzi}, {Dubois},
  \& {Mayer}}]{teyssier11}
{Teyssier}, R., {Moore}, B., {Martizzi}, D., {Dubois}, Y., \& {Mayer}, L. 2011,
  \mnras, 414, 195

\bibitem[{{Thomas} {et~al.}(2020){Thomas}, {Dave}, {Jarvis}, \&
  {Angles-Alcazar}}]{2020arXiv201011225T}
{Thomas}, N., {Dave}, R., {Jarvis}, M.~J., \& {Angles-Alcazar}, D. 2020, arXiv
  e-prints, arXiv:2010.11225

\bibitem[{{Tremblay} {et~al.}(2016){Tremblay}, {Oonk}, {Combes}, {Salom{\'e}},
  {O'Dea}, {Baum}, {Voit}, {Donahue}, {McNamara}, {Davis}, {McDonald}, {Edge},
  {Clarke}, {Galv{\'a}n-Madrid}, {Bremer}, {Edwards}, {Fabian}, {Hamer}, {Li},
  {Maury}, {Russell}, {Quillen}, {Urry}, {Sanders}, \&
  {Wise}}]{2016Natur.534..218T}
{Tremblay}, G.~R., {Oonk}, J.~B.~R., {Combes}, F., {et~al.} 2016, \nat, 534,
  218

\bibitem[{{Tremmel} {et~al.}(2017){Tremmel}, {Karcher}, {Governato},
  {Volonteri}, {Quinn}, {Pontzen}, {Anderson}, \&
  {Bellovary}}]{2017MNRAS.470.1121T}
{Tremmel}, M., {Karcher}, M., {Governato}, F., {et~al.} 2017, \mnras, 470, 1121

\bibitem[{{Turner} \& {Shabala}(2015)}]{2015ApJ...806...59T}
{Turner}, R.~J. \& {Shabala}, S.~S. 2015, \apj, 806, 59

\bibitem[{{Uchiyama} {et~al.}(2007){Uchiyama}, {Aharonian}, {Tanaka},
  {Takahashi}, \& {Maeda}}]{2007Natur.449..576U}
{Uchiyama}, Y., {Aharonian}, F.~A., {Tanaka}, T., {Takahashi}, T., \& {Maeda},
  Y. 2007, \nat, 449, 576

\bibitem[{{Vachaspati}(2020)}]{2020arXiv201010525V}
{Vachaspati}, T. 2020, arXiv e-prints, arXiv:2010.10525

\bibitem[{{van Weeren} {et~al.}(2017){van Weeren}, {Andrade-Santos}, {Dawson},
  {Golovich}, {Lal}, {Kang}, {Ryu}, {Br{\`\i}ggen}, {Ogrean}, {Forman},
  {Jones}, {Placco}, {Santucci}, {Wittman}, {Jee}, {Kraft}, {Sobral}, {Stroe},
  \& {Fogarty}}]{2017NatAs...1E...5V}
{van Weeren}, R.~J., {Andrade-Santos}, F., {Dawson}, W.~A., {et~al.} 2017,
  Nature Astronomy, 1, 0005

\bibitem[{{van Weeren} {et~al.}(2019){van Weeren}, {de Gasperin}, {Akamatsu},
  {Br{\"u}ggen}, {Feretti}, {Kang}, {Stroe}, \&
  {Zandanel}}]{2019SSRv..215...16V}
{van Weeren}, R.~J., {de Gasperin}, F., {Akamatsu}, H., {et~al.} 2019, \ssr,
  215, 16

\bibitem[{{Vardoulaki} {et~al.}(2020){Vardoulaki}, {Jim{\'e}nez Andrade},
  {Delvecchio}, {Smol{\v{c}}i{\'c}}, {Schinnerer}, {Sargent}, {Gozaliasl},
  {Finoguenov}, {Bondi}, {Zamorani}, {Badescu}, {Leslie}, {Ceraj},
  {Tisani{\'c}}, {Karim}, {Magnelli}, {Bertoldi}, {Romano-Diaz}, \&
  {Harrington}}]{vard20}
{Vardoulaki}, E., {Jim{\'e}nez Andrade}, E.~F., {Delvecchio}, I., {et~al.}
  2020, arXiv e-prints, arXiv:2009.10721

\bibitem[{{Vardoulaki} {et~al.}(2019){Vardoulaki}, {Jim{\'e}nez Andrade},
  {Karim}, {Novak}, {Leslie}, {Tisani{\'c}}, {Smol{\v{c}}i{\'c}}, {Schinnerer},
  {Sargent}, {Bondi}, {Zamorani}, {Magnelli}, {Bertoldi}, {Herrera Ruiz},
  {Mooley}, {Delhaize}, {Myers}, {Marchesi}, {Koekemoer}, {Gozaliasl},
  {Finoguenov}, {Middleberg}, \& {Ciliegi}}]{vard19}
{Vardoulaki}, E., {Jim{\'e}nez Andrade}, E.~F., {Karim}, A., {et~al.} 2019,
  \aap, 627, A142

\bibitem[{Vazza {et~al.}(2017)Vazza, Brueggen, Gheller, Hackstein, Wittor, \&
  Hinz}]{va17cqg}
Vazza, F., Brueggen, M., Gheller, C., {et~al.} 2017, Classical and Quantum
  Gravity

\bibitem[{{Vazza} \& {Br{\"u}ggen}(2014)}]{va14relics}
{Vazza}, F. \& {Br{\"u}ggen}, M. 2014, \mnras, 437, 2291

\bibitem[{{Vazza} {et~al.}(2013){Vazza}, {Br{\"u}ggen}, \&
  {Gheller}}]{va13feedback}
{Vazza}, F., {Br{\"u}ggen}, M., \& {Gheller}, C. 2013, \mnras, 428, 2366

\bibitem[{{Vazza} {et~al.}(2018){Vazza}, {Brunetti}, {Br{\"u}ggen}, \&
  {Bonafede}}]{va18mhd}
{Vazza}, F., {Brunetti}, G., {Br{\"u}ggen}, M., \& {Bonafede}, A. 2018, \mnras,
  474, 1672

\bibitem[{{Vazza} {et~al.}(2009){Vazza}, {Brunetti}, \& {Gheller}}]{va09shocks}
{Vazza}, F., {Brunetti}, G., \& {Gheller}, C. 2009, \mnras, 395, 1333

\bibitem[{{Vazza} {et~al.}(2011){Vazza}, {Brunetti}, {Gheller}, {Brunino}, \&
  {Br{\"u}ggen}}]{va11turbo}
{Vazza}, F., {Brunetti}, G., {Gheller}, C., {Brunino}, R., \& {Br{\"u}ggen}, M.
  2011, \aap, 529, A17+

\bibitem[{{Vazza} {et~al.}(2015){Vazza}, {Eckert}, {Br{\"u}ggen}, \&
  {Huber}}]{va15relics}
{Vazza}, F., {Eckert}, D., {Br{\"u}ggen}, M., \& {Huber}, B. 2015, \mnras, 451,
  2198

\bibitem[{{Vazza} {et~al.}(2010){Vazza}, {Gheller}, \&
  {Brunetti}}]{va10tracers}
{Vazza}, F., {Gheller}, C., \& {Brunetti}, G. 2010, \aap, 513, A32+

\bibitem[{{Vazza} {et~al.}(2017){Vazza}, {Jones}, {Br{\"u}ggen}, {Brunetti},
  {Gheller}, {Porter}, \& {Ryu}}]{va17turb}
{Vazza}, F., {Jones}, T.~W., {Br{\"u}ggen}, M., {et~al.} 2017, \mnras, 464, 210

\bibitem[{{Voit} {et~al.}(2020){Voit}, {Bryan}, {Prasad}, {Frisbie}, {Li},
  {Donahue}, {O'Shea}, {Sun}, \& {Werner}}]{2020ApJ...899...70V}
{Voit}, G.~M., {Bryan}, G.~L., {Prasad}, D., {et~al.} 2020, \apj, 899, 70

\bibitem[{{V{\"o}lk} \& {Atoyan}(1999)}]{volk99}
{V{\"o}lk}, H.~J. \& {Atoyan}, A.~M. 1999, Astroparticle Physics, 11, 73

\bibitem[{{Wilber} {et~al.}(2018){Wilber}, {Br{\"u}ggen}, {Bonafede}, {Savini},
  {Shimwell}, {van Weeren}, {Rafferty}, {Mechev}, {Intema}, {Andrade-Santos},
  {Clarke}, {Mahony}, {Morganti}, {Prand oni}, {Brunetti}, {R{\"o}ttgering},
  {Mandal}, {de Gasperin}, \& {Hoeft}}]{2018MNRAS.473.3536W}
{Wilber}, A., {Br{\"u}ggen}, M., {Bonafede}, A., {et~al.} 2018, \mnras, 473,
  3536

\bibitem[{Wittor(2017)}]{wiPHD}
Wittor, D. 2017, PhD thesis, Universit\"at Hamburg, Von-Melle-Park 3, 20146
  Hamburg

\bibitem[{{Wittor} \& {Gaspari}(2020)}]{2020MNRAS.498.4983W}
{Wittor}, D. \& {Gaspari}, M. 2020, \mnras, 498, 4983

\bibitem[{{Wittor} {et~al.}(2017{\natexlab{a}}){Wittor}, {Jones}, {Vazza}, \&
  {Br{\"u}ggen}}]{wi17b}
{Wittor}, D., {Jones}, T., {Vazza}, F., \& {Br{\"u}ggen}, M.
  2017{\natexlab{a}}, \mnras, 471, 3212

\bibitem[{{Wittor} {et~al.}(2016){Wittor}, {Vazza}, \& {Br{\"u}ggen}}]{wi16}
{Wittor}, D., {Vazza}, F., \& {Br{\"u}ggen}, M. 2016, Galaxies, 4, 71

\bibitem[{{Wittor} {et~al.}(2017{\natexlab{b}}){Wittor}, {Vazza}, \&
  {Br{\"u}ggen}}]{wi17}
{Wittor}, D., {Vazza}, F., \& {Br{\"u}ggen}, M. 2017{\natexlab{b}}, \mnras,
  464, 4448

\bibitem[{{Wittor} {et~al.}(2020){Wittor}, {Vazza}, {Ryu}, \&
  {Kang}}]{wittor20}
{Wittor}, D., {Vazza}, F., {Ryu}, D., \& {Kang}, H. 2020, \mnras, 495, L112

\bibitem[{{Xie} {et~al.}(2020){Xie}, {van Weeren}, {Lovisari},
  {Andrade-Santos}, {Botteon}, {Br{\"u}ggen}, {Bulbul}, {Churazov}, {Clarke},
  {Forman}, {Intema}, {Jones}, {Kraft}, {Lal}, {Mroczkowski}, \&
  {Zitrin}}]{Xie2020}
{Xie}, C., {van Weeren}, R.~J., {Lovisari}, L., {et~al.} 2020, arXiv e-prints,
  arXiv:2001.04725

\bibitem[{{Xu} {et~al.}(2009){Xu}, {Li}, {Collins}, {Li}, \& {Norman}}]{xu09}
{Xu}, H., {Li}, H., {Collins}, D.~C., {Li}, S., \& {Norman}, M.~L. 2009, \apjl,
  698, L14

\bibitem[{{Xu} {et~al.}(2011){Xu}, {Li}, {Collins}, {Li}, \&
  {Norman}}]{2011ApJ...739...77X}
{Xu}, H., {Li}, H., {Collins}, D.~C., {Li}, S., \& {Norman}, M.~L. 2011, \apj,
  739, 77

\bibitem[{{Xu} {et~al.}(2020){Xu}, {Spitkovsky}, \&
  {Caprioli}}]{2020ApJ...897L..41X}
{Xu}, R., {Spitkovsky}, A., \& {Caprioli}, D. 2020, \apjl, 897, L41

\bibitem[{{Xu} \& {Zhang}(2017)}]{2017ApJ...846L..28X}
{Xu}, S. \& {Zhang}, B. 2017, \apjl, 846, L28

\bibitem[{{Yang} {et~al.}(2019){Yang}, {Gaspari}, \&
  {Marlow}}]{2019ApJ...871....6Y}
{Yang}, H. Y.~K., {Gaspari}, M., \& {Marlow}, C. 2019, \apj, 871, 6

\bibitem[{{Yang} {et~al.}(2012){Yang}, {Sutter}, \&
  {Ricker}}]{2012MNRAS.427.1614Y}
{Yang}, H.-Y.~K., {Sutter}, P.~M., \& {Ricker}, P.~M. 2012, \mnras, 427, 1614

\bibitem[{{ZuHone} {et~al.}(2010){ZuHone}, {Markevitch}, \&
  {Johnson}}]{2010ApJ...717..908Z}
{ZuHone}, J.~A., {Markevitch}, M., \& {Johnson}, R.~E. 2010, \apj, 717, 908

\end{thebibliography}
\appendix

\section{Visual impression of the two runs.}

\begin{figure*}
\begin{center}
\includegraphics[width=0.245\textwidth]{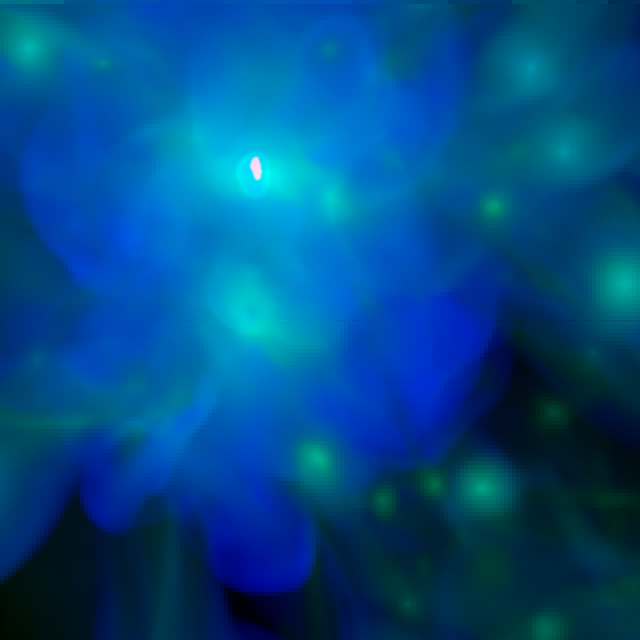}
\includegraphics[width=0.245\textwidth]{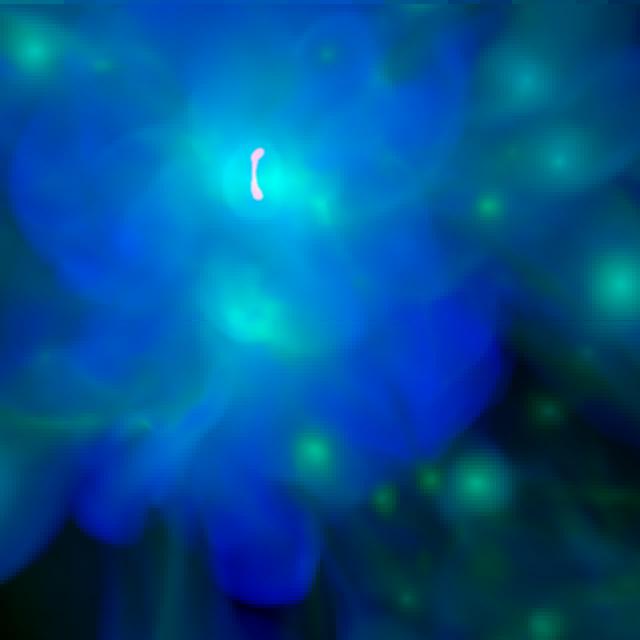}
\includegraphics[width=0.245\textwidth]{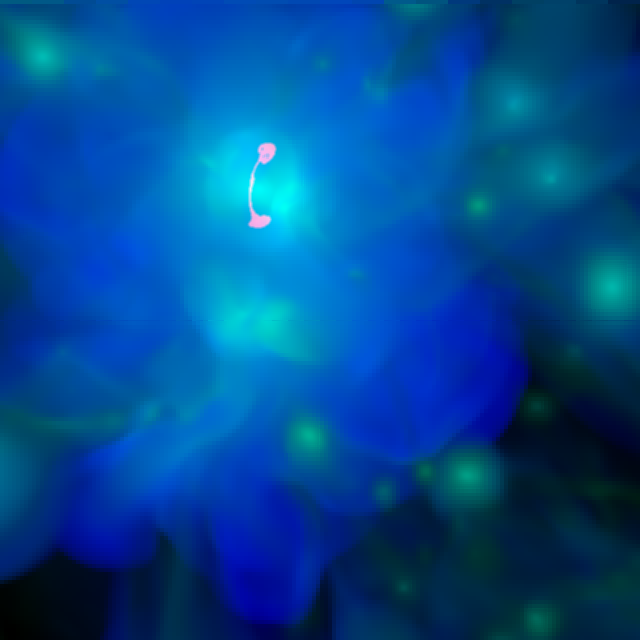}
\includegraphics[width=0.245\textwidth]{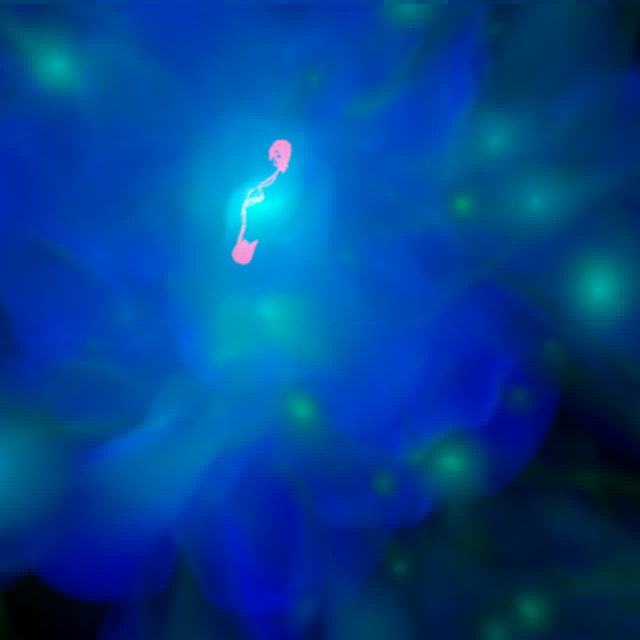}
\includegraphics[width=0.245\textwidth]{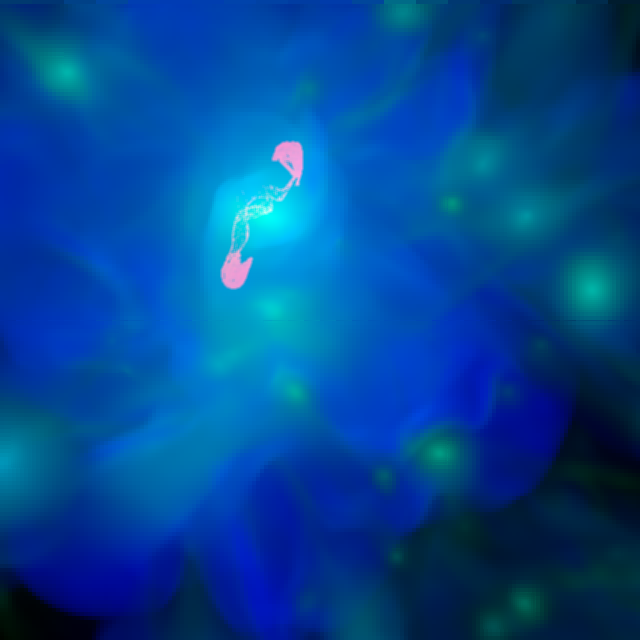}
\includegraphics[width=0.245\textwidth]{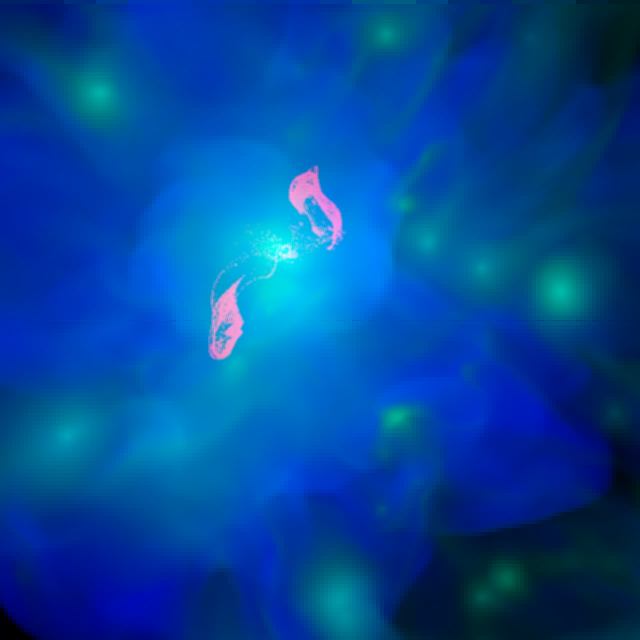}
\includegraphics[width=0.245\textwidth]{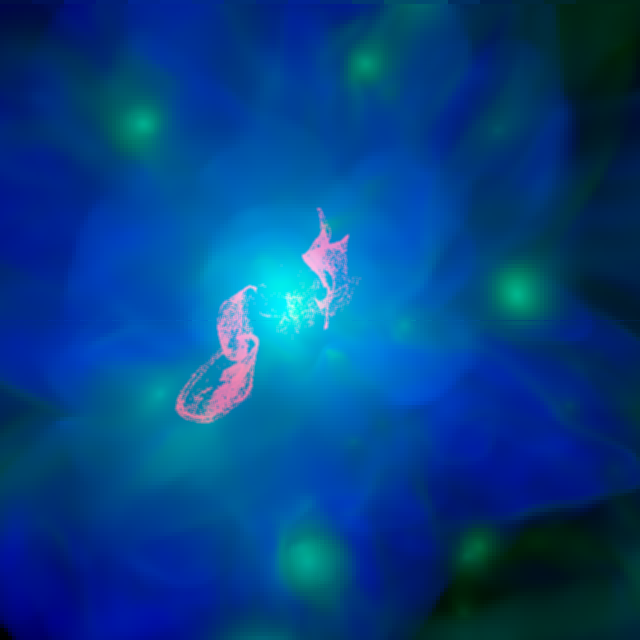}
\includegraphics[width=0.245\textwidth]{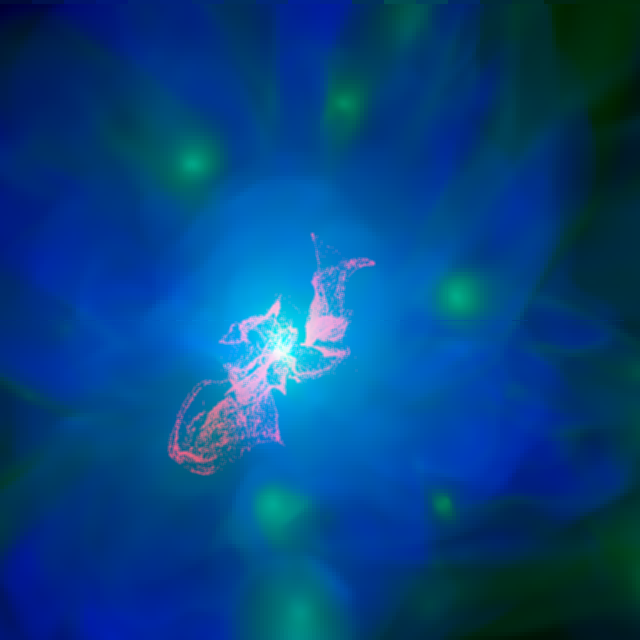}
\end{center}
\caption{Rendering of the evolution of radio emission (pink colours), X-ray emission (green) and gas temperature (blue) for roughly equally spaced timesteps from $z=0.49$ to $z=0.1$ in our Run2 simulation. Each image has a side of 5.5 Mpc (comoving).}
\label{fig:movie1}
\end{figure*}

\begin{figure*}
\begin{center}
\includegraphics[width=0.245\textwidth]{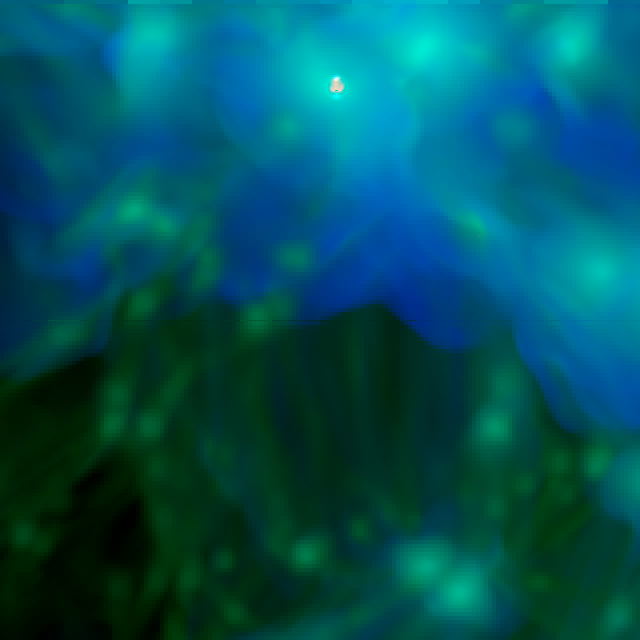}
\includegraphics[width=0.245\textwidth]{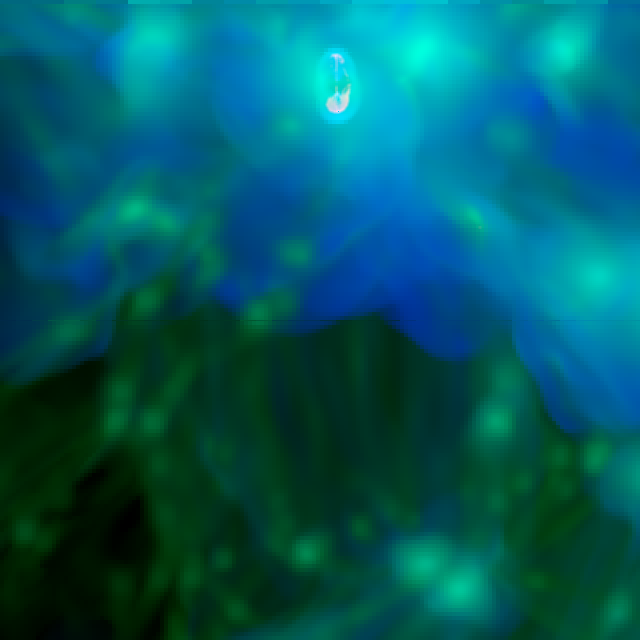}
\includegraphics[width=0.245\textwidth]{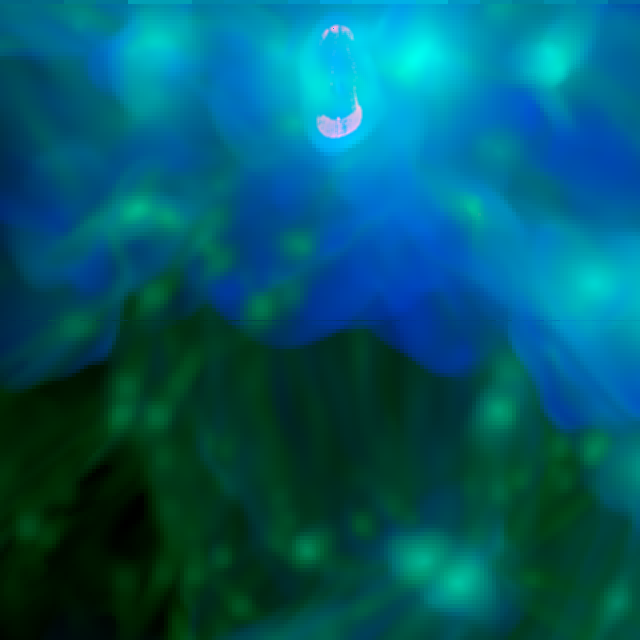}
\includegraphics[width=0.245\textwidth]{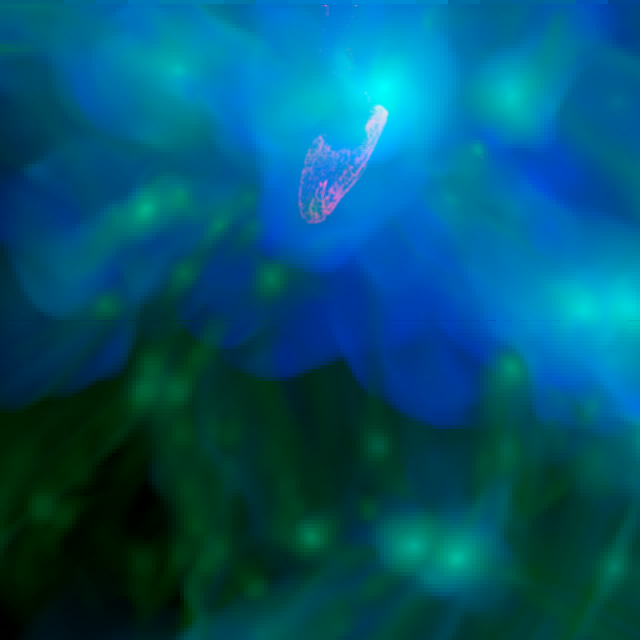}
\includegraphics[width=0.245\textwidth]{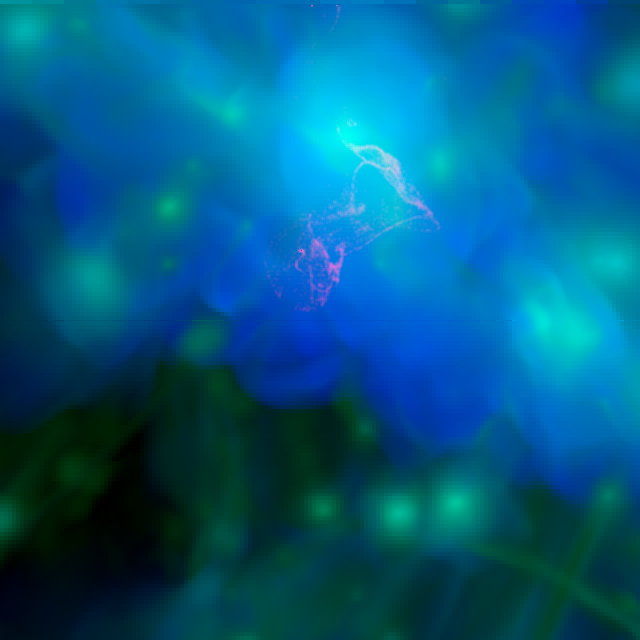}
\includegraphics[width=0.245\textwidth]{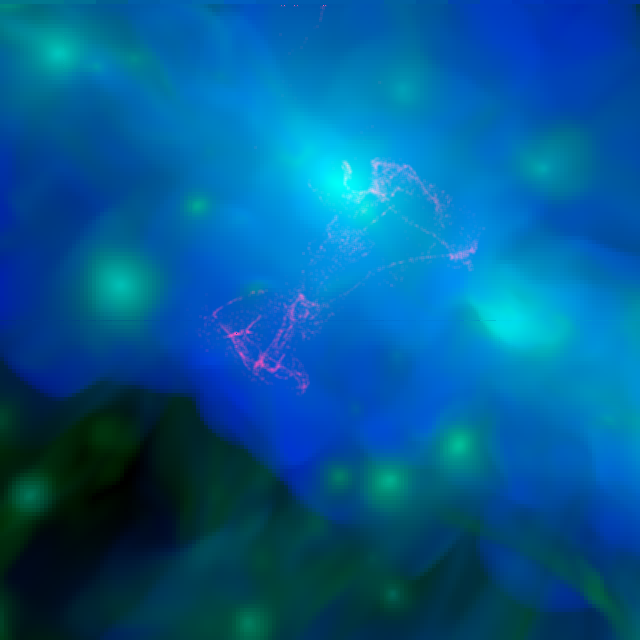}
\includegraphics[width=0.245\textwidth]{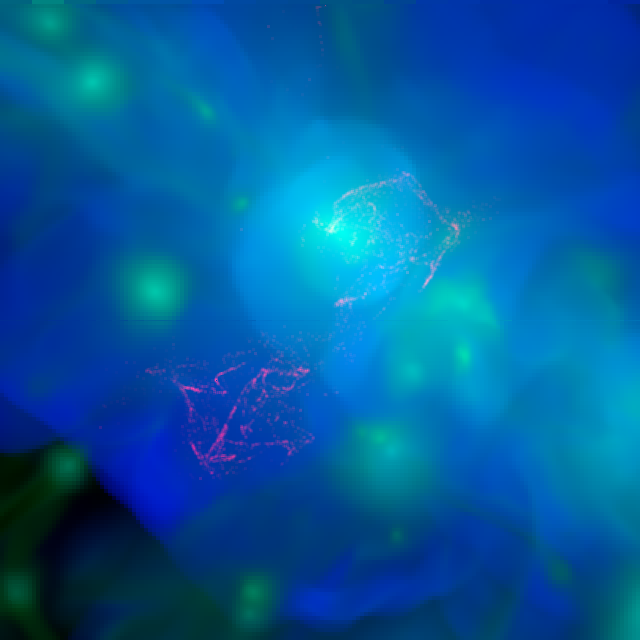}
\includegraphics[width=0.245\textwidth]{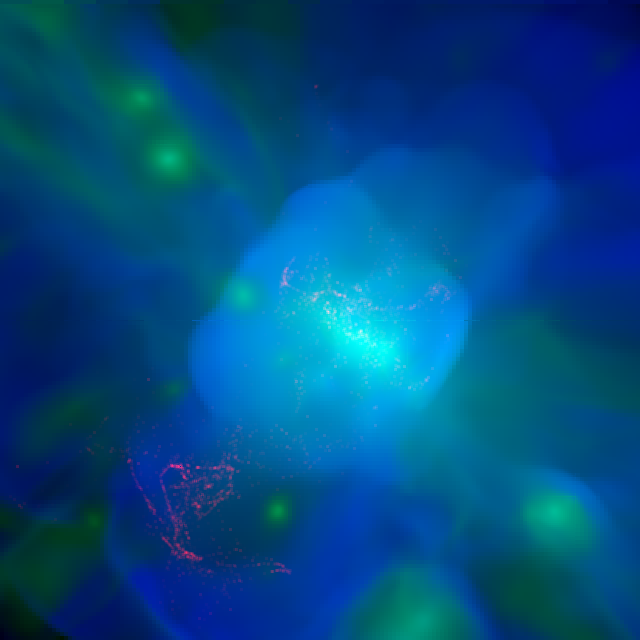}
\end{center}
\caption{Rendering of the evolution of radio emission (pink colours), X-ray emission (green) and gas temperature (blue) for roughly equally spaced timesteps from $z=0.99$ to $z=0.1$ in our Run1 simulation. Each image has a side of 5.5 Mpc (comoving).}
\label{fig:movie2}
\end{figure*}

Figure \ref{fig:movie1} and Figure \ref{fig:movie2} give the visual impression of the refined evolution of Run1 and Run2, using RGB colour coding of different information: radio emission (pink colours), X-ray emission (green) and gas temperature (blue), in all cases applying logarithmic stretching and 
adjusting the range of values of the colour bar to the changing maximum and minimum value of the fields on display,  in order to enhances the visibility of structures. The full movies of the two simulations can be found at https://vimeo.com/490397871 and https://vimeo.com/490399056 . 

\end{document}